%% file: master_ROPP.tex
\documentclass[a4paper]{elsarticle}
\usepackage[utf8]{inputenc}
\usepackage[english]{translator}
\usepackage{graphicx}  
\usepackage{floatrow}
\usepackage{nicefrac}
\usepackage{tabularx}
\usepackage{sidecap}
\usepackage{wrapfig}
\usepackage{lmodern}
\usepackage{url}
\usepackage{hyphenat} 
\usepackage{floatflt, verbatim} 
\usepackage{xspace}             
\usepackage{subfigure}          
\usepackage{multirow}           
\usepackage{pdflscape}          

\usepackage{amsmath, amssymb} 
\usepackage{accents}
\usepackage{xfrac}  

\usepackage{rotating}
\usepackage{afterpage}
\usepackage{upgreek}
\usepackage{wrapfig}
\usepackage{sidecap}

\newcommand{\glq}{\textquotesingle}
\newcommand{\grq}{\textquotesingle}

\newcommand{\degree}{\ensuremath{^\circ}}

\usepackage[plainpages=false,pdfpagelabels,breaklinks=true]{hyperref}

\begin{document}

\begin{frontmatter}
\title{A review of advances in pixel detectors for experiments \\ with high rate and radiation}

\author[]{Maurice~Garcia-Sciveres$^1$ and Norbert Wermes$^2$}
\address{$^1$Lawrence Berkeley National Laboratory, Berkeley, U.S.}
\address{$^2$University of Bonn, Bonn, Germany}
\ead{mgs@lbl.gov, wermes@uni-bonn.de}


\begin{abstract}
The Large Hadron Collider (LHC) experiments ATLAS and CMS have established hybrid pixel detectors
as the instrument of choice for particle tracking and vertexing in high rate and radiation environments,
as they operate close to the LHC interaction points. With the High Luminosity-LHC upgrade now in sight, for which the
tracking detectors will be completely replaced, new generations of pixel detectors are being devised.
They have to address enormous challenges in terms of data throughput and radiation levels, ionizing and non-ionizing, that harm the sensing and readout parts of pixel detectors alike.
Advances in microelectronics and microprocessing technologies now enable large scale detector designs with
unprecedented performance in measurement precision (space and time), radiation hard sensors and readout chips, hybridization techniques, lightweight supports, and fully monolithic approaches to meet these challenges.
This paper reviews the world-wide effort on these developments.
\end{abstract}


\end{frontmatter}

\tableofcontents

\include{introduction}
\include{resolution}
\include{hybridpixels}
\include{fechip}
\include{supports}
\include{cmos}
\include{conclusions}
\include{acknowledge}


\begingroup
\raggedright 
   \bibliographystyle{unsrt}
    \bibliography{master_ROPP} 
\endgroup

\end{document}

%% file: introduction.tex
\section{Pixel detectors at the heart of particle physics experiments: demands, challenges and concepts}\label{sec:intro}



This article reviews recent advances and future directions in pixel detectors to measure high energy charged particle trajectories in high rate and radiation experiments. A general treatment of pixel detectors can be found in~\cite{pixel_book}. 
This review focuses on the on-detector elements, and does not include data acquisition or downstream data processing such 
as track reconstruction. As there are recent reviews on radiation damage of silicon sensors~\cite{Moll-radecs} and 
on mechanics and cooling of such detectors~\cite{Viehhauser:2015} these topics receive an abbreviated treatment here. 
The current state of the art is represented by the detectors in operation or under construction at the Large Hadron Collider at CERN, in the ATLAS, CMS, LHC-b, and NA62 experiments. 

The ATLAS experiment installed a 3-layer hybrid pixel detector in 2007~\cite{ATLAS-pixel-paper_2008}
and an additional layer at lower radius, inserted within the original detector envelope and therefore called Insertable B-Layer (IBL),
in 2014~\cite{IBL-Hugging:2010bj}. The IBL pioneered the use of 3D silicon sensors (in a limited acceptance range) and introduced a new readout integrated circuit (ROIC or readout chip) called FE-I4~\cite{FE-I4} with several of the features needed for future high rate detectors.

The CMS experiment  installed a 3-layer hybrid pixel detector in 2008~\cite{CMS-pixel-paper_2008}, and has replaced it with a new,
4-layer detector in 2017~\cite{CMS-TDR-pixel-upgrade}. The new detector has lower inner radius,
significantly lower mass which improves precision, and higher data rate capability needed to cope with increased accelerator luminosity.
The upgrade baseline ROIC was redesigned for the outer 3 layers,
replacing analog signal readout with on-chip ADCs and digital readout at higher rate~\cite{CMS-Phase1-ROIC}.
A different ROIC has been designed specifically for the inner layer to handle higher rate and radiation,
while keeping to the original footprint.  To reduce the mass of the services,
radiation hard voltage regulators~\cite{FEAST-DCDC}
have been introduced just outside the acceptance, but still inside the CMS inner barrel volume.

The LHC-b experiment is replacing their strip vertex detector with a 26 plane pixel detector (fixed target geometry)~\cite{LHCb-velo-upgrade}.
The detector will use triggerless readout at the full 40\,MHz collision rate of the LHC.
The high data volume will be handled by up to four 5\,Gbps serial outputs per ROIC~\cite{Velopix-ASIC-2015},
transmitted over high bandwidth copper cables outside the physics acceptance.

The NA62 experiment has installed a fixed target geometry hybrid pixel detector called Gigatracker to measure timing with
high resolution (200\,ps) as well as position in each pixel~\cite{AglieriRinella:2017kqh}. 
This is a small area detector, but is
pioneering the use of per-pixel timing in particle tracking.

The challenges that drive ongoing development can be broadly categorized into scale, intensity, and performance.
Particle physics programs demand pixel detectors with larger outer radius to cope with higher rates, and greater 
length along the beam direction to increase acceptance.
Significant development is therefore focused on how to produce pixel detectors with less effort and lower cost.
The two main directions are lower cost production of hybrid pixels and diode sensors 
(chapter~\ref{sec:hybridpixels}), and development of monolithic technology in CMOS foundries (chapter~\ref{sec:cmos}).

Increasing collider intensity places two main demands on pixel detectors: the ability to store and process greater hit rate per unit area
(chapter~\ref{sec:fechip}), and higher radiation tolerance (chapters~\ref{sec:hybridpixels},~\ref{sec:fechip},~\ref{sec:cmos}).
All hits ``raining'' on a pixel detector unit area must be time-stamped and stored for a trigger latency interval.
This means more memory per unit area in the ROIC, which directly translates to a need for a technology with higher logic density (i.e. smaller feature size, following Moore's Law).
Note that the need for higher logic density is not a function of pixel size, but of hit rate per unit area.
Smaller pixels are needed to maintain efficiency (to avoid pileup of hits in a single pixel), for resolution 
(chapter~\ref{sec:resolution}),
and for radiation tolerance (to keep the leakage current per pixel small).
Higher intensity also means that collision events all \glq look alike\grq, because every event has a large number of superimposed low energy scatters (pileup), and may or may not have a hard scatter of interest as well, typically with less energy than the sum of the underlying pileup. As distinguishing events of interest becomes more difficult, triggers must increase rate (for a given signal acceptance more background will pass the trigger)
and/or use more information, including tracking. Pixel detectors must therefore output much more data 
(section~\ref{sec:datatransmission}).

Achieving larger scale and coping with higher intensity are necessary to increase the physics reach of particle physics experiments. The basic performance of a detector can also improve the physics reach, and therefore, the challenges of scaling and intensity are compounded by the desire to increase performance. Given signal size, sensor capacitance, and device specific transconductance in the ROIC, 
the pixel front end achievable performance is determined, and actual ROIC's come close to this limit. 
From a single pixel perspective, the analog performance limit given a power budget and ROIC and sensor technologies 
is well understood and achieved by the implemented circuits.   
New developments to increase performance are therefore looking beyond the basic model of a pixel detector as a collection of individual 
pixels with an output of 3D space points. Correlations between pixels and  measurements with internal degrees of freedom (in addition to the usual spatial coordinates) can be used to improve pattern recognition, suppress background, and even provide input to a trigger system. The main degrees of freedom under study are timing, direction, and cluster shape and charge distribution
(see chapter~\ref{sec:resolution}).

%% file: resolution.tex
\section{Space--time point resolution}\label{sec:resolution}
\subsection{Demands and current directions}
The basic detection mechanism of silicon detectors is the generation and drift of mobile charges (e/h) in a depleted silicon junction. This charge cloud has a rich spatial and temporal structure with some dependence on incident particle type and trajectory as well as existing electric and magnetic fields in the silicon. Silicon tracking detectors have typically had granularities in space and time greater than or equal to the charge cloud deposit, resulting in one 3-D space point per particle crossing a sensor with a fairly coarse arrival time stamp,
sufficient to associate the point with a given accelerator collision event. Measurement of the magnitude of the collected charge has also been generally available for pixel detectors, and has been used to improve the 3-D space point precision through interpolation as well as for particle identification through specific ionization measurement.

In current detector development we are starting to see increased space and time granularity, in order to measure and make use of the structure of charge deposits. There are many interesting applications of this extra information beyond pure space points.
We will give below a few present or anticipated examples:
multi-track to cluster association using machine learning, a
angular information from cluster length,
3-D cluster shapes using charge arrival time,
and disentangling of multiple interactions from sub-nanosecond hit timing.

To be fully exploited, the trend towards fine granularity in space and time, to resolve the charge deposit structure and even the sequence of particle impacts within a detector, must be accompanied by an evolution of the pattern recognition and track fitting algorithms used. Interestingly, algorithm evolution seems to lag detector development. Currently used techniques are still largely based on space points, with extra information added for specific tasks after space point reconstruction. In this chapter we review the basics of space point reconstruction and extend
towards new space-time measurement directions as far as they are known and published today. But we expect to soon see modifications of the conventional, well known formulas to include angle, direction, and time measurements.

\hfill\break
In general, the tasks of pixel detectors in HEP experiments can be listed as follows:
\begin{enumerate}
  \item Pattern recognition and identification of particle tracks at large background and pile-up levels
  \item Measurement of primary and secondary vertices;
  \item Multi-track separation and vertex identification in the core of (boosted) jets;
  \item Momentum measurement of particles (together with other detectors, like strip detectors);
  \item Measurement of specific ionization.
\end{enumerate}

Small pixel size keeps the pixel occupancy down at high particle rates (important for item 1) and also leads to good hit resolution.
Space point resolutions in the order of 10\,$\upmu$m or less have been routinely achieved at least in one dimension.
Achieving the best possible resolution is of utmost importance to cope with the above challenges.

Recent developments (e.g.~\cite{ATLAS-TIDE_2015}) also address processing of multi-hit pixel clusters as complex objects to exploit directional information and improve the two track resolution, in particular inside the core of jets.

Time resolutions at the 10--100 ps level have not been realized so far with semiconductor tracking detectors.
If achieved this will add another measurement dimension which can be exploited for example to distinguish
calorimeter jets coming from a hard interaction from those with large pile-up contributions
by using the arrival time distribution (being broad for pile-up jets).
Precise timing might also allow coincidence measurements e.g.\,between tracks in the forward and backward regions of a detector.

\subsection{Space point and direction measurements}\label{sec:space-point}
\begin{figure}
	\centering
		\includegraphics[width=0.80\textwidth]{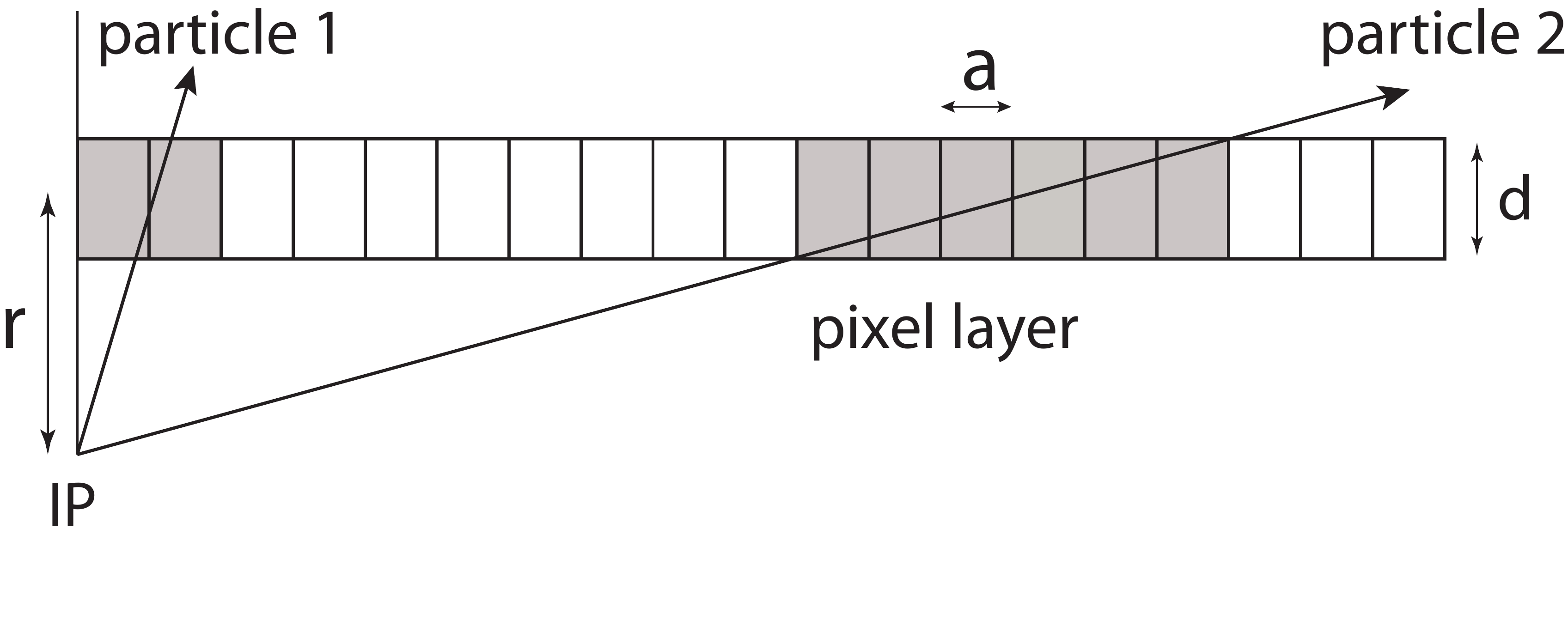}
	\caption{\label{fig:hit_clusters} Pixel hit clusters for tracks under different incident angles.}
\end{figure}

Figure~\ref{fig:hit_clusters} sketches two typical situations for pixel hit clusters. Case (1) represents
the case of a particle impinging close to perpendicularly to the pixel module leading to hit clusters of typically
one or two pixels. Case (2) represents tracks impinging at steep angles, thus producing larger hit clusters
with some directional information when properly treated by reconstruction algorithms, in particular when
such clusters appear in several detector layers along a track. Exploiting the hit information this way
could become important for hit assignment and track recognition in high track density at the LHC (see for example \cite{ATLAS-TIDE_2015}).

\subsubsection{Point measurements}
The precision of a space-point measurement enters the momentum resolution in a track measurement with $N$
detector layers as given by the Gluckstern formula~\cite{gluckstern}:
\begin{equation}\label{eq:gluckstern}
 \frac{\sigma_{p_T}}{p_T} = \left( \frac{p_T}{ 0.3|z|}
 \frac{\sigma_{\rm point}}{L^2 B}
 \sqrt{\frac{720}{N+4}}\, \right)   \oplus  \left( \frac{\sigma_{p_T}}{p_T} \right)_{\rm MS}\,
\end{equation}
where $p_T$ is transverse momentum in GeV/c, $L$ is the radial length in m, $B$ magnetic field in T,
$z$ is the particle electric charge in elementary units,
$\sigma_{\rm point}$ is the point resolution of the detectors in m, and $N$ is assumed to be large in this approximation.
Important for a precise momentum measurement is the point resolution, but also (quadratically)
the total length $L$ of the tracker and the bending field $B$.
The multiple scattering (MS) contribution for a number of detector layers $N$ can be written as
\begin{equation}\label{eq:sigma_pt_MS}
 \left( \frac{\sigma_{p_T}}{p_T} \right)_{\rm MS} =
 \frac{0.0136}{0.3\,\beta\,B L} \sqrt{\frac{(N-1)x/{\rm sin}\theta}{X_0}}\sqrt{C_N}\, , \qquad  [L] = {\rm  m}, \ [B] ={\rm T} \,
\end{equation}
where $L$ is the tracker length projected onto the plane perpendicular to the magnetic field, and $(x/{\rm sin}\,\theta)/X_0$ is the total material thickness traversed by a particle incident with polar angle $\theta$
with respect to the beam, in units of the radiation length. $C_N$ is a factor depending on the number of layers: $C_N = 2.5$ for the minimum of three layers to measure a circle; it approaches $C_N = 1.33$
for $N\to \infty$ (continuous scattering).

Even though the hard collisions at the LHC produce high energy jets, low momentum tracks around and below 1\,GeV/c transverse momentum play an important role, especially in the forward direction.
For example, efficient jet tagging - jets in general and b-jets in particular - and suppression of pile-up contributions by primary vertex identification suffer from imprecise detection of low momentum particles.
For the HL-LHC, low $p_T$ tracks will become even more important as pile-up increases. This renders low mass (low $x/X_0$) extremely important.

Similarly, the precision of secondary vertex measurement with an $N$ layer tracker can be expressed by the impact parameter resolution (here in linear extrapolation):
\begin{equation}\label{eq:vertex_resolution}
\sigma_{d_0} \approx \frac{\sigma_{\rm point}}{\sqrt{N}}\sqrt{1+\frac{12(N-1)}{(N+1)}\, \left(\frac{r}{L}\right)^2}\, \ \oplus \
\theta_0\, r_{pv}\sqrt{\frac{N(2N-1)}{6(N-1)^2}}
\end{equation}
where the first term results from the extrapolation from the tracker to the
primary vertex with $r/L$ being the ratio of the extrapolation distance to the tracker length.
The point resolution enters linearly. For a pixel detector with four layers as in ATLAS, at radii between 3.3\,cm and 12.3\,cm and with a point resolution of about 10\,$\upmu$m, this yields $\sigma_{d_0} \approx 12.5\,\upmu$m without multiple scattering.

The second term is due to multiple scattering approximated by assuming extrapolation from the first layer to the primary vertex, the slope of which is smeared by multiple scattering, with $\theta_0$ being the multiple scattering angle \cite{pdg-2014}:
\begin{equation} \label{eq:tracking:delta_d0_Nlayers_equi}
	\theta_0 \approx \frac{0.0136\,\rm GeV/c}{\beta p} \sqrt{x/X_0}\, ,
\end{equation}
and $r_{pv}$ the distance of the first pixel layer to the primary interaction vertex.
For a 4-layer geometry like in ATLAS and a material thickness of typically around 3\%\,$X_0$ the multiple
scattering contribution to the $d_0$ resolution yields
\begin{equation}
	\sigma_{d_0}^{\rm scat} \approx \frac{90\,{\rm \upmu m\, GeV/c}}{p}\, .
\end{equation}

Spatial resolution also impacts pattern recognition performance in complex interplay with
pixel occupancy, material thickness, and layout.
However, pattern recognition algorithms are outside the scope of this review.

\subsubsection{Space-point reconstruction methods}
In designing a pixel detector one selects the (initial) amount of charge sharing.
The tuning parameter for a given planar sensor thickness is the ratio of the Lorentz angle
(the deviation from perpendicular to the sensor surface of the  drift path of charges in a magnetic field \cite{Bartsch:2002tv}),
to the tilt angle of the module with respect to perpendicular particle passage.
While charge sharing between pixels allows for better spatial resolution by charge interpolation, a signal decrease caused by irradiation during the detector's lifetime demands minimal charge sharing. Whether collected charge causes a hit also depends on the pixel threshold. Noise deteriorates the precision of reconstruction and causes spurious hits.

For the present LHC pixel detectors with relatively large pixels, covering mainly central rapidity and not having too steep incidence angles, the number of hits in a cluster from a single track is rarely greater than 2.
At large incidence angles in the forward regions of a barrel pixel detector or in the future with smaller pixel sizes, the cluster distribution can become much larger and this information
can also be exploited (see below).

Resolutions achievable with classical space point reconstruction methods can be classified as follows:
\begin{itemize}
  \item For single hit clusters - independent of having binary (yes/no) or analog origin - resolution is given by the well-known pitch/$\sqrt{12}$ RMS resolution assuming a flat prior distribution of track position within the pixel (most conservative assumption). For an arbitrary given cluster size and track direction the resolution still is constant and is determined by the RMS of the prior track distribution that can produce a cluster of this size. For large clusters, high spatial resolution is possible even with binary readout, due to the averaging of many pixels.
  \item Analog hit information can be obtained at the expense of a larger total data volume (see e.g.\,\cite{ATLAS-pixel-paper_2008}\cite{Hemperek:2009xra} for ATLAS) or by processing the analog pulse
  off the module as is done in CMS \cite{CMS-pixel-paper_2008}. In this case the reconstructed hit position $x_{rec}$ can be obtained e.g. by the centre of gravity method
        \begin{equation}\label{eq:cog_with_noise}
            x_{rec} = \frac{\sum (S_i + n_i) \, x_i}{\sum (S_i + n_i)}
        \end{equation}
      where the $S_i$ and $n_i$ are the signal and noise fractional weights, respectively, and $x_i$ is the center of each
individual pixel in the cluster. The achievable resolution is:
        \begin{eqnarray}\label{eq:A:sigma_with_noise}
            \sigma_x^2 & = & \sigma_n^2 \left[ \left( \sum_{i=1}^N x_i^2 \right) + N \langle x^2 \rangle \right] + {\cal{O}}(\sigma_n^3)\, ,
        \end{eqnarray}
      where $\sigma_n$ denotes uncorrelated noise and $N$ is the number of hit pixels \cite{NW_book,PeFi_spaceresol}.
  \item If two hit clusters is the most common case for pixel detectors the $\eta$-reconstruction
        method~\cite{belau,Turchetta:1993vu} is optimal for space reconstruction of Gaussian charge clouds, since detector effects are automatically included. For two adjacent left and right electrodes the function
            \begin{equation}\label{eq:A:eta}
                \eta = \frac{S_L}{S_L + S_R} \, ,
            \end{equation}
        constructed from measured signals $S_{L,R}$ is position dependent, $\eta = \eta(x)$, and the hit position is obtained from the inverse $\eta^{-1}$:
        \begin{equation}\label{eq:eta_integral}
            x_{rec} =  \eta^{-1} \left( \frac{S_L}{S_L+S_R} \right) = \frac{a}{N} \int_0^\eta \frac{dN}{d\eta'} d\eta' \, .
        \end{equation}
where $dN/d\eta'$ is the probability distribution (with normalization $N$) of hits vs $\eta'$ resulting from uniform illumination (random $x$ distribution), and $a$ is the channel pitch.
        Including noise $n_{L,R}$ (fractional, i.e. in units of the signal) \eqref{eq:eta_integral} becomes \cite{NW_book}\cite{PeFi_spaceresol}:
        \begin{eqnarray}\label{eq:rec_eta_with_noise}
            x_{rec} \, & = & \, \eta^{-1}\left( \frac{S_L + n_L}{S_L + S_R + n_L + n_R} \right) = \eta^{-1}\left( \frac{\eta(x) + n_L}{1 + n_L + n_R} \right) \nonumber \\
    \, & \approx & \, x + \frac{d\eta^{-1}(s)}{ds}\bigg|_{\eta(x)} \cdot \bigg( \,  n_L (1 - \eta(x)\, ) - n_R \, \eta(x) \,  \bigg) \,
\end{eqnarray}
with $s = \eta(x) + n_L ( 1 - \eta(x)\, ) - n_R \eta(x) $.
Obviously, a steep response function $\eta$ yields the best resolution, given by:
\begin{eqnarray}\label{eq:space_resol_with_eta}
  \sigma_x^2 = \sigma_n^2 \bigg\langle \frac{ 1-2\eta + 2\eta^2 }{ \eta' \,^2} \bigg\rangle + 2 \langle n_L n_R \rangle \cdot \bigg\langle \frac{\eta^2 - \eta }{ \eta' \, ^2} \bigg\rangle \, ,
\end{eqnarray}
where the second term vanishes for uncorrelated noise, $\langle n_L n_R \rangle = 0$.
%
\end{itemize}

By using analog information either directly (CMS) or via a Time over Threshold (ToT) digitization (ATLAS), resolutions
as shown in table~\ref{tab:pixel_resolutions} have been obtained (references in table). The ToT method is
explained in section~\ref{sec:ROIC-fend}.
\begin{table}
	\centering
	\caption{\label{tab:pixel_resolutions} Spatial resolutions obtained with the ATLAS and CMS pixel detectors under LHC conditions.
R/O stands for Readout and ToT for Time over Threshold.}
\begin{tabular}{|c|c|c|c|c|}
\hline
Experiment & R/O method & pitch [$\upmu$m] & inc. angle & $\sigma_x$ [$\upmu$m] \\
\hline
ATLAS \cite{ATL-INDET-PUB-2016-001} & binary/ToT & 250 & 0\degree & 66.5 \\
ATLAS \cite{ATL-INDET-PUB-2016-001} & ToT    & 50  & 0\degree & 10  \\
ATLAS \cite{ATLAS-pixel-paper_2008} & ToT    & 50  & 10\degree & 7 \\
CMS \cite{Chatrchyan:2014fea}   & analog  & 150    & 0\degree & 44 \\
CMS \cite{Chatrchyan:2014fea}   & analog  & 150    & 30\degree & 21 \\
CMS \cite{Chatrchyan:2014fea}   & analog  & 100    & 0\degree & 9.4 \\
\hline
\end{tabular}%
\end{table}

\subsubsection{Limitations to the space point resolution of a pixel detector}
When aiming for ultimate spatial resolutions with hybrid pixel detectors the following limitations are encountered:
\begin{itemize}
    \item Since for hybrid pixels the chip's pixel area must match the sensor's pixel area the smallest pixel size is
    determined by the amount of CMOS electronics needed to amplify, discriminate, and process the hit information in  the area occupied by the pixel cell. Third generation chips for high hit rate achieve about $50\times50\,\upmu$m$^2$ or $25\times 100\,\upmu$m$^2$, limited by logic density as explained in section~\ref{sec:fechip}.
    \item The signal of a traversing MIP (Minimum Ionizing Particle) spreads due to diffusion according to $\sigma_x = \sqrt{2Dx/v_D}$, where
    $D$ is the diffusion constant, $x$ the drift distance to the electrode and $v_D$ the drift velocity. For a typical detector thickness of 150--200$\,\upmu$m, $\sigma_x$ becomes 4--8\,$\upmu$m. Hence pixel pitches well below this (at the same sensor thickness)
    would lead to excessive charge sharing.
    \item Pixel hybridization technology (see section~\ref{sec:hybridpixels}) can currently cope with bump pitches in the order of 25--50\,$\upmu$m. The technological limit for galvanic or
    evaporation methods for the coming decade seems to lie in the order of 5--10\,$\upmu$m \cite{IMEC_Majeed_2015} (see also section~\ref{sec:bonding-techniques}).
\end{itemize}

\subsubsection{In-pixel decoding}\label{sec:in-pixel_decoding}
It has been proposed \cite{Peric:2014faa} to effectively achieve smaller pixel size by encoding and decoding
the pixel sensor cell by means of \glq smart sensors\grq\  when using depletable substrates with a CMOS electronics layer (see also section~\ref{sec:cmos}).
\begin{figure}
   \centering
    \subfigure[Coupling of smart sensor and readout chip]{\raisebox{0.5cm}
        {\includegraphics[width=0.38\textwidth]{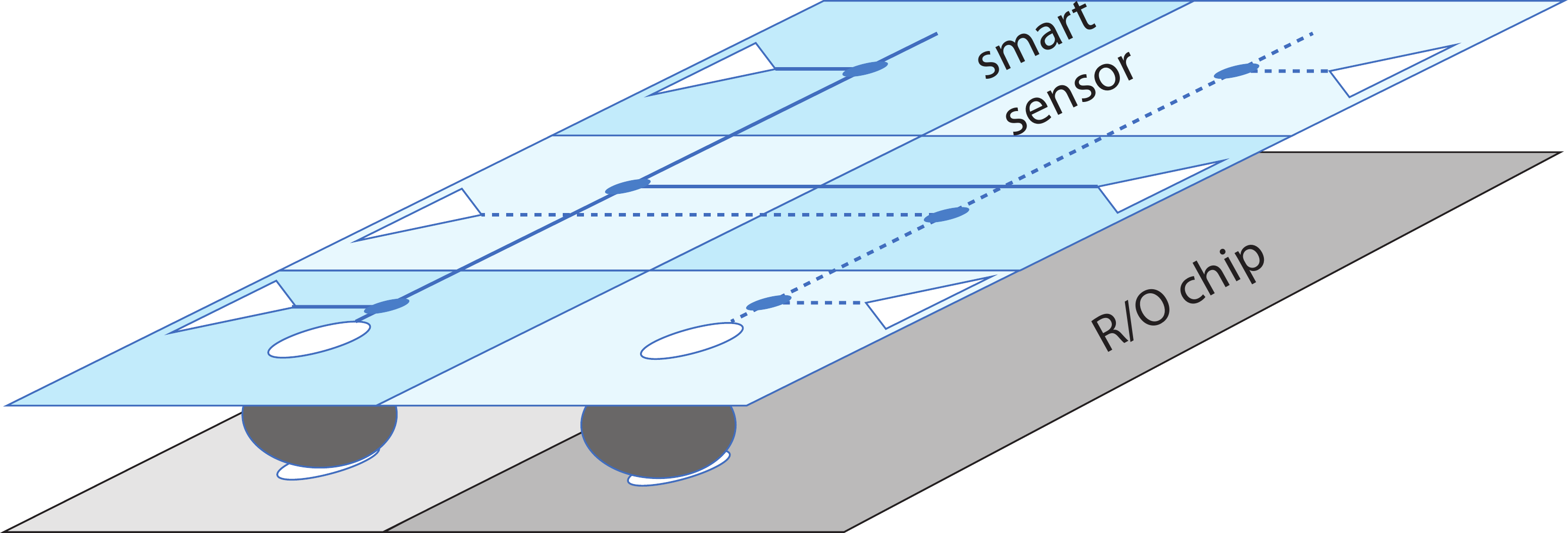}}\label{fig:inpixel_decoding_a}
        }\hskip 0.2cm
    \subfigure[Function principle]{\raisebox{0.0cm}
        {\includegraphics[width=0.55\textwidth]{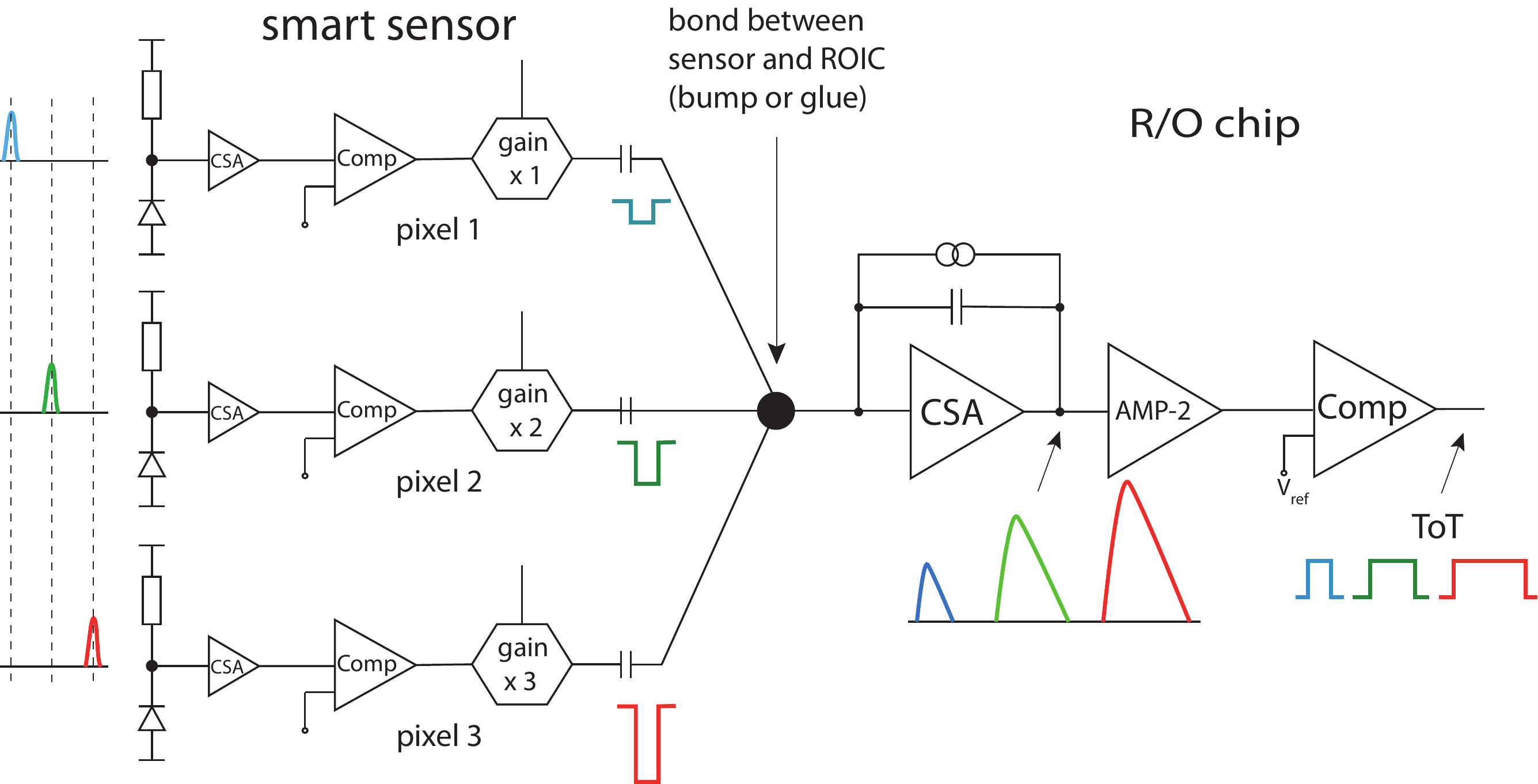}}\label{fig:inpixel_decoding_b}
        }\hskip 0.0cm \vskip 0.0cm
	\caption{\label{fig:inpixel-decoding} In-pixel en- and decoding using a \glq smart sensor\grq\ bonded to a pixel readout chip (here showing 6 smaller sensor pixels bonded to two readout pixels). The sensor pixels contain amplifications with adjustable gains (here $\times$\,1, $\times$\,2, $\times$\,3). The gains are set such that the origin of a hit can be decoded from the output pulse height encoded as ToT.}
\end{figure}

Figure~\ref{fig:inpixel-decoding} shows the principle. Two pixels of the readout chip (e.g. FE-I4
with 50\,$\upmu$m\, $\times$\,250\,$\upmu$m each) match six (sub-)pixels of the smart sensor with sizes 33\,$\upmu$m\,$\times$\,125$\upmu$m. Three sub-pixels
at a time are connected to one of the two FE-I4 inputs via AC coupling, either by a capacitor included in the sensor pixel plus DC connection (bump bonding) to the corresponding FE-I4 chip pixel, or by gluing the sensor chip/wafer to the FE-I4 wafer, in which case the coupling is also capacitive.
The digital (voltage) output pulse of the smart sensor is capacitively coupled into the CSA input of the readout chip.
The three digital output stages of the sensor pixels can be programmed with three different gains (high, medium, low) such that digital pulses of different height are capacitively coupled to the input of FE-I4 depending on which
of the sub-pixels has been hit. Further processing of the pulse by the FE-I4 amplification and discrimination stages
turns different pulse heights at its input into different ToT values at the output. Hence
the sub-pixel hit by the MIP is finally decoded from its ToT value. Test beam measurements demonstrated
the functionality of the method~\cite{Hirono:2016zck,Miucci:2014jxa}.

\subsubsection{Directional information from hit clusters}
\begin{figure}
	\centering
    \subfigure[Cluster to track assignment]{\label{fig:TIDE_cluster_to_track_assignment}
        \includegraphics[width=0.46\textwidth]{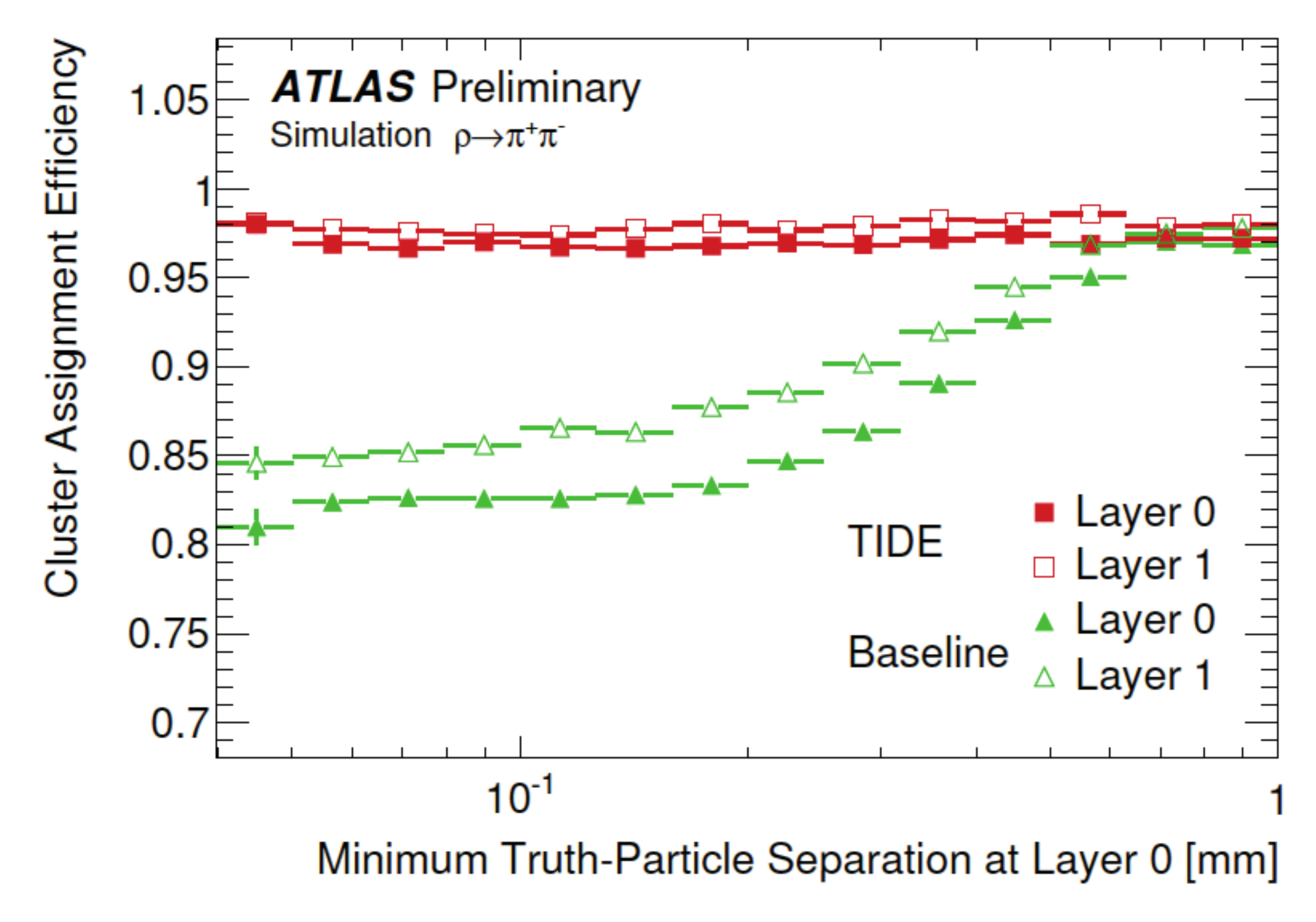}}
            \hskip 0.1cm
	\subfigure[Track reconstruction efficiency inside dense jets]{\label{fig:TIDE_track_eff_in_jetcore}			
        \includegraphics[width=0.48\textwidth]{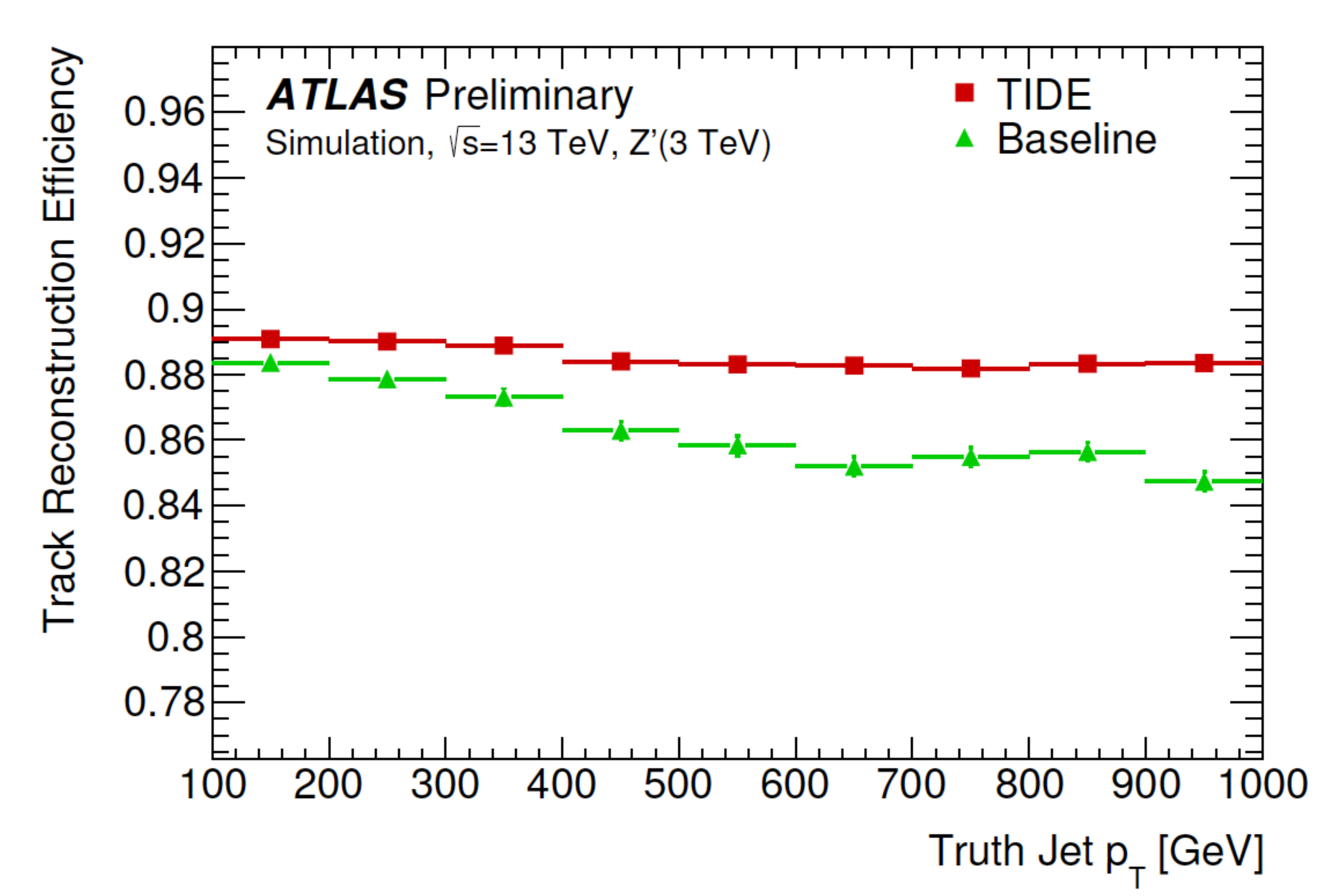}}
\caption{Improvement in hit cluster assignment and tracking efficiency using artificial neural networks to resolve merged clusters
for Tracking In Dense Environments (TIDE), taken from~\cite{ATLAS-TIDE_2015}.
(a) Cluster assignment efficiency as a function of separation of two pion tracks from simulated $\rho^0 \to \pi^+\pi^-$ decays at the innermost pixel detector layers.
(b) Track reconstruction efficiency inside dense jets for high $p_T$ jets.}
\end{figure}

Multivariate algorithms like artificial neural nets (ANN) have now also entered the area of hit/cluster to track
association for improved track reconstruction in high multiplicity environments.
The single hit precision is not the only figure of merit to tune.
One must also optimize the capability of cluster shape
analysis (especially for inclined tracks) and the analysis of pattern information of clusters from dense tracks over several layers.
For example in \cite{ATLAS-TIDE_2015} it is shown how successive employment of ANNs can improve (a) the association of merged clusters to tracks, (b) track reconstruction inside the core of (boosted) jets, and (c) heavy flavour and $\tau$-jet identification.
The amount of charge per pixel, for instance provided by the ToT technique \cite{ATLAS-pixel-paper_2008}, and a precise knowledge of the pixel coordinates is already sufficient as input for the ANN
to identify merged clusters efficiently. The emission of $\delta$-rays and secondary interactions of tracks prevent a perfect ANN performance. The approach benefits from knowing the
incident angle of tracks, estimated from the coordinates of modules with respect to the beam spot, and from correlating information from consecutive layers of the pixel detector \cite{ATLAS-TIDE_2015}.
Figure~\ref{fig:TIDE_cluster_to_track_assignment} demonstrates using simulation the gain in cluster assignment efficiency using this technique on a sample of simulated $\rho^0 \to \pi^+\pi^-$ decays. Figure~\ref{fig:TIDE_track_eff_in_jetcore} shows the improvement in tracking efficiency inside dense jet-cores obtained using the ANN-based improved merger identification.
Improvement in high-$p_T$ heavy flavour identification of 7--13\% (b-tagging and b-jet identification) and of 5\% in 3-prong $\tau$-jets above $p_T = 600$\,GeV is the result \cite{ATLAS-TIDE_2015}.

\subsection{Time measurement}\label{sec:time-point}
The charge collection time, i.e.\,the time until the arrival of the last electron at the pixel electrode,
typically is in the order of 3--10\,ns, depending on sensor thickness and E-field. The induced current pulse at the electrode instantly starts with the particle passage
and consists of the incoherent sum of induced pulses from each carrier's path towards (e$^-$) and away (h$^+$) from the pixel electrodes.
The individual contributions are weighted by the pixel weighting field (see e.g.
\cite{spieler_book,pixel_book,NW_book}) leading to the fact that the carriers' paths close to the pixel implant
have the strongest contributions to the signal. In a typical pixel detector the current pulses are integrated by the
input stage of the subsequent readout chip, a charge sensitive amplifier (CSA). While such a readout method
has low noise
performance\footnote{Typical noise values are on the order of 150\,e$^-$ for the present LHC
detectors~\cite{ATLAS-pixel-paper_2008,CMS-pixel-paper_2008}
and will be around 80\,e$^-$ for future detectors with smaller pixels~\cite{RD53A-specs}.}
it does not allow timing
measurements at or below 100\,ps, values at which timing becomes interesting for high energy particle detection.

For a time coordinate measurement to reach equivalent spatial resolutions in the cm to mm regime,
time resolutions of the order of a few tens of picoseconds are needed. To achieve this with pixel detectors, where charge collection times are in the ns range, is a real challenge. Recently, amplification structures in silicon have been brought forward to cope with these demands \cite{Sadrozinski:2013nja,Cartiglia:2013haa,Cartiglia:2017yyy}. Potential benefits are the suppression of pile-up jets
by recognizing the time-wise association of tracks to a primary vertex.

\subsubsection{LGAD structures}
In order to reach into the picosecond timing regime with silicon detectors, in-silicon amplification can
be employed. Avalanche generating silicon based devices have been developed in the context of photodetectors (see for example \cite{Buzhan:2001xq},\cite{Buzhan:2003ur},\cite{APD_Hamamatsu_UK}).
They can be distinguished by their operation mode, either as linear avalanche photodiodes (APD) operating in a linear
amplification mode or as so-called Geiger APDs operating near the breakdown point thus sensing a light pulse by a large
breakdown pulse with follow-up recovery time. Timing in Geiger-mode operation will be governed by the time jitter
introduced by the breakdown process (multiplication jitter), similar to the case for gaseous detectors where fast timing is addressed for example with RPCs (resistive plate chambers)~\cite{RiegLipp:2004gc}.
Another approach has been chosen by \cite{Sadrozinski:2013nja}, namely by operating avalanche silicon diodes at a low gain operation point, so-called \emph{Low Gain Avalanche Diodes}, LGADs.

\begin{figure}
	\centering
        \includegraphics[width=1.0\textwidth]{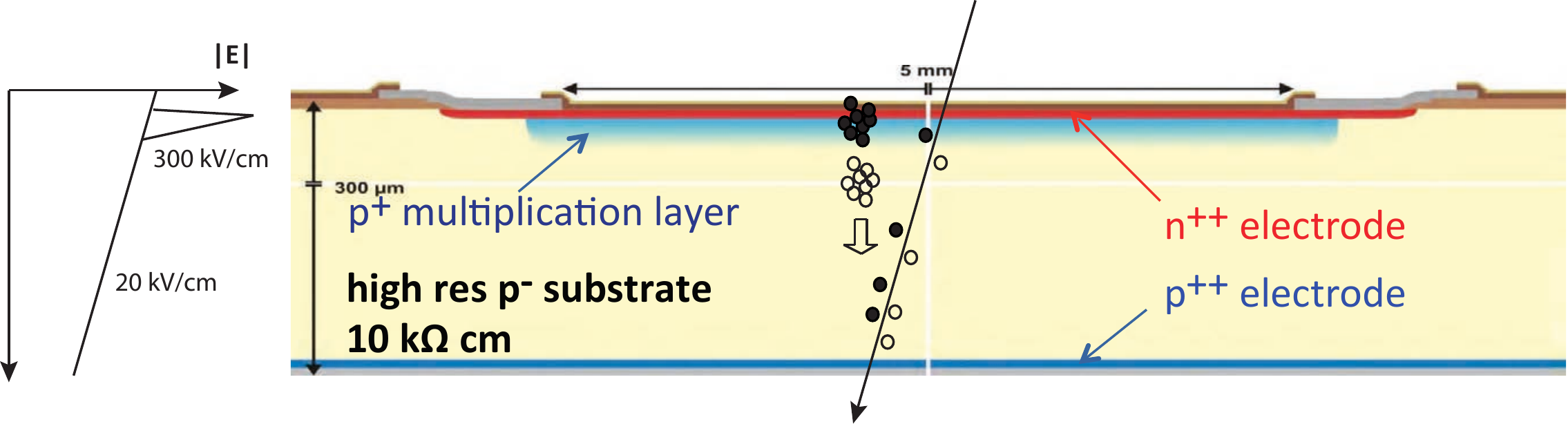}
	\caption{Cross section through an LGAD structure using a high ohmic p-type bulk and featuring a metallurgical junction underneath the n$^{++}$ electrode acting as the amplification layer (adapted from \cite{Cartiglia:2015iua}). The shape of the electric field strength is indicated on the left. \label{fig:LGAD-structure}}
\end{figure}

In order to create precise time stamps, fast and large signals and low noise are needed. The \emph{slew rate} can serve as a figure of merit, defined as
\begin{equation}\label{eq:slew-rate}
  \frac{dV}{dt} \approx \frac{\rm signal\, height}{t_{rise}}\, .
\end{equation}
Figure~\ref{fig:LGAD-structure} shows a principal cross section of an LGAD structure \cite{Cartiglia:2015iua}. In a $p^-$ bulk, faced on both surfaces by the usual very highly doped\footnote{We here adopt the notation p$^{++}$ for $N_A \gg 10^{16}/$cm$^{3}$, likewise for  $n^{++}$.} $n^{++}$ and $p^{++}$ electrodes rendering the structure of the detector, an additional, highly doped $p^{+}$ layer ($N_A \approx {\cal{O}}(10^{16}/$cm$^{3}$)) is implanted immediately underneath the $n^{++}$ electrode, thus creating a very local high
electric field of up to 300\,kV/cm. This is the multiplication layer amplifying all arriving electrons
by a factor of the order of 10-20, thereby creating secondary e/h-pairs.
\begin{figure}
	\centering
        \includegraphics[width=0.8\textwidth]{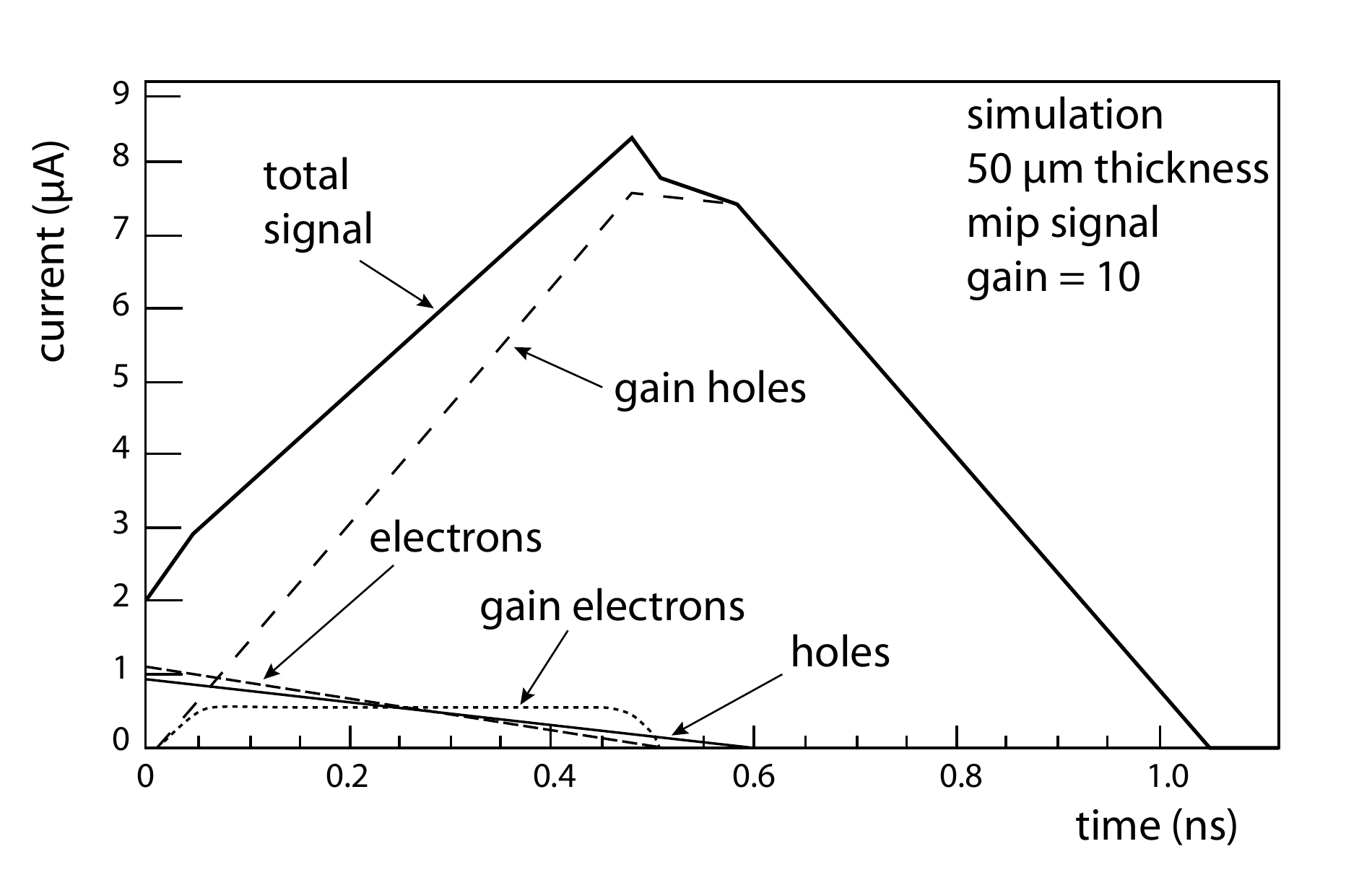}
	\caption{LGAD signal pulse detailing the contributions from electrons and holes before and after amplification following weighting field simulations \cite{Cenna:2015}. The contributions from both electron and hole signals are shown before and after amplification as well as the total signal (adapted from \cite{Cartiglia:2015iua}).\label{fig:LGAD-signal_shape}}
\end{figure}
%

The signal induced on the electrodes therefore has several components (see fig.~\ref{fig:LGAD-signal_shape}):
\begin{itemize}
  \item Electrons and holes drifting from the point of creation along the track towards their respective collection electrodes. The individual e/h parts of this contribution are small before amplification takes place and end when the last carriers have arrived at their respective electrodes.
  \item Amplification electrons created in the multiplication layer reach the top electrode (fig.~\ref{fig:LGAD-structure}) almost instantaneously and their contribution to the induced signal current $i_S (t)$ is close to negligible, because - according to the Shockley-Ramo theorem~\cite{shockley,ramo} -
      $$
        i_S (t) = e \vec{v}_D (t) \cdot \vec{E}_w
      $$
      with $E_w$ = weighting field and $v_D$ = drift velocity, there is almost no current contribution since the drift duration for electrons is quasi non-existent and ending before the next primary electrons reach the amplification layer.
  \item By contrast, the multiplied holes drifting from the amplification layer towards the backside contribute to $i_S (t)$ as long as they are drifting. Their contribution adds up, provided that the weighting field is  formed such that it does not suppress contributions from distant holes too much. This implies that the pixel/pad implant widths must be large compared to the implant distances and should be in the same order of magnitude as the detector thickness.
\end{itemize}
As can be seen from fig.~\ref{fig:LGAD-signal_shape}, a large slew rate and hence fast timing requires fast (primary) electron arrival at the amplification zone together with fast (amplified) hole movement away. Short bunching and fast movement in high fields ($\sim$\,20\,kV/cm) benefit from thin sensors.

The time resolution has several contributions \cite{Cartiglia:2016sjr}:
\begin{eqnarray}\label{eq:LGAD_time_resolution}
     \sigma_t^2 & = & \underbrace{\left( \frac{V_{th}}{dV/dt}\bigg|_{rms} \right)^2 } + \underbrace{\left( \frac{\rm Noise}{dV/dt}\right)^2} +\,\sigma^2_{\rm arrival} + \sigma^2_{\rm dist} + {\sigma^2_{\rm TDC} } \\
       && \hskip 0.5cm {\sigma_\textrm{time walk}^2}   \hskip 1.3cm {\sigma_\textrm{noise}^2}
\nonumber
\end{eqnarray}
The first term represents time walk introduced by different signal pulse heights due to different energy deposits
coming from Landau fluctuations in the energy loss process and its resulting pulse height fluctuations at the
discriminator input. Minimization of this term is possible e.g.\,by employing a constant fraction discrimination or a ToT correction architecture in the readout.
The second term is noise jitter, i.e.{\ }time fluctuations due to noisy signals.
Both contributions are made small when large slew rates are achieved.
An irreducible contribution comes from fluctuations resulting from non-uniform depositions of charge along the particle track moving towards the amplification junction, hence causing an intrinsic jitter in the arrival time $\sigma_{\rm arrival}$ (third term). Fluctuations from secondary ionizations and from the amplification process enter here as well.
The thinner the detector the less disturbing is this effect.
An additional 4$^{th}$ contribution is signal distortion due to non-uniform weighting field regions and variations in (non-saturated) drift velocities.
The final term is time fluctuations due to uncertainties in the time
digitization. It is assumed that the latter can be made negligible (below 10\,ps) with GHz TDCs.

The key ingredients for fast timing and small time jitter are thin detector substrates (${\cal{O}}$(50\,$\upmu$m)) providing three essential benefits:
\begin{itemize}
  \item larger slew rate ($\times$1.6 slew rate increase going from d\,=\,100\,$\upmu$m to 50\,$\upmu$m at a gain of 20),
  \item better bulk radiation hardness,
  \item smaller arrival time fluctuations due to a shorter charge deposition path.
\end{itemize}

\begin{figure}
   \centering
    \subfigure[dV/dt versus thickness]{\raisebox{0.1cm}
        {\includegraphics[width=0.40\textwidth]{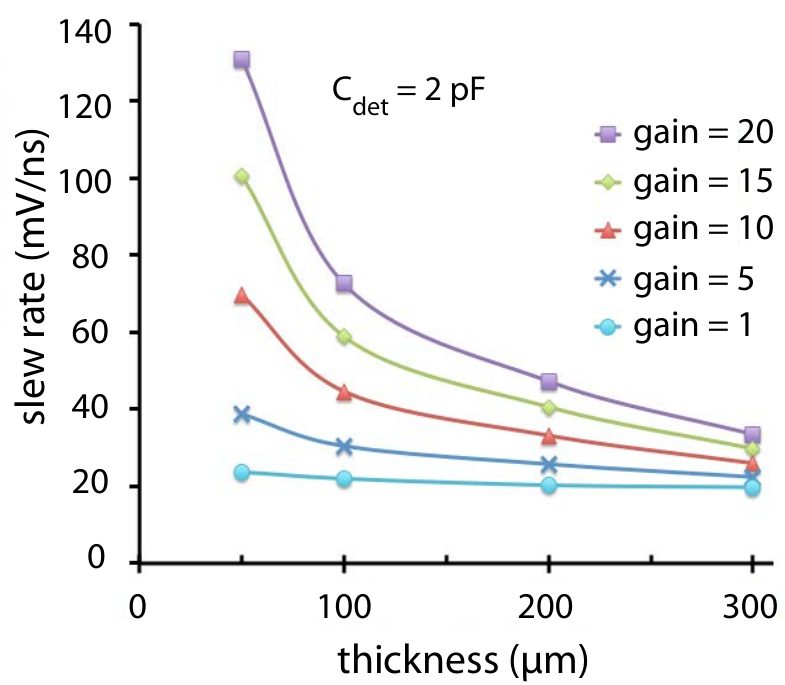}}\label{fig:LGAD_features_summary}
        }\hskip 0.3cm
    \subfigure[Time resolution]{\raisebox{0.2cm}
        {\includegraphics[width=0.55\textwidth]{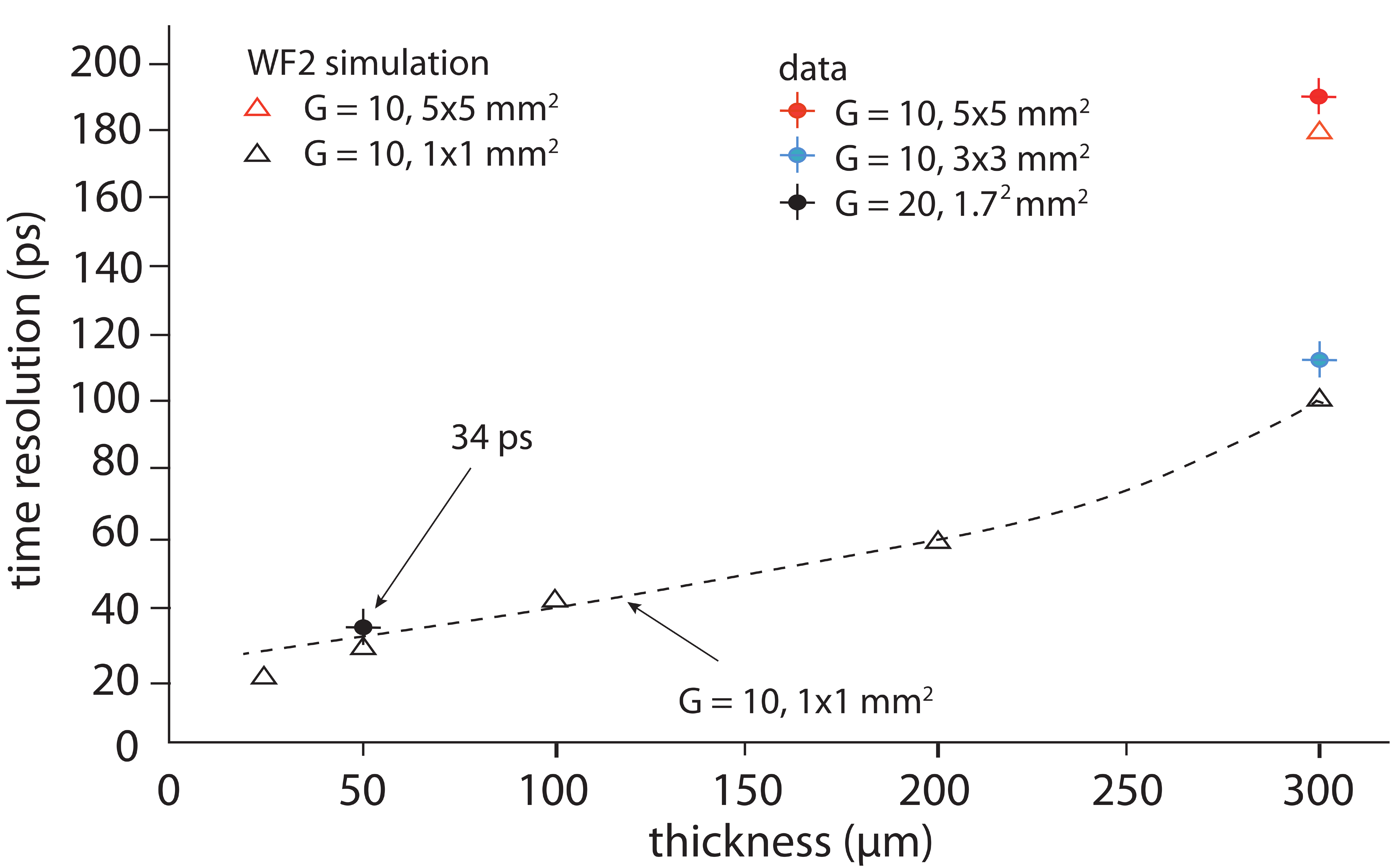}}\label{fig:LGAD_results}
        }\hskip 0.0cm \vskip 0.5cm
\caption{LGAD weighting field simulations and measurements: (a) slew rate as a function of detector thickness for different amplification gains (simulation); (b) comparison of time resolutions from weighting field simulations (WF2) and from test beam measurements using constant fraction discrimination. Open triangles are simulation points, crossed circles are measured resolutions without errors given (bars do not indicate the measurement errors). Note the different pad sizes entering the figure. The WF2 simulations for 1x1$\upmu$m$^2$ are underlined by the dashed line only to guide the eye. The data are taken from \cite{Cartiglia:2015iua},\cite{Sadrozinski:2016xxe},\cite{Cartiglia:2016voy},\cite{Cartiglia:2017yyy}, and \cite{Cartiglia:2016sjr}.}
\end{figure}

In fig.~\ref{fig:LGAD_features_summary} the slew rate is plotted against the diode thickness for different gains assuming an input capacitance of 2\,pF \cite{Cartiglia:2015iua}.
It is evident that large slew rates can be obtained with thin detectors and sufficient gain ($\gtrsim$\,10). Much higher gains compromise the S/N ratio, because bulk shot noise is also amplified by the amplification structure (called excess noise) such that there is an optimal gain for maximum signal-to-noise (see for instance \cite{APD_Hamamatsu_UK} or \cite{Cartiglia:2016sjr}).

Prototype LGAD structures have been fabricated \cite{Cartiglia:2015iua,Sadrozinski:2016xxe} with relatively large pads (from 8$\times$8\,mm$^2$ to 1$\times$1\,mm$^2$), rather than pixels, mainly to validate the underlying models and simulations of the achievable time resolution. Figure~\ref{fig:LGAD_results} shows the predicted time resolution by simulations \cite{Cenna:2015} as a function of the detector thickness. At the right hand side measured
data from 300\,$\upmu$m thick LGAD pad detectors are compared with simulations. Extrapolation to thin detectors
yields time resolutions in the order of 30 ps.
Very thin detectors have only recently become available. In a test beam measurement with 180\,GeV pions, a 50$\,\upmu$m LGAD sensor with 1.2$\times$1.2\,mm$^2$ pads has been characterized, achieving 34\,ps time resolution at a bias voltage of 200\,V \cite{Cartiglia:2016voy}, in excellent agreement with predictions from weighting field simulations (fig.~\ref{fig:LGAD_results}).

A major concern for the use of LGADs at the HL-LHC is their performance in high radiation areas. Apart from the usual effects caused by radiation in silicon detectors like leakage current increase and deteriorated charge collection efficiency, effective doping changes play a sensitive role. Since the metallurgical junction providing the amplification gain requires p-doping concentrations
of order of 10$^{16}$cm$^{-3}$, radiation induced acceptor removal will have detrimental effects upon the gain.
Detailed radiation studies are currently ongoing \cite{Cartiglia:2016sjr}, including very thin sensors ($\lesssim$\,50\,$\upmu$m) and devices with gallium substituted as p-dopant in the amplification implant instead of boron, to reduce interstitial capture that scales
with atomic mass \cite{RD50_home}. Another proposed method is to use carbon-enriched wafers with (abundant) carbon being
trapped by silicon interstitials rather than boron.

Currently LGADs are not yet discussed for pixel tracking detectors but rather as precision timing layers, aiming to distinguish primary vertex positions (in $z$) from each other (see e.g.{\ }\cite{Mannelli:2017yai}).
CMS-TOTEM is planning to use LGADs in Roman Pot stations \cite{Arcidiacono:2017jek}.

%% file: hybridpixels.tex
\section{Pixel sensors and hybridization}\label{sec:hybridpixels}
\begin{figure}
    \centering
    \subfigure[Hybrid pixel]{\raisebox{0.5cm}
        {\includegraphics[width=0.37\textwidth]{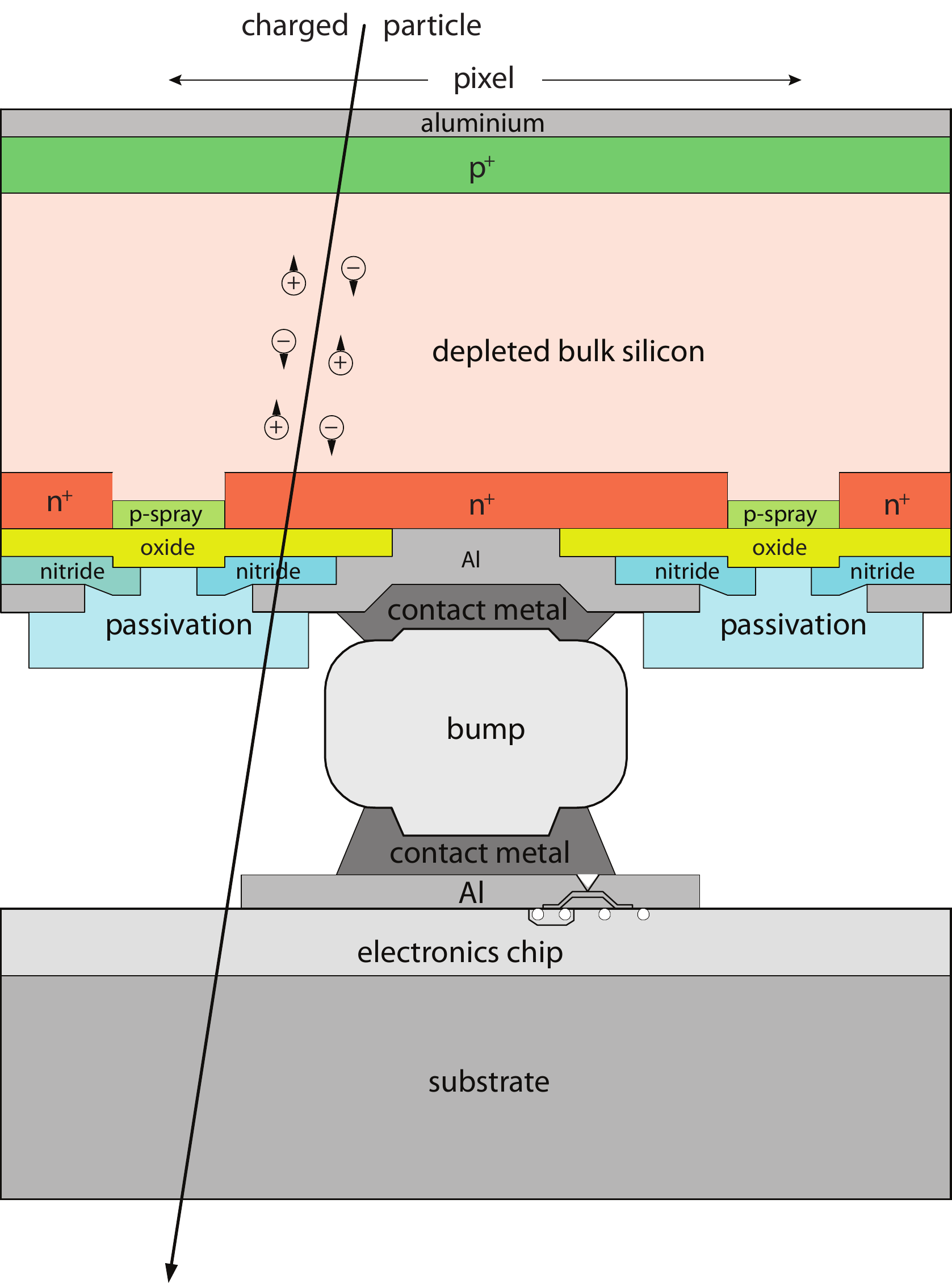}}}
    \hskip 0.5cm
    \subfigure[Pixel matrix]{\raisebox{0.8cm}
        {\includegraphics[width=0.56\textwidth]{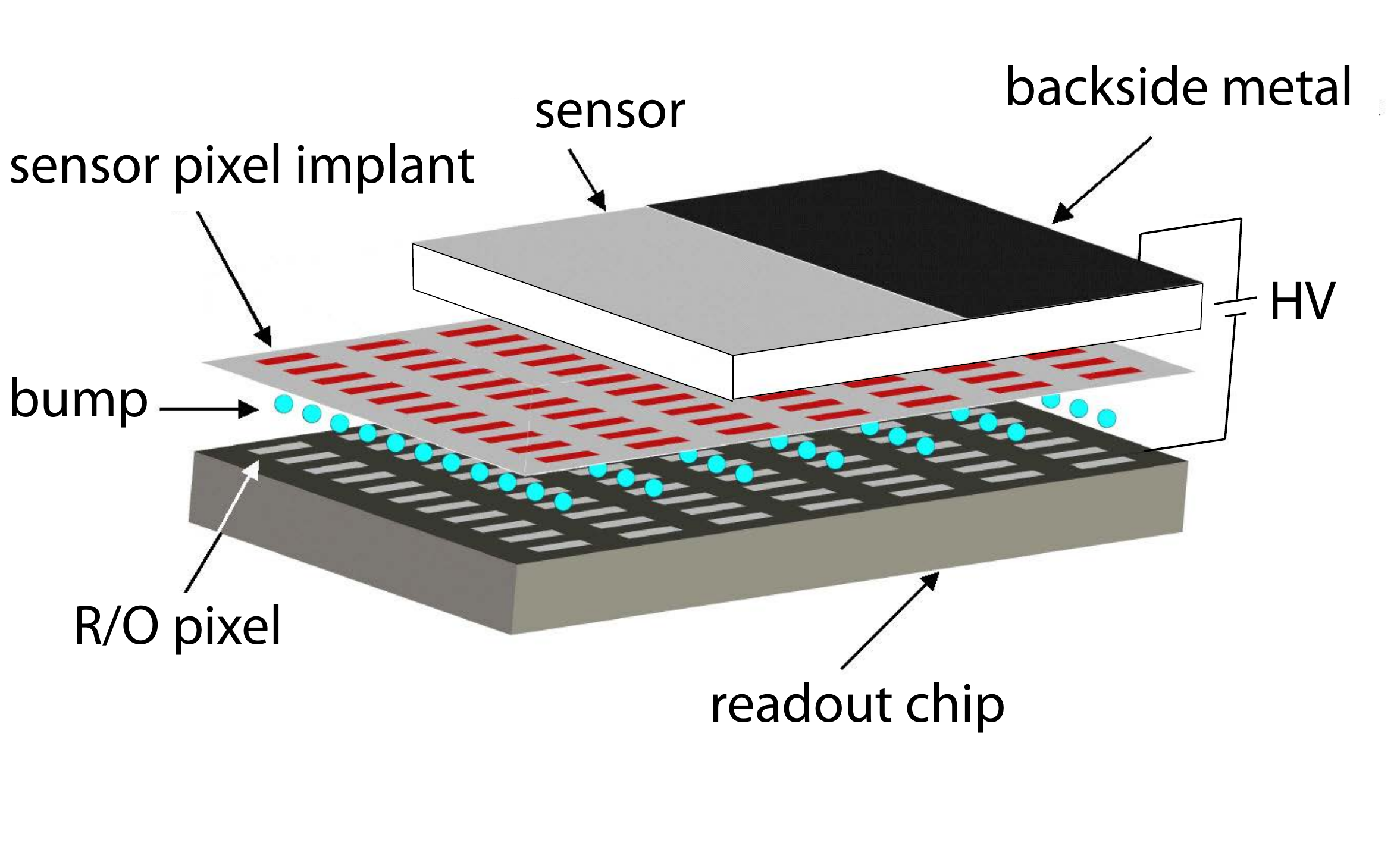}}}
    \caption{Hybrid pixel detector: (a) Layout of an individual pixel cell having a sensor and an electronics cell in 1$\mbox{--}$1 correspondence; (b) a hybrid pixel matrix; sensor and electronics chip have pixels of the same size, bonded to each other by means of bump contacts.}
    \label{fig:hybrid_pixels}
\end{figure}

\glq Hybrid pixels\grq\  including a sensor (electronically passive) and a readout chip as separate entities (fig.~\ref{fig:hybrid_pixels}) currently are the state-of-the-art technology for large scale pixel detectors in most particle physics experiments \cite{pixel_book}. For both parts, the radiation requirements at LHC-type experiments constitute a challenge which is addressed by dedicated R{\&}D and process engineering techniques.
In this chapter we describe passive sensor types used in hybrid pixel assemblies, to be contrasted to CMOS
active monolithic sensors that integrate electronics circuitry and sensor, described in chapter~\ref{sec:cmos}.
The mechanical and electrical mating of sensor and readout  chip, called hybridization, requires state-of-the-art processing technologies, also described in this chapter (section~\ref{sec:bonding-techniques}). The hybridization mating partner,
the readout chip is described in detail in chapter~\ref{sec:fechip}.

\bigskip
After the success of planar sensors (section~\ref{sec:sensors_planar}) in LHC run-1,
having planar pixel implants at one side of the sensor
and being fabricated using 4$''$ sensor wafer technologies,
the desire for large detector modules at affordable cost has fueled the move to 6$''$ and possibly even 8$''$ wafer sensor technologies.
In line with this a better understanding of radiation damage effects and radiation hard sensor device engineering became a research branch on its own, coordinated within the CERN RD50 collaboration \cite{RD50_home}.

Radiation hard device engineering includes using silicon with increased oxygen content supplied in the
growth process \cite{Lindstrom2001,Harkonen:2004iu,Pintilie200952} and operation at low-temperature, both employed to reduce the damage's impact on the detector performance. Furthermore, so-called 3D-silicon sensors \cite{3D-parker1997} (see section~\ref{sec:sensors_3D}) have been developed having vertically structured electrodes
fabricated within the volume of the silicon substrate (fig.~\ref{fig:3D-Si_sensors}).
Their main feature is a better ratio of ionization thickness to charge collection distance than for planar sensors.
After an operation performance demonstration in the first pixel upgrade of ATLAS (Insertable B-Layer, IBL \cite{IBL-Hugging:2010bj}), an innermost pixel barrel layer at a distance of 3.5\,cm from the interaction point,  equipped with 25\% 3D-Si sensors (at high $\eta$) and 75\% planar pixel sensors, both sensor types now compete for the HL-LHC upgrade as pixel sensors for the radiation hot areas near the interaction point.

A challenge for both sensor types imposed by the radiation fluences of $>$10$^{16}$\,neq/cm$^{2}$ is the small foreseen sensor thickness (100$\mbox{--}$150\,$\upmu$m) and the correspondingly increased handling and hybridization difficulty.

Another advance in cost reduction of sensor wafers has been made possible by the availability high quality p-substrate material ($\gtrsim\,$2\,k$\Omega$cm). With n-in-p type sensors cost-efficient single sided processing suffices for fabrication while maintaining electron collection (n-type pixels). During the initial LHC pixel development period (late 1990s) high ohmic p-type sensor wafers were not favoured, largely for reasons of historical development\footnote{Single sided Si sensor development had begun using n-substrate material with hole collection on p-electrodes. For radiation tolerance reasons electron collection became favored leading to $n^+$-in-$n$
structures with double sided processing. For n-in-p sensors it then has taken time to develop n-side multi-guard-ring structures and reliable HV bias isolation from the electronics before high quality p-type sensors were fabricated.}, and $n^+$-in-$n$ planar pixel sensors have been used instead, requiring processing steps on both sides of the wafer.

With CMOS technology vendors becoming more open to smaller market customers, the use of high ohmic
8$''$ wafers fabricated in (much lower cost) CMOS processing lines become options for (planar) sensor options
(see section~\ref{sec:passive_CMOS}).

Currently, therefore, the trend in hybrid pixel module development goes to large readout chips and
thin sensors, preferentially using p-type bulk material, all being beneficial for cost, radiation tolerance,
and detector mass, while the handling and hybridization demands increase. The latter together with the increased interest for large area coverage using pixel detectors also revived collaboration with industry to develop lower cost hybridization techniques with thinner wafers/chips.

\subsection{Planar pixel sensors}\label{sec:sensors_planar}
\begin{figure}
    \centering
    \subfigure[$n^+$ in $n$ design: 2-sided]{\raisebox{0.0cm}
        {\includegraphics[width=0.35\textwidth]{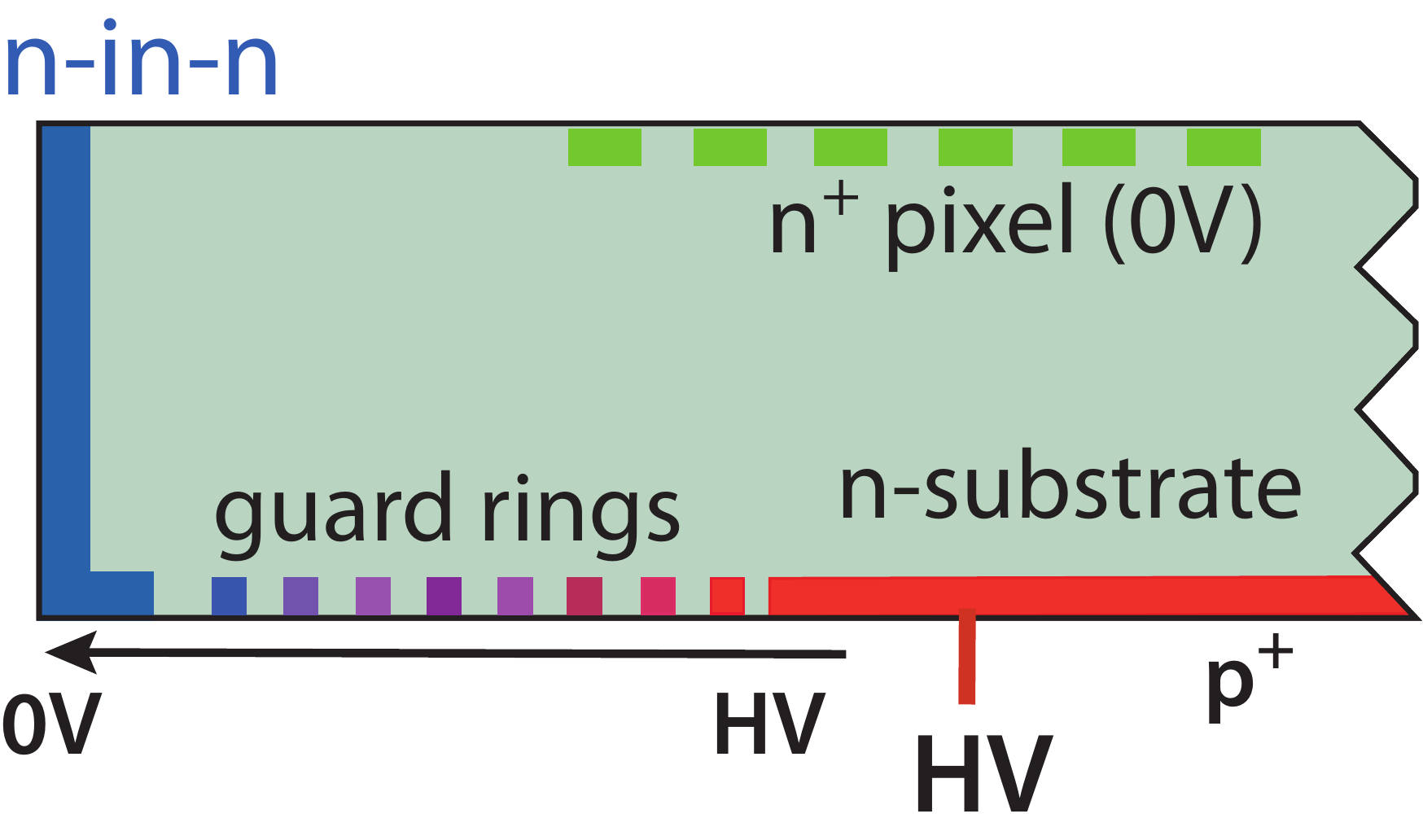}}\label{fig:planar_sensor_design_a}
        }\hskip 1.0cm
    \subfigure[$n^+$ in $p$ design: 1-sided]{\raisebox{0.0cm}
        {\includegraphics[width=0.50\textwidth]{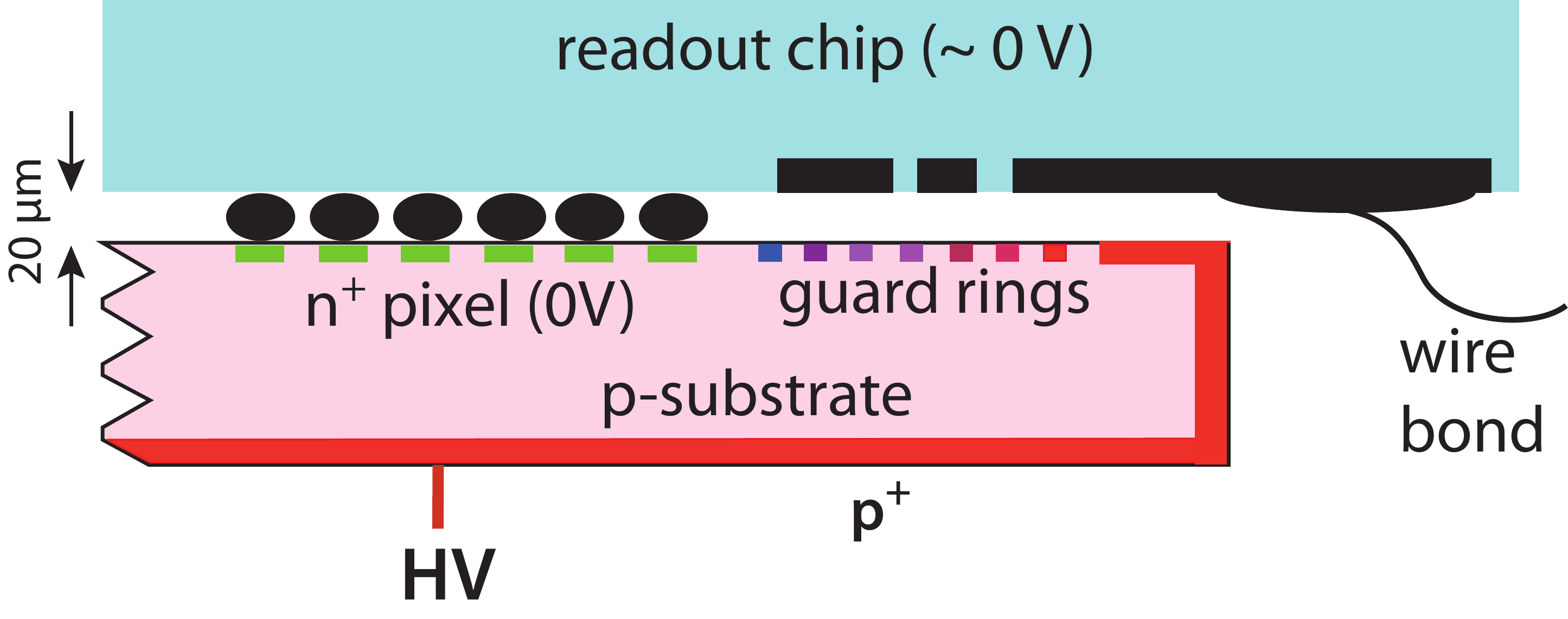}}\label{fig:planar_sensor_design_b}
        }\hskip 0.0cm \vskip 0.5cm
    \subfigure[Hit efficiency]{\raisebox{0.0cm}
        {\includegraphics[width=0.60\textwidth]{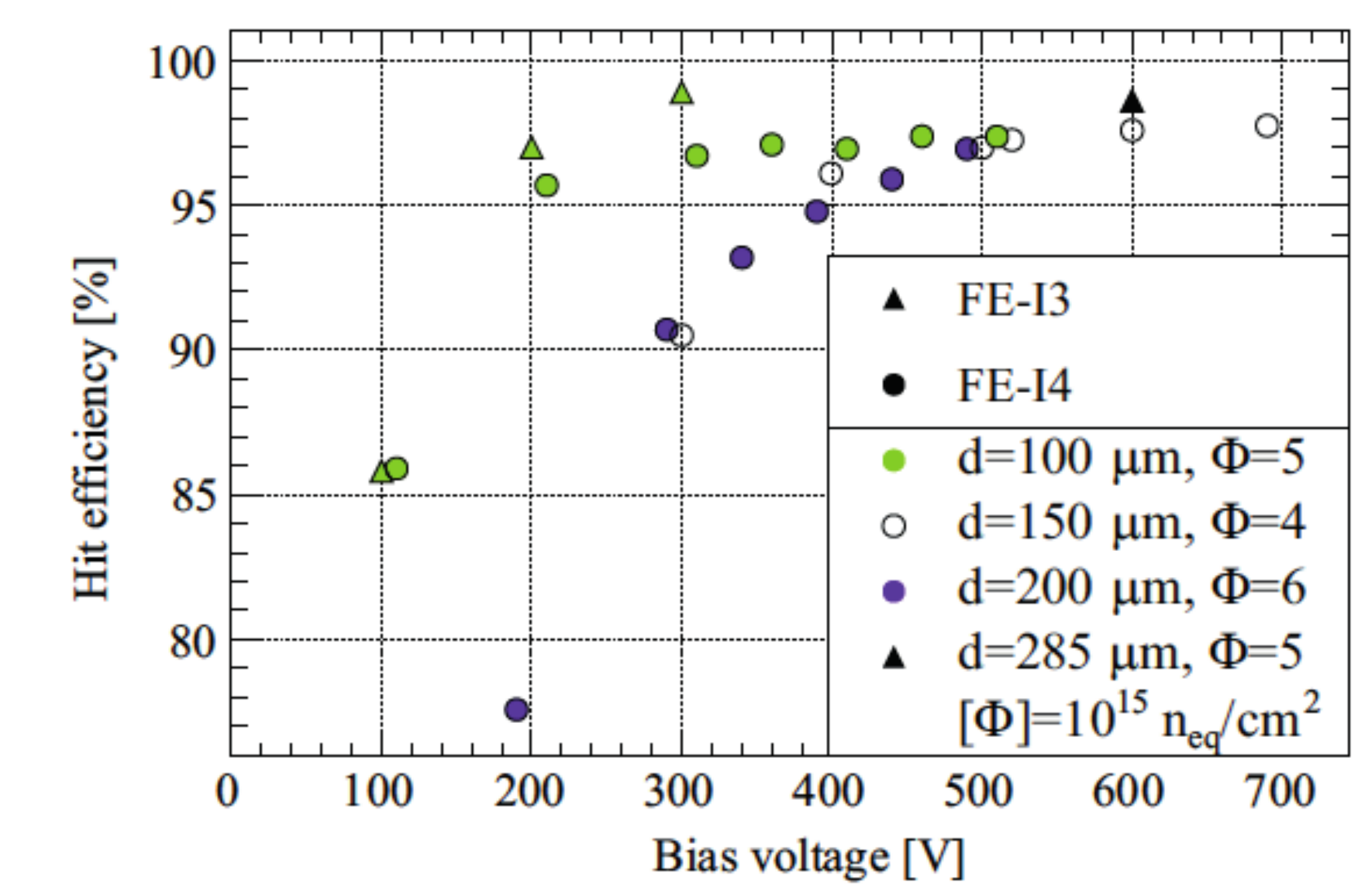}}\label{fig:planar_sensor_performance}}\hskip 0.0cm
    \caption{Planar sensors: (a) LHC conventional $n^+$-in-$n$ design requiring 2-sided wafer processing; (b) $n^+$-in-$p$ design with single-sided processing (bump) bonded to a readout chip. Note the small distance between
    readout chip and guard ring implants; (c) $n^+$-in-$p$ sensor hit efficiency as a function of bias voltage for different fluences and various thicknesses \cite{Macchiolo:2016xzi}.}\label{fig:planar_sensors}
\end{figure}

The proven standard of planar pixel sensors at the LHC experiments has $n^+$ pixel implants on $n$-type substrate material as shown in fig.~\ref{fig:planar_sensor_design_a}. The fabrication demands double-sided processing.
The negative bias potential on the backside is brought down to zero potential at the
sensor edge by a many guard ring structure implanted on the backside.
The pn-diode is initially on the unstructured backside of the sensor, changing to the electrode side after radiation has turned the bulk's $n^-$ doping into effectively p-type (type inversion). For typical substrate resistivities of \mbox{2--5}\,k$\Omega$cm, this happens after fluences of some $10^{12}$\,neq/cm$^{2}$. After type inversion the depletion zone grows from the electrode side into the bulk guaranteeing that signals induced by moving charges can be sensed on the pixels even if the substrate is no longer fully depleted, i.e. not depleted at the side opposite to the pixels. To prevent an electron accumulation layer from short circuiting the $n^+$-in-$n$ pixels before type inversion, shallow $p$-doping is used in between
the pixel electrodes, with a smoothly changing doping profile preventing high field corners.
Details are given in \cite{pixel_book}.

With increasing radiation fluence the bias voltage needed for sufficient depletion must be increased, eventually
reaching values of more than 500\,V at fluences around and beyond $10^{15}$\,neq/cm$^{2}$. For the innermost layers close to the interaction point
this corresponds to detector lifetimes of only a few years at current LHC luminosity.

The development of planar sensors for the LHC high luminosity upgrade has therefore concentrated on tailored designs
guaranteeing high fields and sufficiently large depletion depths after fluences of up to and above $10^{16}$\,neq/cm$^{2}$.
This has been achieved using n$^+$ pixel implants in p-substrate material
and thin (100-150$\,\upmu$m) sensors operated at bias voltages of 500-700\,V \cite{Macchiolo:2016xzi}.
While providing smaller signals per MIP, the benefits of thin sensors are higher electric fields as well as shorter and faster electron collection for a given bias voltage and hence better radiation tolerance. Thin 6$''$ or even 8$''$ sensor wafer production is enabled by techniques employing SOI or Si-Si handling wafers,
or by thinning (e.g. by cavity etching) and forming a back side ohmic contact at low temperature after the front side processing is complete.
Currently the limit in thickness is considered to be around 50\,$\upmu$m \cite{Macchiolo_2016}.
%

The guard ring designs, bringing down the potential from the bias implants (HV) to the pixel implants (0\,V), play an important role in optimizing the breakdown behaviour of sensors after irradiation. The number of rings, implant distance and size, as well as metal overhangs for vertical field suppression are the parameters to optimize. Optimal performance regarding radiation tolerance has been obtained \cite{Macchiolo_2016} with 10--11 rings with metal overhangs on both sides of the implant over a total length of 350\,$\upmu$m. The distance from the edge to the active part of the sensor is $\gtrsim$400\,$\upmu$m. Other designs trying to minimize this distance make use of etched and doped edges, called active edges, reaching values down to about 50$\,\upmu$m \cite{Terzo:2014sua}.

Figure~\ref{fig:planar_sensor_design_a}
shows a planar pixel design as used in present LHC hybrid pixel detectors (200\,$\upmu$m thick, $n^+$-in-$n$ design) compared to a thin $n^+$-in-$p$ design (fig.~\ref{fig:planar_sensor_design_b}). Note that, if cost saving single sided processing is to be used, the guard ring structure needs to be placed on the electrode side and is thus in very close proximity ($\lesssim 20\,\upmu$m) to the readout  chip, a challenge for the design. Dedicated passivation with BCB (benzocyclobuthene) or parylene, however, appears to be able to solve this issue \cite{Terzo_PhD_2015}.
Figure~\ref{fig:planar_sensor_performance} shows that high hit efficiencies can be achieved
after irradiation with thin sensors for bias voltages in excess of 500\,V.

\subsubsection{Use of 8$\,''$ wafers and CMOS foundries}\label{sec:passive_CMOS}
\begin{figure}
    \centering
        {\includegraphics[width=1\textwidth]{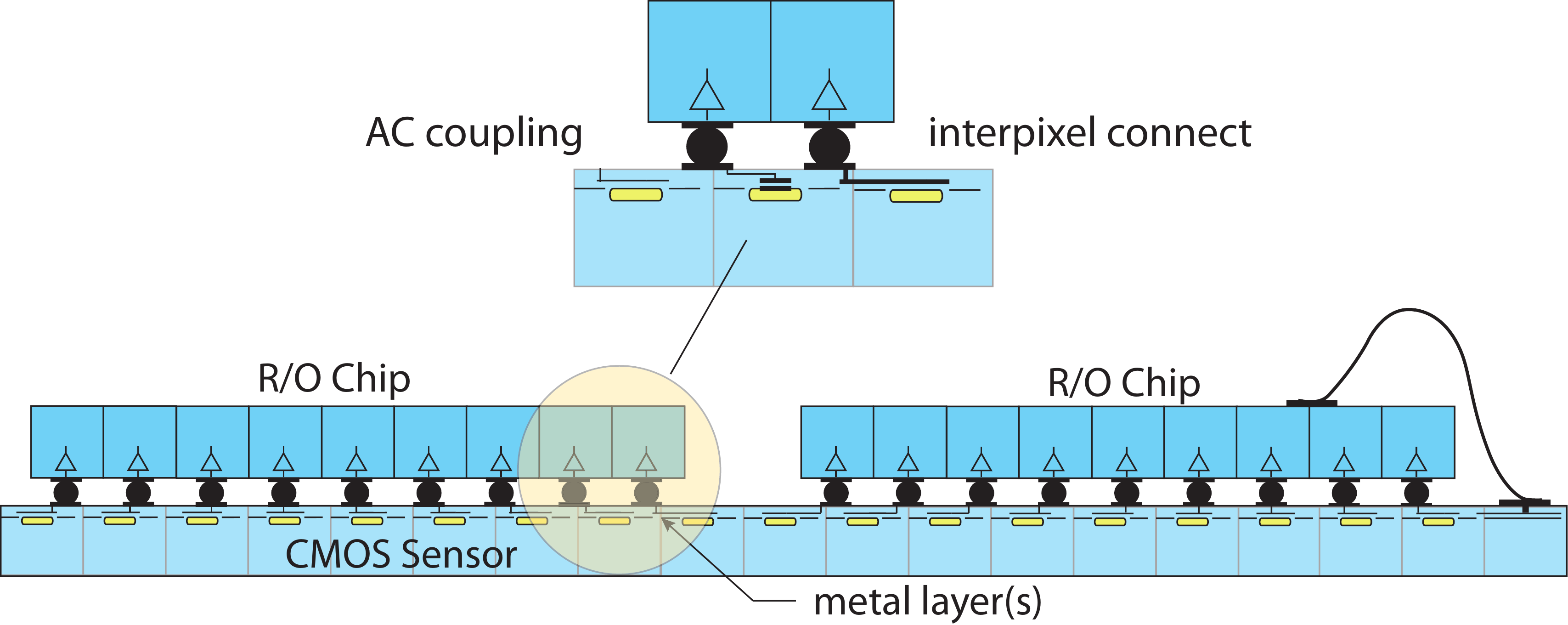}}
    \caption{Hybrid pixel module using passive CMOS pixel sensors (schematic). Depleted sensor employing CMOS technology with 1-2 metal planes that can be used for (i) AC coupling and (ii) rerouting. The insert details these features. The yellow area in every pixel denotes the charge collection node.}
    \label{fig:Passive_CMOS}
\end{figure}

\begin{figure}
    \centering
        \includegraphics[width=0.7\textwidth]{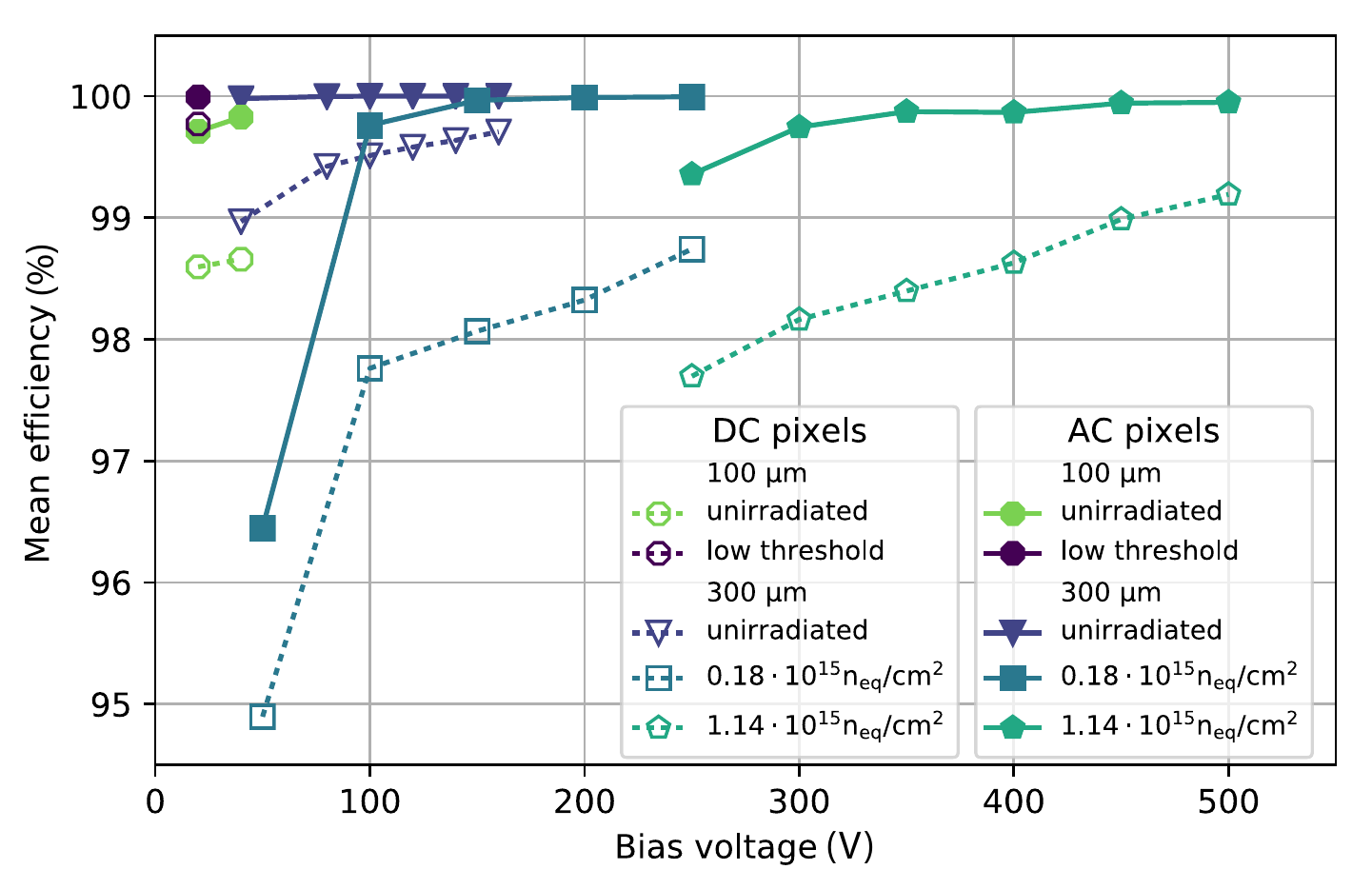}
    \caption{Hit efficiency measured in 3.2 GeV electron test beams~\cite{passCMOS_Bonn:2017}
of passive CMOS pixel sensors as a function of bias voltage, unirradiated and for two fluence levels.
For low threshold, unirradiated, there is only one point for each thickness at minimal bias voltage.
\label{fig:Passive_CMOS_efficiency}}
\end{figure}

It is also possible to fabricate planar sensors on 8$''$ wafers using high ohmic substrates either
in dedicated sensor fabrication facilities or by employing CMOS foundries, as has been successfully demonstrated for strip and pixel sensors \cite{Dragicevic:2014xja,Bergauer:2016uoq}. This has the following advantages compared to the standard small volume sensor fabrication:
\begin{itemize}
  \item Large volume production lines can be used with price advantages, suited in particular for large area detectors, e.g. in outer tracking layers at collider experiments.
  \item Wafer sizes of 8$''$ or 12$''$ are a commercial standard allowing large sensors when stitching over reticle boundaries is applied. Stitching is a method to cross the boundaries between reticles of a wafer with metal lines and implant areas, widely applied in the digital imaging industry \cite{patent_reticle_stitching_1997}. Current reticle sizes are about 25$\times$25\,mm$^2$.
      By putting individual building blocks of a design separately (rather than together) on the reticle/mask one can join them together over the reticle's boundary by special lithographic programming, thus allowing IC fabrication of sensor sizes much larger than the reticle area of the CMOS process.
  \item The wafers can be purchased with solder bumps already provided by the CMOS vendor (e.g. so-called C4 bumps \cite{C4-IBM}, pitch $\gtrsim 150\,\mu$m).
  \item One or two metal layers can be exploited for AC coupling of the sensor cell to the amplifier and for connection redistribution, e.g. to connect amplifier inputs to pixels in areas
  not covered 1:1 by readout  chip cells (regions in between readout  chips).
\end{itemize}
A sketch illustrating some features employing passive CMOS sensors is shown in fig.~\ref{fig:Passive_CMOS}.
Measurements on passive CMOS pixel sensors, 100\,$\mu$m and 300\,$\mu$m thick,
irradiated to fluences of 1.1$\times 10^{15}$\,neq/cm$^2$
have shown lab and test beam performance equal to those of planar sensors fabricated in dedicated sensor production lines
\cite{passCMOS_Bonn:2017}. Figure~\ref{fig:Passive_CMOS_efficiency} shows the mean hit efficiency as a function of bias voltage
measured with 3.2 GeV electrons for DC and AC coupled passive CMOS sensors bonded to the ATLAS readout chip FE-I4.
In particular the AC-coupled devices show excellent performance without any efficiency losses. The DC devices are a bit less efficient due to the area taken by the punch-through dot for biasing of the pixel implants.
%

\subsection{3D-Si sensors}\label{sec:sensors_3D}
\begin{figure}
    \centering
    \subfigure[3D-Si: penetrating columns]{\raisebox{0.0cm}
        {\includegraphics[width=0.43\textwidth]{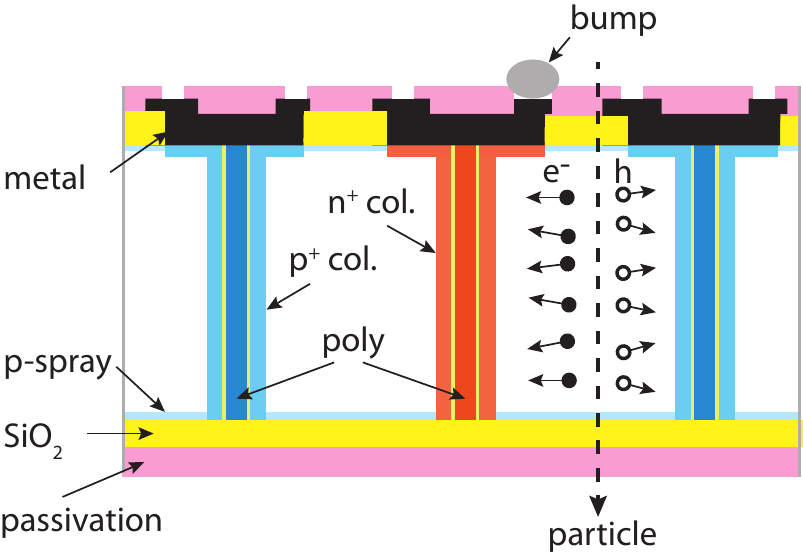}}\label{fig:3D-Si_through}}\hskip 0.5cm
    \subfigure[Partial 3D-Si (double sided)]{\raisebox{0.0cm}
        {\includegraphics[width=0.43\textwidth]{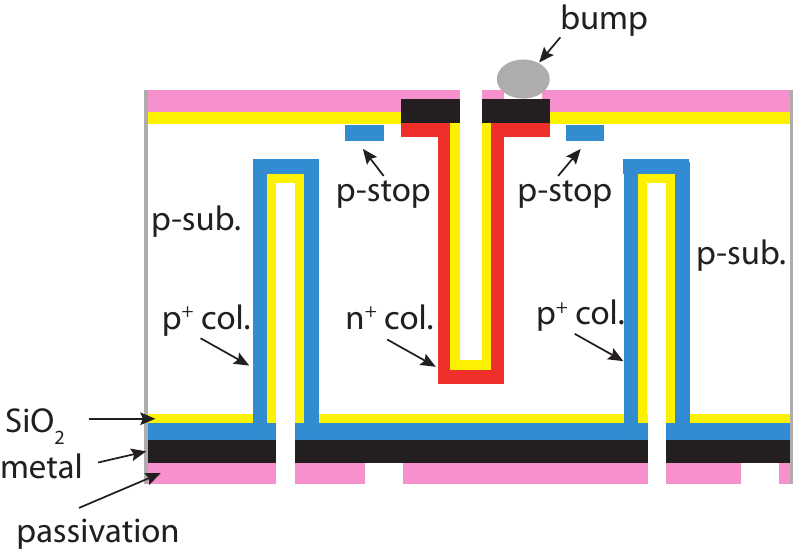}}\label{fig:3D-Si_open}}\hskip 0.0cm
    \subfigure[HL-LHC design]{\raisebox{0.0cm}
        {\includegraphics[width=0.73\textwidth]{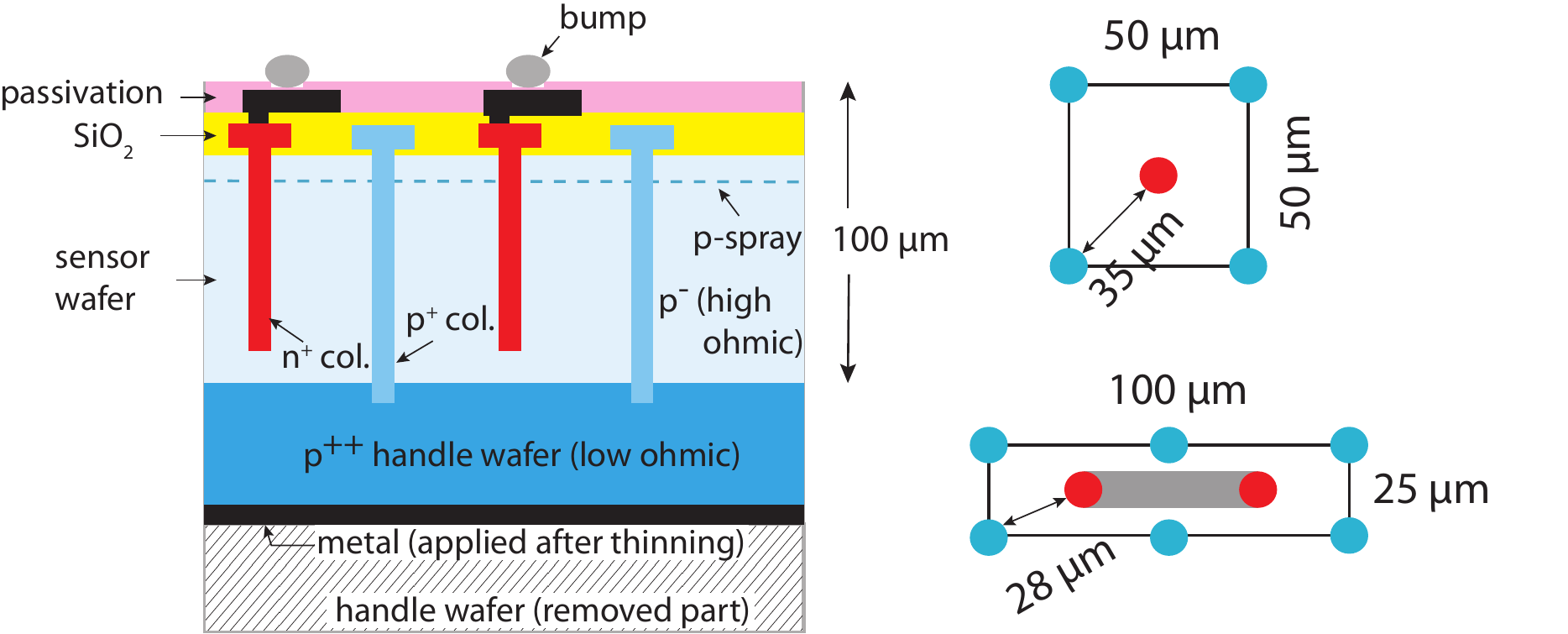}}\label{fig:3D-Si_new}}\hskip 0.0cm
    \caption{3D-Si sensors: (a) Design (single sided) with columns going completely through the sensor bulk \cite{DaVia:2005wa}; (b) double sided design with columns entering from both sides, but not reaching through (adapted from \cite{Pellegrini:2008zza});
(c) thin design optimized for HL-LHC (adapted from \cite{DallaBetta:2016iua}) with
 two top view sketches for $50\times 50\,\upmu$m$^2$ and $25\times 100\,\upmu$m$^2$ pixel sizes, respectively. \label{fig:3D-Si_sensors}}
\end{figure}

So-called 3D-silicon sensors have been developed since the late 1990s \cite{3D-parker1997,DaVia:2013nn} featuring columnar electrode implants driven into the
Si substrate perpendicular to the sensor surface (fig.~\ref{fig:3D-Si_sensors}). The electrode distance is made smaller (50$\upmu$m) than the typical
sensor thickness (200-250$\upmu$m), thus rendering a shorter average drift distance for particles impinging on the sensor face than in the case of planar sensors (compare fig.~\ref{fig:hybrid_pixels}(a) to fig.~\ref{fig:3D-Si_through}). In addition, high drift fields are obtained with still moderate bias voltages. Both these facts result in an increased radiation tolerance due to a reduced trapping probability.

The 3D-Si technique has been developed over many years. The first structures were fabricated at Stanford (later also at Oslo) \cite{DaVia:2005wa} using single sided processing with columns reaching completely through the bulk
(called \glq full-3D\grq and shown in fig.~\ref{fig:3D-Si_through}).
Further development by CNM \cite{Pellegrini:2013nq} and FBK \cite{FBK_3D:2013} of sensors used in the ATLAS IBL detector,
resulted in double-sided 3D designs with columns entering the bulk from both sides,
either in full-3D or in partial-3D (shown in fig.~\ref{fig:3D-Si_open}).
The process fabricates about 10\,$\upmu$m diameter columns by etching, followed by a 1\,$\upmu$m polysilicon layer covering the inside of the etched holes, then passivated by a wet oxide \cite{Pellegrini:2013nq}.
In addition the sensor edges can be fabricated with active edge implants thus rendering sensors with an unrivaled active area fraction \cite{Betta:2013xva}.
More details on etching holes into silicon can be found in section~\ref{sec:3D-integration}.

Within the ATLAS IBL detector 3D-Si pixel sensors have been proven to operate well in a running experiment
\cite{IBL-Hugging:2010bj}. After two years of operation the performance of 3D-Si pixel modules in terms of operation characteristics (signal, noise, threshold settings, in-time efficiency) are on par with planar pixel modules operated with significantly higher voltages \cite{LaRosa:2016nbd}.

Current developments motivated by HL-LHC demands \cite{Lange:2016jbm,Pellegrini:2013nq,DallaBetta:2016_oct,DallaBetta:2016_review} target the following goals to optimize radiation hardness, granularity, material budget, and processing costs \cite{DallaBetta:2016_review}: (a) thin sensors ($\sim$100\,$\upmu$m) on 6$''$ wafers, (b) narrower electrodes ($\sim$5\,$\upmu$m), (c) shorter electrode spacing ($\sim$30\,$\upmu$m), and (d) very slim ($\sim$50\,$\upmu$m) or active edges. Single sided processing is preferred providing cost benefits. An advanced design \cite{DallaBetta:2016_review} is shown in fig.~\ref{fig:3D-Si_new}. A thin, highly resistive (p-type) sensor wafer is supported by a low ohmic (p$^{++}$) handle wafer that can be backside thinned after processing.
While the p$^+$ columns are deep etched through to the handle wafer where they receive their electric potential, the n$^+$ columns stop about 15\,$\upmu$m short from the handle wafer.
In addition to cost and yield advantaged of single-sided processing, studies have shown that the trade-off between signal efficiency and breakdown performance favors partial depth n-columns (not extending all the way through the thickness)~\cite{dbetta2017}.
At the top surface isolation of the n-columns is achieved by a p-spray layer preventing the electron accumulation layer underneath the oxide from creating shorts.
Sensors are designed to meet the currently planned pixel area sizes of $50\times 50\,\upmu$m$^2$ or $25\times 100\,\upmu$m$^2$,
as shown on the right of fig.~\ref{fig:3D-Si_new}.
The performance of such designs has been shown to yield high breakdown voltages before and after irradiation \cite{Sultan:2016vzg}.

\begin{figure}
    \centering
        \includegraphics[width=0.6\textwidth]{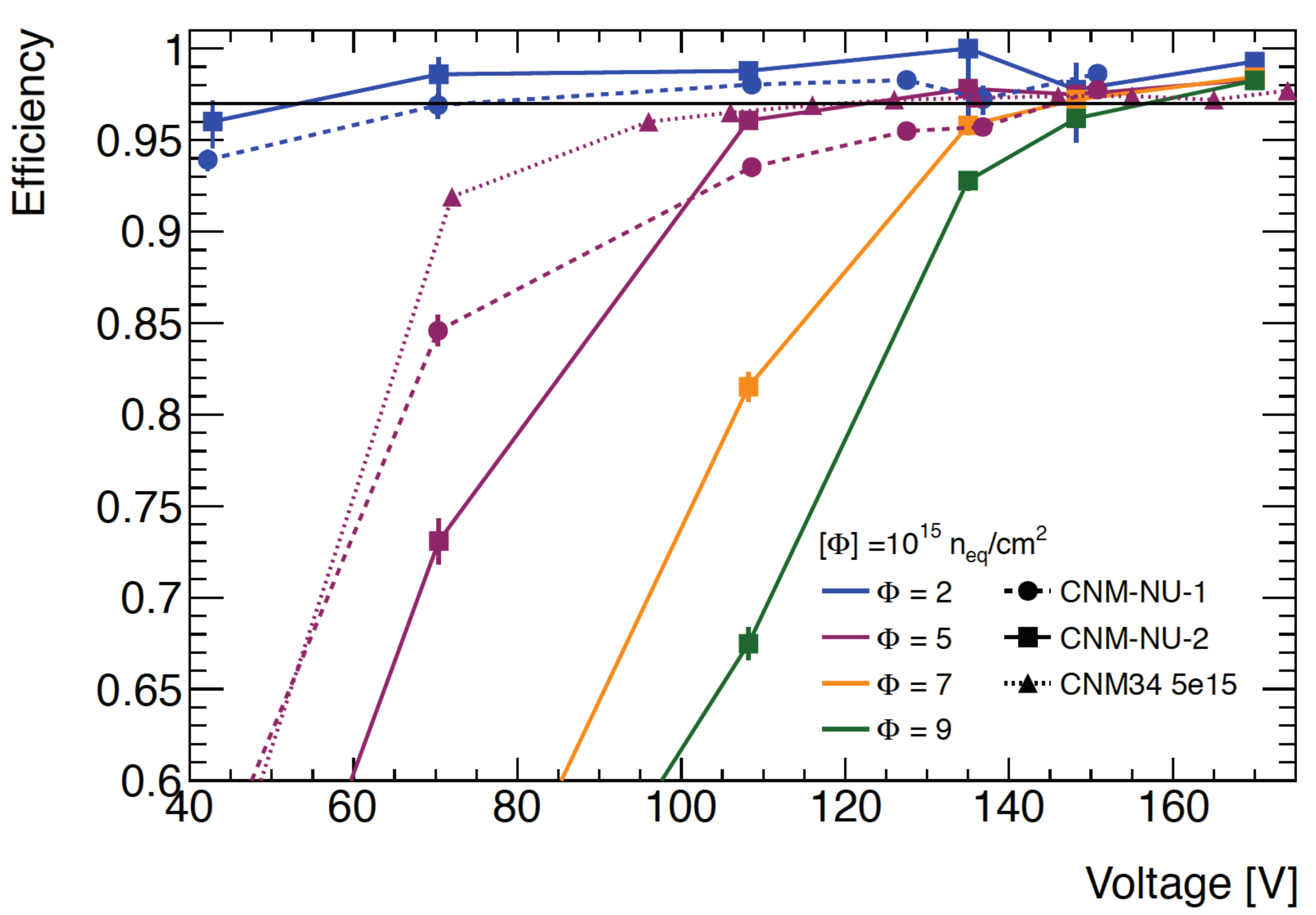}
    \caption{3D-Si sensors: hit efficiency as a function of bias voltage for
    different fluences and design variants \cite{Lange:2016jbm}.\label{fig:3D-Si_performance}}
\end{figure}

The hit efficiency obtained with 3D-Si structures designed by CNM \cite{Lange:2016jbm} is demonstrated in
fig.~\ref{fig:3D-Si_performance} for fluences of up to 0.9$\times 10^{16}$neq/cm$^2$ \cite{Lange:2016jbm}.
A $>$97\% efficiency plateau is reached with comparatively low bias voltages of 150\,V even at the highest fluences, with the missing 3\% being largely due to tracks with straight incidence into the column structures, a case not possible for tracks from actual LHC collisions. Smaller implant pitches with lower voltages for the same field strength also reduce power dissipation due to leakage currents after irradiation. This creates some safety margin against thermal runaway, not completely negligible even at the moderate bias voltages of 3D-Si sensors \cite{jlange2017}.
%

3D-Si sensors therefore are a strong contender for hybrid pixel modules for the innermost pixel detector layers at the HL-LHC. It should be noted, however, that
the fabrication process does is currently low volume. making it unlikely to cover large areas ($>1$\,m$^2$) with this technology choice.

\subsection{Bonding techniques}\label{sec:bonding-techniques}
Wafer- and chip-bonding as well as 3D-integration technologies are in wide\-spread demand in the semiconductor industry
sector of IC module assembly and stacking.
Many techniques exist that are suitable for a large variety of applications and given fabrication constraints. A large amount of literature is available, reviewed and summarized for example in  \cite{bonding_book_Ramm,3D_integration_book_Garrou_1&2,3D_integration_book_Garrou_3}.

For hybrid pixel detectors, the mating of readout chips to substrate (sensor) plates or substrate wafers is the main application of bumping/flip-chipping and 3D-integration techniques. The overriding demands in these applications are small capacitance additions to the preamplifier by the bond connection, good yield with $<$\,10$^{-4}$ defect rate (open or short), good contacts ($<$\,100\,m$\Omega$), and robustness against temperature cycling (--\,40\degree C to +60\degree C).
The need for small pitches between bond connections imposed a very high demand on industrial standards when pixel {R\&}D for the LHC started. At the time the required pitch of 50\,$\upmu$m was about 15 years
ahead of industrial demands. This pitch has become a standard today and requirements for current hybrid pixel {R\&}D often target pitches less than 50\,$\upmu$m.

Bump or Under Bump Metalization (UBM, see below) application usually is a wafer-scale process. Flip-chipping, on the other hand,
is conventionally employed to mate bumped readout  chips to (bumped) sensor plates\footnote{Depending on the techniques one of the mating partners can have bumps as well, e.g. for \emph{In-In} flip-chipping, or can just have UBM for the mating process, e.g. for solder bump flip-chipping.}.
For the existing detectors at the LHC experiments and likely also for the upcoming upgrades,
\emph{eutectic soldering}\footnote{Eutectic systems are solid mixtures that form a superlattice by striking a unique atomic percentage ratio and thus have the same melting point.} \cite{Oppermann_bonding_2012} and \emph{In-In}
thermocompression bonding are the methods that have been preferred over other techniques
(see e.g.\,\cite{pixel_book}) including anodic bonding, fusion bonding, and adhesive bonding (see below).
In current techniques ICs are usually bonded to sensor plates or wafers (chip-to-wafer bonding, C2W). Wafer-to-wafer bonding (W2W) is a cost attractive future possibility for pixel detectors that might become interesting for applications in connection with further advances in vertical electrical interconnection in so-called 3D integration techniques.
In particular \emph{through silicon vias} (TSV) open the possibility to provide elegant and space efficient
electrical contacts for hybrid pixel module fabrication when employing W2W or C2W bonding techniques. TSVs also allow
reaching through to the backside of a chip (or sensor). This can be exploited to use the chip's backside metal for redistribution of readout or service lines (redistribution layer, RDL).

\subsubsection{Solder bumps and bonding}\label{sec:solder-bonding}
One can perhaps subdivide the bonding technologies relevant in this review into (a) technologies requiring intermediate media to perform the bonding like for example eutectic bonding by means of solder or adhesive bonding using glue layers and (b) direct bonding, either metal-to-metal or silicon oxide-to-oxide. When intermediate materials are used no requirements on special surface treatments exist except for flatness.
After UBM (and bump) fabrication, oxide removal is needed before flip-chipping can be done.
The bonds are strong and post-processing - like thinning - is often possible. The lack of electrical connections (in the case of adhesive bonding) or price and pitch constraints (for solder bonding) are drawbacks.

\begin{figure}
    \centering
    \subfigure[Solder bumping and flip/chip]{\raisebox{0.0cm}
        {\includegraphics[width=0.60\textwidth]{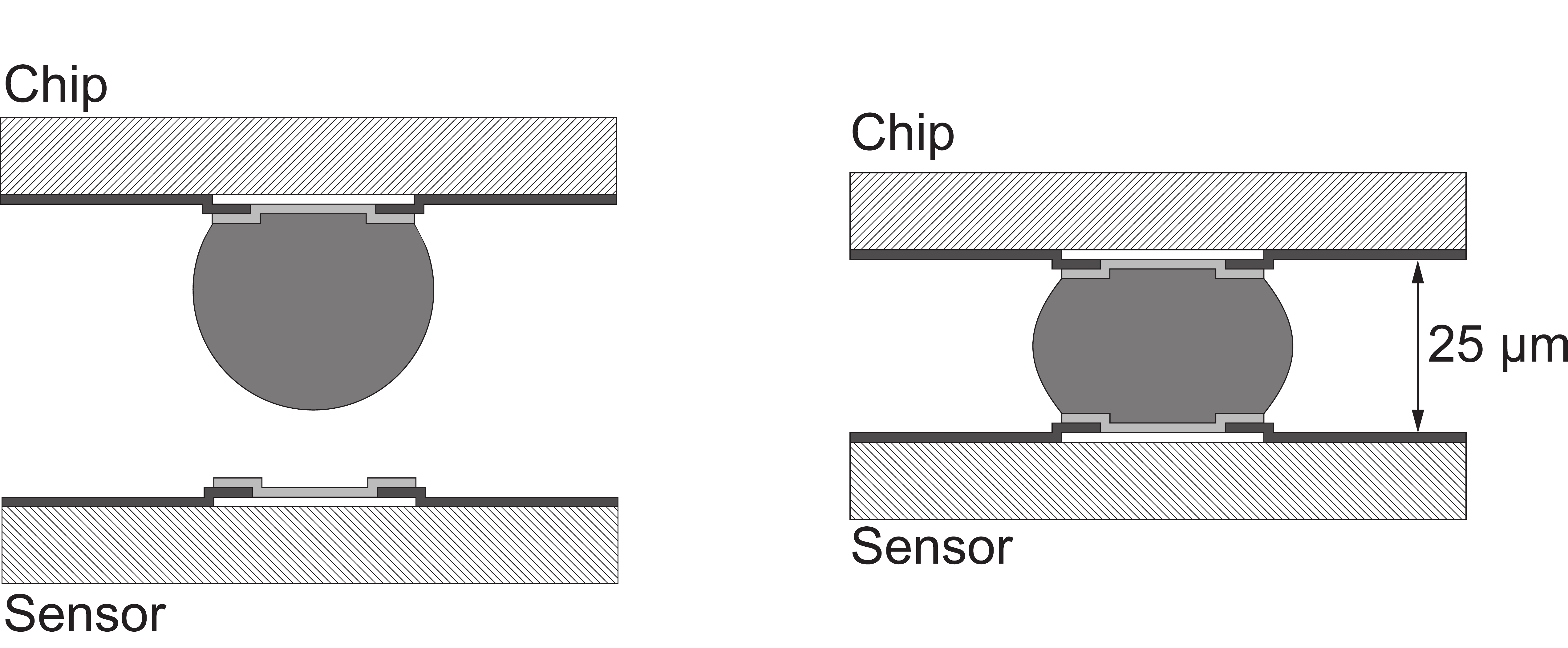}}\label{fig:solder_bumps}}\hskip 0.5cm
    \subfigure[Photograph]{\raisebox{0.0cm}
        {\includegraphics[width=0.35\textwidth]{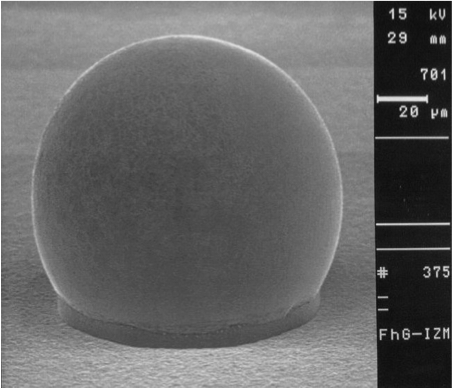}}\label{fig:solder_photo}}\hskip 0.0cm
    \caption{Eutectic solder bumps and the flip-chipping process: (a) Solder bump attached to a readout  chip is bonded to the Under Bump Metalization of the pixel sensor; (b) SEM micrograph of a SnAg
     bump (courtesy of Fraunhofer IZM Berlin) \label{fig:solder_flipchip}.}
\end{figure}

The bonding technique currently used most often is the electrochemical application of solder micro bumps to either of the mating parts:
chip or sensor (fig.~\ref{fig:solder_flipchip}).
A description of the method and the process steps is given in \cite{pixel_book}.
Solder bumping was introduced in 1969 by IBM in the C4-process (Controlled Collapse Chip Connection)~\cite{C4-IBM}.
The process is still in use by CMOS vendors
and is often offered with CMOS wafer production as an add-on. The pitch of the applied bumps is limited to 170\,$\upmu$m or larger, often too large for fine-pitch, high granularity pixel detectors in particle physics.
However, price and reliability arguments render C4-bumping still very attractive for low cost, large area applications at LHC upgrades, as is for example addressed in section~\ref{sec:sensors_planar}.

Eutectic solder bumping has meanwhile been developed with high connection density down to pitches of 25\,$\upmu$m \cite{IZM_Dietrich_2007} and bump dimensions of (15\,$\upmu$m)$^3$. High yield is obtained with Sn/Pb as well as with lead free Sn/Ag or Au/Sn alloys \cite{IZM_Klein_2008}. The ultimate pitch limit is believed to be around 5--10\,$\upmu$m \cite{IMEC_Majeed_2015}.
Readout ICs typically have Al or Cu pads for bump connection bounded by a passivation layer for chip protection. Typically a thin (100\;nm) Ti:W layer is first sputtered onto the wafers to promote adhesion and provide a barrier to prevent Cu atoms from diffusing into the pad metal (Al). Then a Cu or Au plating base ($>$\,150\,nm) is applied for the electrochemical contact followed by a well wettable Cu layer (1--5$\,\upmu$m). This layer stack is the UBM and is applied to both mating parts. In a wafer-based process cylinders of an electrodeposit are now grown onto the UBM of one part and turned into a spherical shape by reflow.
Bonding of both parts is so far mostly done after dicing of bumped readout ICs and sensor plates or wafers supplied with UBM in a flip-chipping process and subsequent re-melting of the solder bumps.
Note that this reflow process provides self-alignment of the mated parts.

Eutectic solder bumping is regarded as present day's workhorse bonding technique for hybrid pixel detectors. For the ATLAS pixel detector very high bond yields of more than 99\% have been obtained (i.e. number of
modules not rejected due to bonding issues).
\subsubsection{In-In bonding}
\begin{figure}
    \centering
    \subfigure[Indium bumps (flat)]{\raisebox{0.0cm}
        {\includegraphics[width=0.45\textwidth]{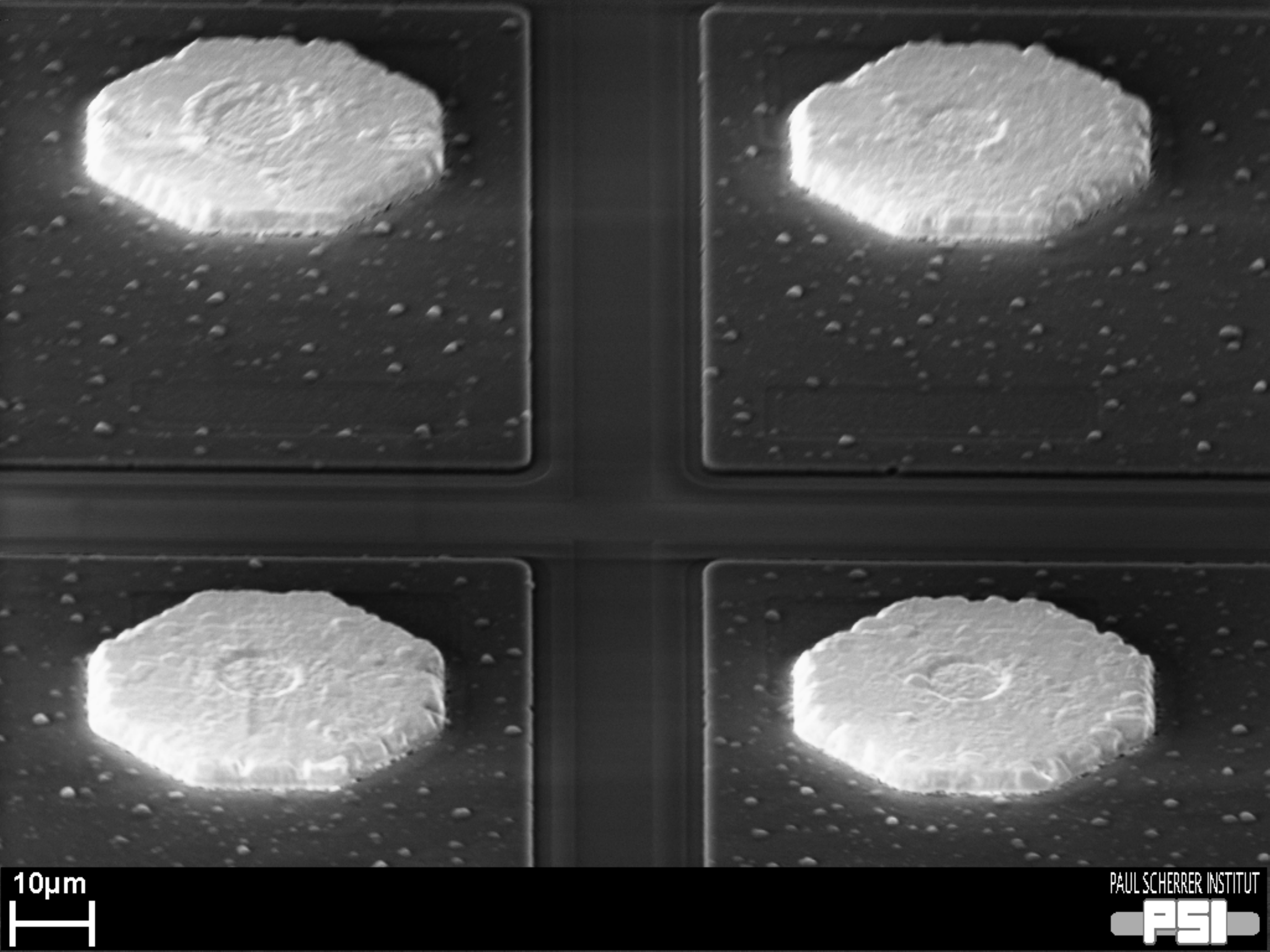}}\label{fig:In-In_bumps_flat}}\hskip 1.0cm
    \subfigure[Photograph]{\raisebox{0.0cm}
        {\includegraphics[width=0.45\textwidth]{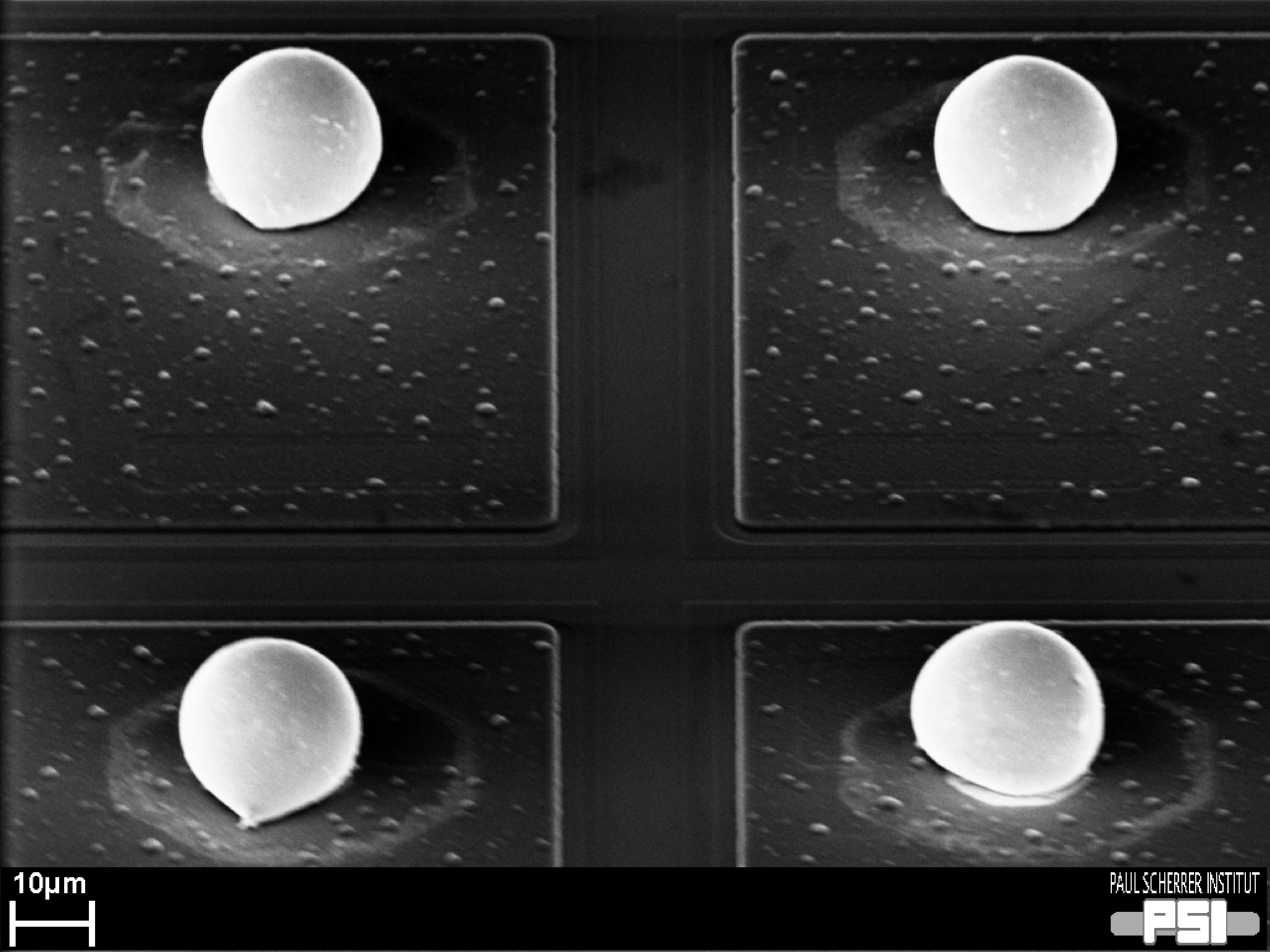}}\label{fig:In-In-_bumps_spheres}}\hskip 0.0cm
    \caption{In-In bonding: indium bumps produced before (a) and after (b) reflow to a sphere (from \cite{pixel_book}, courtesy of S. Ritter, PSI) \label{fig:In-In-bumping}.}
\end{figure}

In-In bonding has also been used in LHC pixel detectors for a large quantity of pixel modules. The technique usually employs vapour deposition\footnote{electroplating of indium is also possible}
of indium through openings in the lift-off masks deposited on the wafer (see e.g.\,\cite{pixel_book}).
The mask is pulled off from the wafer by a wet lift-off process, requiring the bumps to be fairly flat and thin ($\approx$10\,$\upmu$m, fig.~\ref{fig:In-In_bumps_flat}).
Also a UBM is necessary for this process, usually Ti-Pt-(Au) is used.
The applied indium bumps are then bonded by thermocompression at about 100\,{\degree}C.
The advantages of the technique are its ease of application and the low temperature requirements. The bonds are, however, comparatively fragile, and the fabricated modules
tend to have a lower mechanical damage threshold than solder bumped assemblies.
In the Run-1 ATLAS pixel detector about 50\% of the modules were produced using In-In bonding, showing high reliability with acceptance yields around 90$\%$.
For the CMS pixel detector a reflow step has been introduced after bump deposition \cite{Broennimann:2005qv}, turning the In bumps into spherical shapes (fig.~\ref{fig:In-In-_bumps_spheres}). The two spheres are merged
in a second reflow at 180\,\degree C. This provides a larger distance between chip and sensor, but it introduces self-alignment and stronger bonds.

\subsubsection{Adhesive bonding}
Recently adhesive wafer-to-wafer bonding has been discussed in the context of active CMOS sensors, capacitively coupled to readout chips where the capacitive coupling is provided between the metal pads
of sensor and chip, respectively, separated by the adhesive bonding layer.
Adhesive bonding has been claimed to reduce production cost by avoiding the UBM, bump deposition, and reflow steps.
This has, however, not yet been demonstrated and is not at all obvious. Planarity requirements are stricter and requirements for alignment for chip-to-chip/sensor placement are the same as for standard bump bonding, the latter being the cost driver, not to mention the need for (cost intensive) electrical connections of the mated parts if provided by TSVs.

In the context of capacitive coupling of \,\glq smart sensors\grq\ to readout ICs (CCPD) adhesive bonding has been investigated regarding thickness control and uniformity of the adhesive as well as
radiation tolerance \cite{Gaudiello:2016fvw}. Good glue uniformity at the micrometer level has been achieved, however further evaluation of its competitiveness is needed, in particular whether
electrical connections can be provided by cheap techniques like e.g.\,wire bonding, thus, however, abandoning the goal of compact module shapes.

\subsubsection{Cu-Cu direct bonding}
Metal-metal adhesion (Cu-Cu or Au-Au) of flat and polished surfaces has been known for a long time.
The bonding forces principally involve capillary, Van der Waals, and electrostatic forces, as well as solid bridges caused by impurities and hydrogen bonds between OH groups.

Two mirror-polished wafers are put into contact and held together by adhesive forces without any intermediate material
\cite{bonding_book_Ramm}. The process completes by Cu atom diffusion between the
two Cu layers. Thermocompression (diffusion) bonding is simplest, but the required temperatures are usually too high for typical sensors and CMOS chips used in pixel detector applications.

Especially interesting is surface-activated Cu-Cu bonding as it works at ambient temperature. Surface-activation in this context means increasing the bonding force by surface treatment, e.g. by augmenting the number of hydrogen bonds (hydrophilicity) or by generating new types of chemical bonds. The treatment methods comprise wet chemical processes and
(oxygen) plasma etching. The bonds are electrically conductive, an advantage for any wafer-to-wafer
but also chip-to-wafer bonding project.

Requirements for good bonding are native oxide and other remnant removal, excellent chemical-mechanical polishing (CMP), and highly planar parts (wafers) with sufficient total Cu fraction. Very fine pitch
($<$\,4\,$\upmu$m) and low capacitance contacts are possible. The demands on surface cleanliness and flatness are the biggest drawbacks. Wafers must be processed soon after fabrication. Wafers must be very planar
and sufficiently stiff which might compromise very thin solutions. At present the technique is also still fairly
expensive.

\subsubsection{Oxide-oxide direct bonding}
Under room temperature conditions silicon wafers are covered with an oxide layer that can be used for direct wafer bonding, a method commonly used for SOI wafer production.
Most common is \glq hydrophilic\grq\ wafer bonding in which the wafers are, after cleaning, rinsed or stored in deionized water. Water
is bonded via Si-OH groups on the silica surface. The bonding can be done at room temperature but the bonding strength increases with temperature (and time).
With the addition of an electric field one speaks of anodic bonding, which can tolerate rougher surfaces.

Required is high quality CMP polishing and extremely good surface cleaning to avoid large bonding voids. The bonds are non-electrical. Hence TSVs or other electrical connections are needed for
pixel assemblies. For this process wafer-to-wafer bonding is much easier than chip-to-wafer or chip-to-chip.

\subsubsection{Solid-liquid interdiffusion bonding (SLID)}
\begin{figure}
    \centering
        \includegraphics[width=0.80\textwidth]{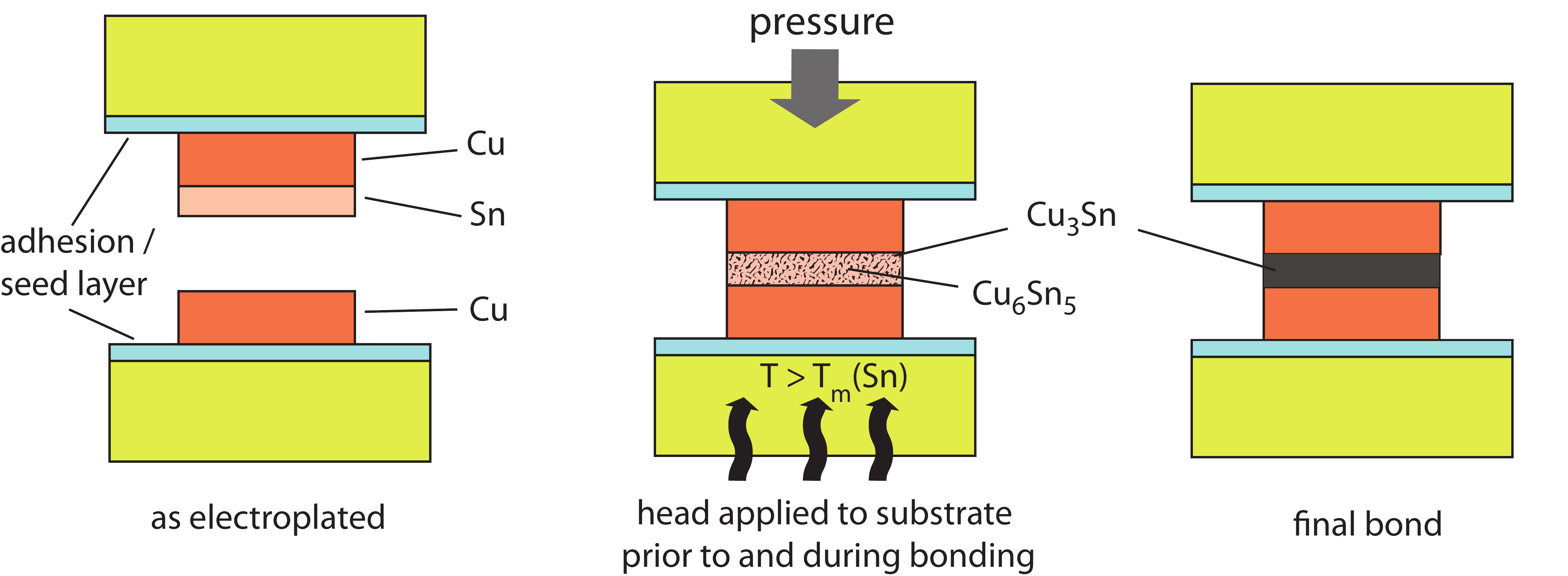}
         \caption{Process steps of Solid-Liquid Interdiffusion bonding (SLID): (a) structure with metal stacks before bonding, (b) alloy forming step under pressure at 240-320\,{\degree}C, (c) resulting bond connection. \label{fig:SLID}}
\end{figure}

SLID bonding uses an intermetallic alloy formation and represents an alternative fine-pitch bonding technique based on thin (often eutectic) Cu-Sn connects \cite{Munding_2008,bonding_book_Ramm}.
Between two Cu layers of several micron thickness a thin (3\,$\upmu$m) Sn layer is applied on one of the Cu layers (fig.~\ref{fig:SLID}). They are brought in contact and heated to a temperature of around 240--320\,\degree C. At this temperature the tin diffuses into the copper forming the Cu-Sn alloy. As the melting point of this alloy is around 600\,\degree C multiple layers can be stacked and connected without melting the previously formed SLID connections. Furthermore, the process is also a flux-free bonding alternative with pitches much below 20\,$\upmu$m, whereas the number of processing steps is about the same as for
solder bonding (the reflow step is not needed).
Technically, for SLID, pressure and temperature are required during the bonding step. This renders sequential bonding of multi-chip pixel modules necessary (chip after chip), whereas bonding by
reflow does not require pressure (c.f. page~\pageref{sec:solder-bonding}) and can be done in parallel for a number of chips bonded to a sensor plate.

Despite their similarity SLID and soldering are fundamentally different processes with distinct and unique properties. A major difference is that soldering is reversible, whereas SLID is
irreversible and only melts at temperatures much higher than the SLID process temperature.
This implies also that no reworking is possible.
Some process similarities to metal-metal thermocompression bonding also exist.
A detailed comparison is given in \cite{bonding_book_Ramm}.
SLID is particularly suited for wafer-to-wafer and chip-to-wafer bonding.

\subsection{Wafer thinning}
Thin wafers are a goal not only for low-mass pixel detectors in particle detection, but also for a number of integration processes, in particular for Through-Silicon Vias. While thick wafers are better for handling, other process steps like for example etching benefit from small wafer thickness. However, dedicated \glq handling wafers\grq\  are needed
to deal with flatness and bowing issues. Often handling wafers are of the SOI (Silicon-on-Insulator) type, because the
Si-SiO$_2$ interface transition offers a sharp etch stop with the oxide also acting as a sacrificial layer after ion etching.

Wafers are thinned by (backside) grinding. The active (front) side is first protected by tape. Grinding in two steps (coarse/fine) is
performed using grinding wheels with diamond grain sizes from 1-8\,$\upmu$m (fine) to 20-80\,$\upmu$m (coarse) diameter. After the thinning process, breaking strength, warpage, and bow are characterization parameters that depend on the grinding tool characteristics and the grinding speed. More details are given in \cite{3D_integration_book_Garrou_1&2}.

\subsection{3D-integration and through-silicon vias}\label{sec:3D-integration}
\begin{figure}
    \centering
    \subfigure[Device layers connected by TSVs]{\raisebox{0.0cm}
        {\includegraphics[width=0.45\textwidth]{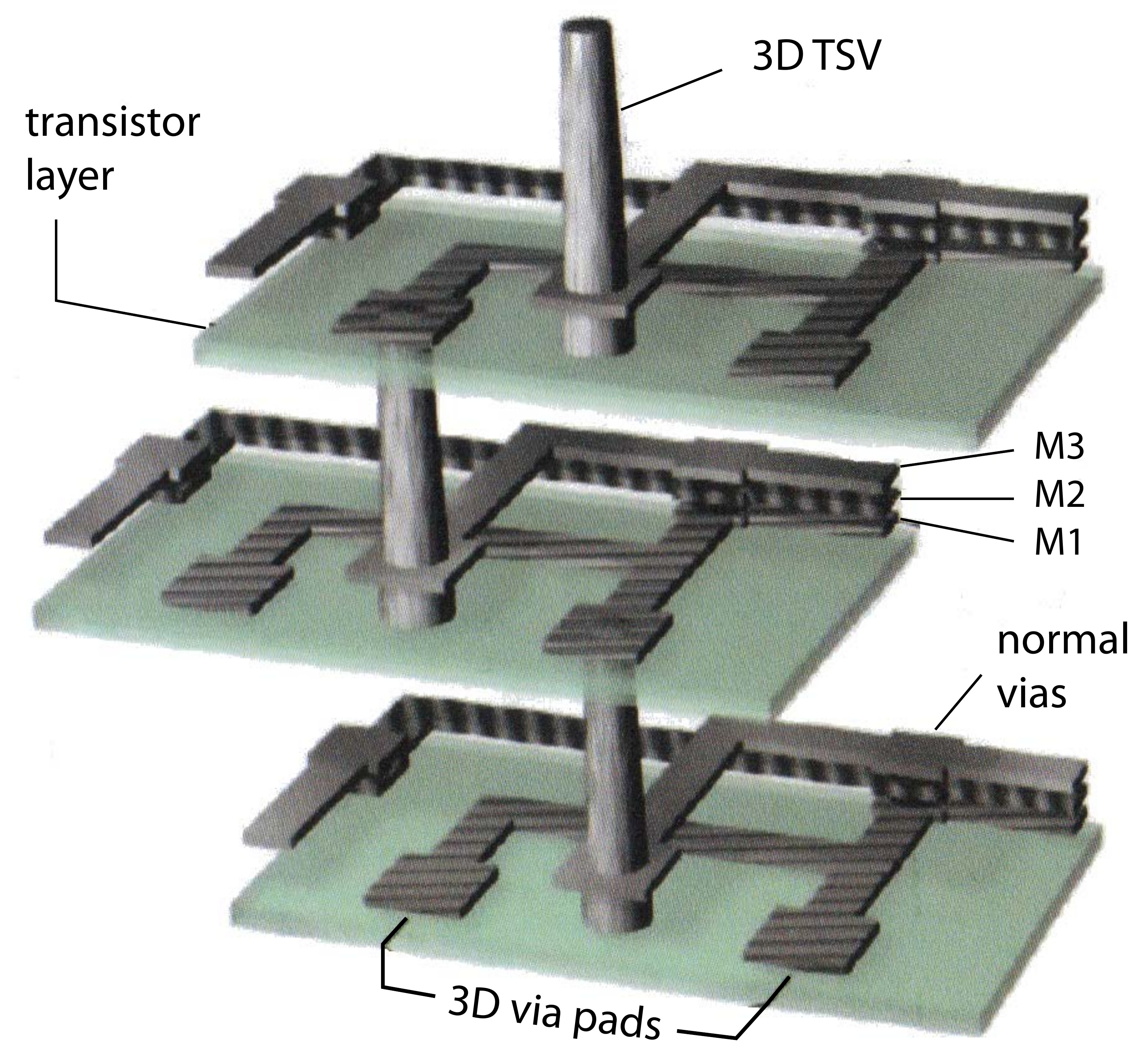}}\label{fig:TSV-principle}}\hskip 1.0cm
    \subfigure[Bonding of two wafers with electrical connection via a SuperContact TSV.]{\raisebox{0.0cm}
        {\includegraphics[width=0.45\textwidth]{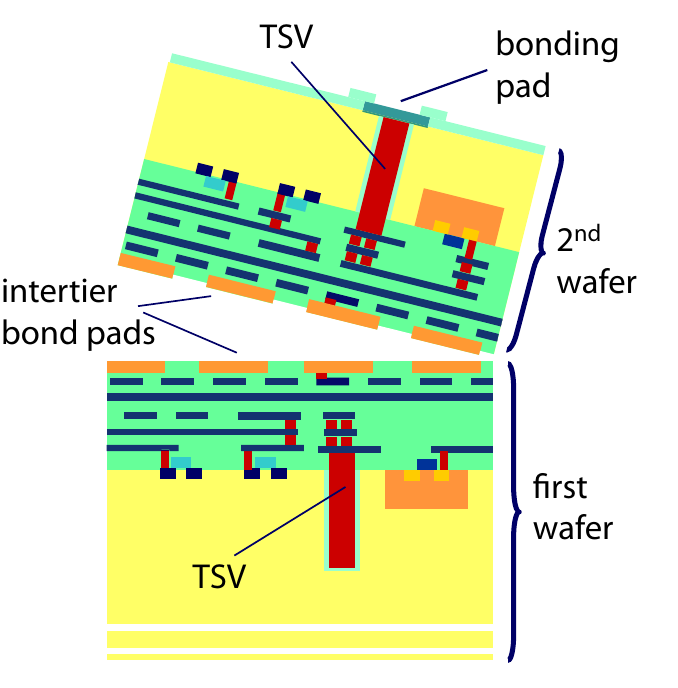}}\label{fig:3D-integration_principle}}\hskip 0.0cm
    \caption{3D-integration principle: (a) TSVs connecting different electronics device layers
(adapted from \cite{3D_integration_book_Garrou_1&2}); (b) bonding of two wafers with electrical
    SuperContacts to the outside (Tezzaron; adapted from \cite{PattiR_2008}).\label{fig:TSV-concept}}
\end{figure}
%
Extending electronics integration into the third dimension is -  apart from feature size shrinking - regarded as the second route to ever increasing circuit density.
3D-stacking of several electronic device layers (tiers) is thus an eminent field of industrial
research. Further advantages are reduced power consumption due to smaller connections and smaller involved capacitances as well as larger I/O bandwidth and more functionality at lower cost.
Vertical vias running through layers of silicon (TSVs) are a key ingredient for 3D-stacking \cite{3D_integration_book_Garrou_1&2}. Tiers can be different CMOS layers but also electronics and sensor layers or layers interfacing to an optical signal transport.
Figure~\ref{fig:TSV-principle} is a sketch illustrating the principle of 3D-interconnection via TSVs.

Besides TSV fabrication, wafer thinning and aligned wafer bonding are the key follow-up processes for 3D-integration.
Direct bonding, especially Cu-Cu fusion or SLID (see above) are the preferred bonding techniques for 3D-stacking, providing strong fine-pitch bonds that are also conductive.

For future pixel detectors in particle physics, 3D-integration and TSVs are of interest if
(a) wafer scale production becomes practical and/or (b) if TSVs allow
better connectivity of detector modules by avoiding space consuming wire bonds and providing better buttable modules.
In addition TSVs can reach through to the backside of
a chip, allowing redistribution of electrical connections on the chip's backside metal is addressed.
This in turn saves the material normally added by flex interconnect circuits.
\pagebreak

\subsubsection{TSV fabrication}
\begin{figure}
    \centering
        \includegraphics[width=0.80\textwidth]{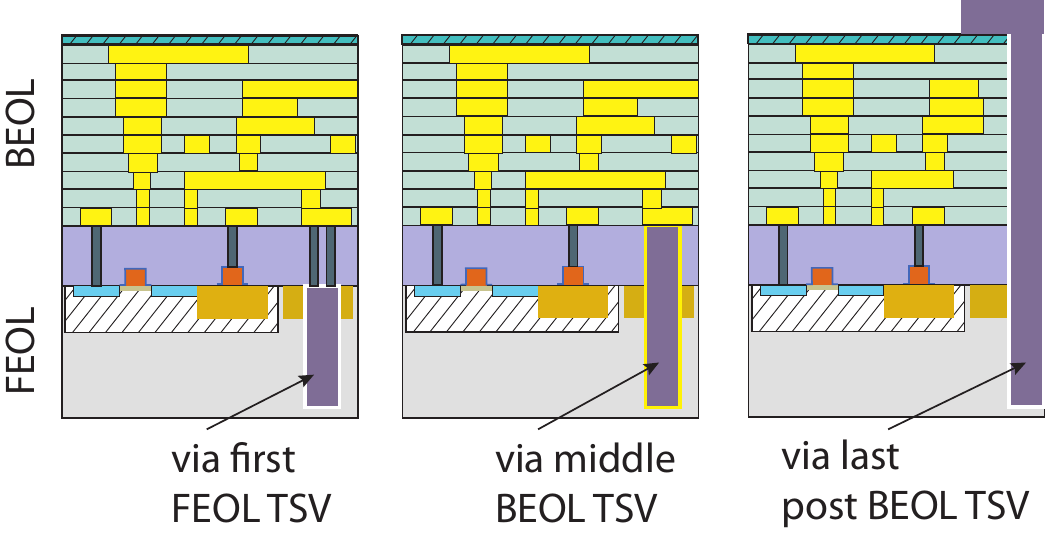}
         \caption{Different types of vias. (left) \emph{via-first}: fabrication before or during FEOL, (middle) \emph{via-middle}: fabrication after FEOL during BEOL, (right) \emph{via last} fabrication after IC completion (post-BEOL) (adapted from \cite{TSV_types_2011}). \label{fig:TSV_types}}
\end{figure}
%
%
\begin{figure}
            \includegraphics[width=0.4\textwidth]{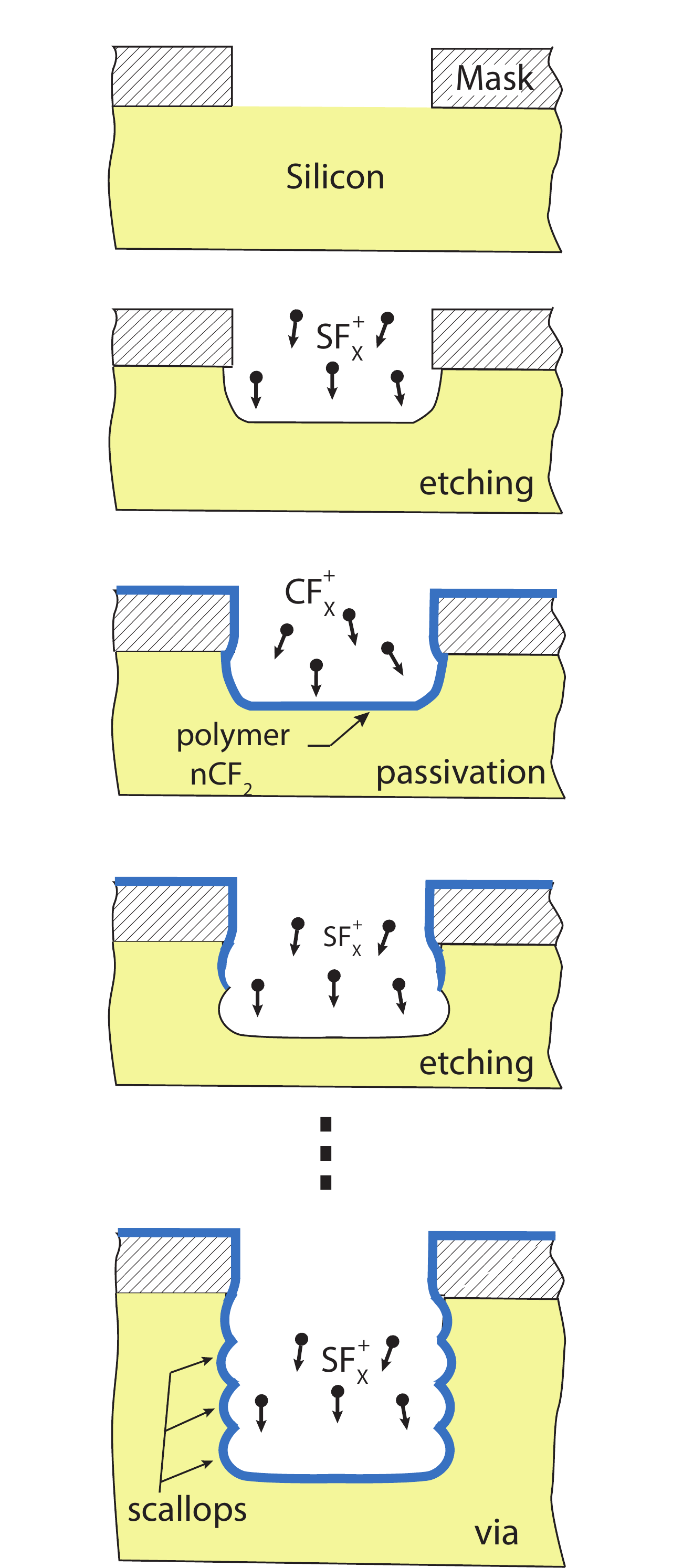}
    \caption{Deep Reactive Ion Etching (DRIE) in the Bosch process \cite{Bosch-process}. From top to bottom a sequence of ion etching and passivation is executed (see text).} \label{fig:Bosch-process}
\end{figure}
%
In IC fabrication all processes before the first wiring metal are called \emph{front end of line} (FEOL), whereas the \emph{back end of line} (BEOL) begins with this
first metal processing and ends with the last IC processing step.
Different types of TSVs are distinguished depending on the point in the process flow at which they are fabricated \cite{mjwolf2017}, i.e.
\begin{itemize}
  \item \emph{via-first} or FEOL via (polysilicon) applied before or during FEOL fabrication.
  \item \emph{via-middle} or BEOL via (metal) applied after FEOL during BEOL fabrication
  \item \emph{via-last} or post-BEOL via(metal) applied after IC fabrication {post-BEOL}. Here one distinguishes also \emph{front side via-last} as in fig.~\ref{fig:TSV_types}\,(right) and \emph{backside via-last}, depending on whether the via is etched from the front through the BEOL stack (difficult) or from the back through (thinned) silicon substrate.
\end{itemize}
Note that the conducting material for FEOL TSVs must be doped polysilicon for reasons of thermal and material compatibility, having the disadvantage of higher resistivity compared to metals.
For BEOL and post-BEOL vias tungsten and copper are most common.
Figure~\ref{fig:TSV_types} shows these distinct TSV types schematically.

Important TSV specifications are diameter, pitch, and aspect ratio (length-diameter ratio). Present typical ranges are given in
table~\ref{tab:TSV_params}.

The main process for TSV production is \emph{deep reactive ion etching} (DRIE, invented by Bosch \cite{Bosch-process}). The process employs a directional repetitive sequence of ion etching and wall passivation resulting in anisotropic etching of the silicon bulk. The principle is shown in fig.~\ref{fig:Bosch-process}. The alternating cycles (typically lasting about 5\,s) are: etching under a bias voltage with SF$_6$ or NF$_3$ in argon atmosphere to form gaseous SiF$_x^+$ products - passivation of the surfaces with C$_3$F$_6$, C$_4$F$_8$, or CHF$_3$ in argon forming a protecting fluorocarbon polymer surface layer on sidewalls and bottom. The bias voltage provides a directional orientation of the bombardment. The resulting via has a typical scalloped surface with undercuts that are, however, substantially reduced by advanced DRIE parameters.
\begin{table}
	\centering
	\caption{\label{tab:TSV_params} Typical TSV parameters (2017) \cite{3D_integration_book_Garrou_1&2,mjwolf2017,TFritsch_feb2017}.}
\begin{tabular}{|c|c|c|c|}
\hline
via type & smallest diameter & smallest pitch & typical aspect ratio \\
\hline
via-first           & 3 -- 5  $\upmu$m  & 6 -- 10  $\upmu$m  &     10 : 1       \\
via-middle          & 3 -- 5  $\upmu$m  & 6 -- 10  $\upmu$m  &  5--10 : 1       \\
front side via-last  & 10 -- 20 $\upmu$m & 20 -- 40 $\upmu$m  &  5 : 1       \\
backside via-last   & 5 -- 20 $\upmu$m  & 10 -- 40 $\upmu$m  &  5--12 : 1       \\
\hline
\end{tabular}%
\end{table}

\subsubsection{First HEP experiences with TSVs}
A characteristic of TSV application for HEP pixel detectors is the fact that readout chips or sensors are still comparatively thick (100-200\,$\upmu$m) for via fabrication. On the other hand the required via pitch and density
often is relaxed, for example when only service lines need to be via-connected or when I/O pads can have large pitches.

\begin{figure}
    \centering
    \subfigure[Tapered via: cross section]{\raisebox{0.0cm}
        {\includegraphics[width=0.45\textwidth]{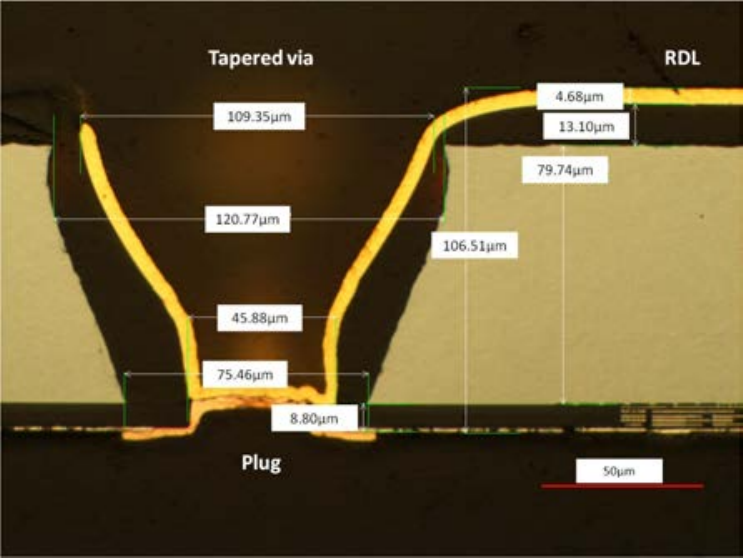}}\label{fig:tapered_via_xsection}}\hskip 1.0cm
    \subfigure[SEM photo]{\raisebox{0.0cm}
        {\includegraphics[width=0.45\textwidth]{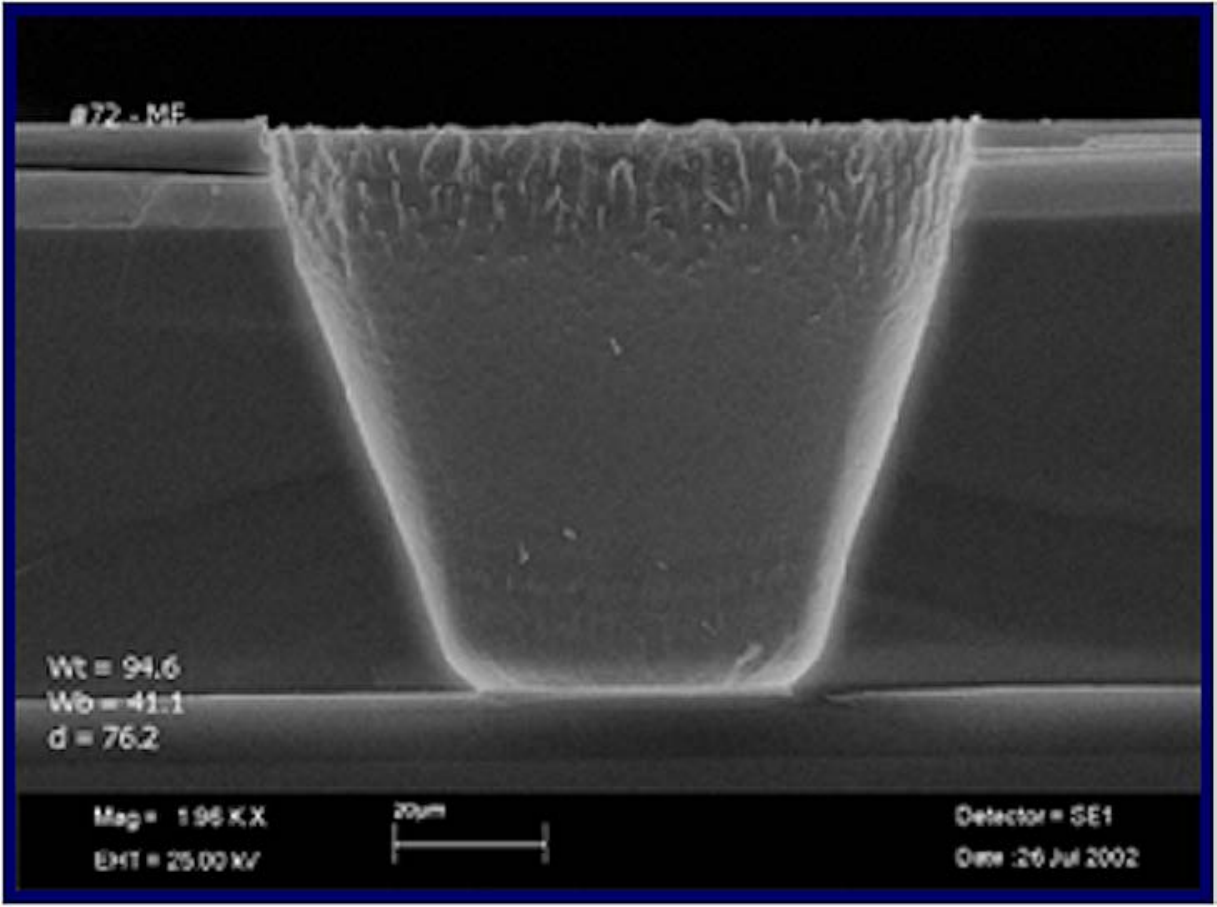}}\label{fig:tapered_via_photo}}\hskip 0.0cm
    \caption{Tapered via demonstrated on an ATLAS FE-I3 pixel module: (a) cross section detail; (b) SEM photograph \cite{barbero:2012xyz}.\label{fig:tapered_via}}
\end{figure}

Therefore, the first successful HEP pixel module operation employing TSVs \cite{barbero:2012xyz} was obtained with small aspect ratio ($\sim$1:1) tapered vias as shown in fig.~\ref{fig:tapered_via}, fabricated at Fraunhofer IZM, Berlin.
Chip wafers (FE-I3) of the ATLAS pixel detector production were thinned to about 80\,$\upmu$m thickness. The I/O pads
were accessed through TSVs and redistributed on the backside of the chip. The via geometry can be seen from fig.~\ref{fig:tapered_via_xsection}: 110/45\,$\upmu$m diameters (top/bottom), $\sim$\,$70\degree$ tapering angle, 150\,$\upmu$m via pitch. After wafer dicing a pixel module was assembled by bonding a readout chip containing vias to a pixel sensor using solder bonding.
The pixel module was operated and characterized with radioactive sources without performance loss compared to modules connected via wire bonds instead of vias.

\begin{figure}
    \centering
    \subfigure[Straight via IZM]{\raisebox{0.0cm}
        {\includegraphics[width=0.55\textwidth]{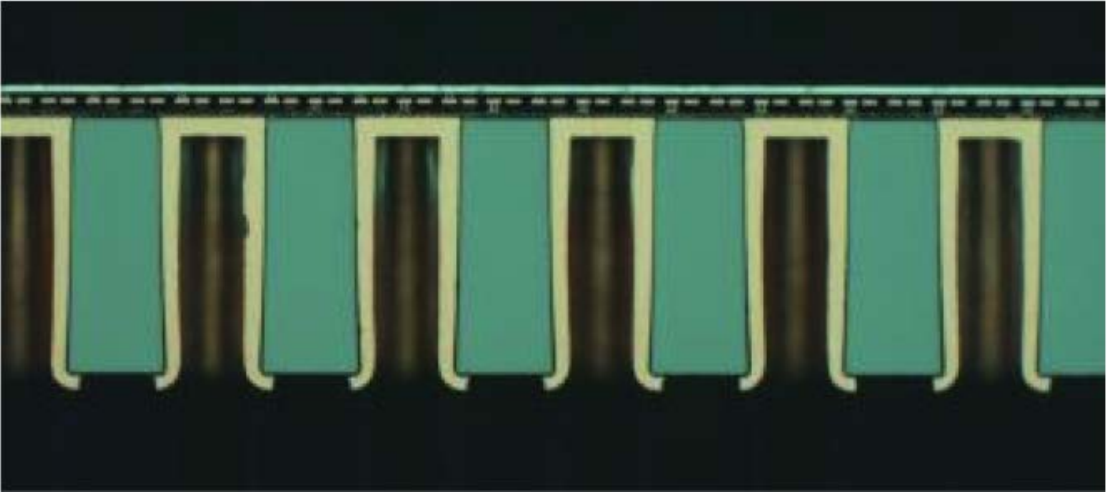}}\label{fig:straight_via_IZM}}\hskip 1.0cm
    \subfigure[Straight via CEA/LETI]{\raisebox{0.0cm}
        {\includegraphics[width=0.35\textwidth]{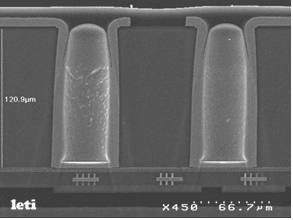}}\label{fig:straight_via_LETI}}
    \caption{Straight via applications used for HEP pixel detector prototypes: (a) IZM straight via
(here in a MEDIPIX3 chip)~\cite{TFritsch_feb2017};
(b) LETI/CEA \cite{CEA-LETI2}.}
\end{figure}

Larger via depths and aspect ratios are required to connect large area chips like the ATLAS FE-I4 chip \cite{FE-I4} having about 2$\times$2\,cm$^2$ area. Without handling supports, such chips show large bows and the thickness must be optimized for
handling, bonding, and TSV yields. A successful approach was achieved using straight copper vias (Fraunhofer IZM, Berlin) with an aspect ratio of 2.7:1 applied to 160\,$\upmu$m thick FE-I4 wafers \cite{IZM_TSVs_Medipix_2016},\cite{TSVs-FEI4:2016}
(fig.~\ref{fig:straight_via_IZM}). Similar efforts are carried out at LETI/CEA \cite{SKuehn:2016} (fig.~\ref{fig:straight_via_LETI}).
A 2$\times$2\,cm$^2$ pixel module was produced along similar lines containing about 25\,500 pixels, and characterized using radioactive sources. To increase the via yield to close to 100\% a new approach using chip wafers thinner than 100$\,\upmu$m with handling wafer support using straight Cu-coated vias is under way.
%

%% file: fechip.tex
\section{Readout integrated circuits, data transmission, and power distribution}\label{sec:fechip}

The ROIC provides the functionality of a hybrid pixel assembly. From the start, the purpose of the hybrid approach has been to allow independent optimization of ROIC and sensor~\cite{Heijne:1988ep}. The ROIC technology development has been enabled by
Moore's Law (the doubling of logic density every 2 years in commercial electronics) 
towards deep submicron commercial CMOS, in order to meet rate and radiation tolerance demands. 
The alternative of stacking circuit layers in the 3$^{\rm rd}$ dimension without reducing feature size has also been explored, 
but has so far been less successful. Data transmission and power distribution are system challenges concerning the ROIC as well 
as other elements such as optical components, encoding/decoding protocols, cables, and cooling.

\subsection{CMOS ROICs}

It is useful to classify CMOS ROICs in terms of generations, which have a rough correlation with the feature size used.
First generation pixel chips are typified by those in the original ATLAS and CMS detectors~\cite{Peric:2006km,Barbero:2004psi46},
Medipix~\cite{Llopart:2003157}, using 0.25\,$\upmu$m feature size CMOS with custom layout techniques for radiation tolerance.
Second generation chips have been developed and fabricated using 130\,nm CMOS and are running in current experiments or
devices~\cite{FE-I4,medipix3rx,timepix3}.
Third generation chips are under development, focusing on 65\,nm CMOS~\cite{RD53A-specs,Valerio:2014clic,Campbell:2016pix4}.
Table~\ref{tab:generations} shows the evolution of memory and readout bandwidth between generations.

\begin{table}[htb]
\centering
\small
\begin{tabular}{ | l | c | c | c | }
\hline
& \textbf{$1^{\rm st}$ gen.} & \textbf{$2^{\rm nd}$ gen.} & \textbf{$3^{\rm nd}$ gen.} \\
\hline
\textbf{\small Hit data storage density} & $<1$~Gb/s/cm$^2$ & 5~Gbps & 40~Gb/s/cm$^2$ \\
\hline
\textbf{\small Chip output bandwidth} & 40-160\,Mb/s & 0.3-1.2\,Gbps & 2-20\,Gb/s  \\
\hline
\end{tabular}
\caption{Hit storage memory and output bandwidth for each generation of pixel readout chip.}
\label{tab:generations}
\end{table}

The basic elements and organization of a hybrid pixel readout chip are common to all generations and applications.
The defining characteristic is the use of a dedicated amplification and discrimination channel per pixel ({\em front end}),
with the ability for all these front ends to operate in parallel, followed by parallel processing of the output signals.
Section~\ref{sec:ROIC-fend} reviews the front end characteristics.
The output of the front end is most commonly digitized in some way on the pixel, but can also be sampled on an analog memory
for later readout as an analog level~\cite{Barbero:2004psi46}.
On-chip digitization allows for higher readout bandwidth with pure digital readout
and is expected for all third generation chips.
The preferred digitization method has so far been ToT (see section~\ref{sec:ROIC-fend}).
We normally refer to a pixel {\em firing} when the amplified signal exceeds the discriminator threshold.
Managing power transients and noise coupling when all pixels are simultaneously {\em live}
(having the ability to fire at any time) is a main system challenge of pixel ROIC design.
Many chips are designed for a given maximum occupancy
(number of pixels firing at once) and therefore only guarantee system stability in applications that do not exceed this occupancy.

While the front end architecture remains similar from one generation to the next, processing of the information produced by the front end,
and the relationship of the front end to the rest of the chip, have evolved significantly. First
generation chips were mainly analog circuits, with some logic and memory to manage hit buffering and readout. The most salient
new feature of second generation chips was the use of synthesized logic side-by-side with analog front ends in the entire pixel
matrix, but with an organization in columns and small digital blocks stepped and repeated. The third generation will go further,
implementing an almost entirely digital chip, with a large number of pixels within a synthesized logic basic unit.
The front ends will be embedded as data sources within a digital fabric entirely surrounding each analog unit.

Logic density is critical to pixel ROIC design,
unlike in most other physics detector applications, where analog performance tends to be of the highest importance.
Achievable logic density is intimately tied to rate capability as will be seen later.
The 1$^{\rm st}$ generation chips had a mainly analog pixel matrix,
with digital operations implemented using full custom circuits (produced with analog design methods).
Any digital processing and data buffering were done in the periphery.
This resulted in a relatively large periphery area
and the need for high bandwidth data transfer from pixels to periphery (see section~\ref{sec:ROIC-architecture}).
The most prominent new development in 2$^{\rm nd}$ generation chips was the use of synthesized logic within the pixel matrix, enabled by a leap in logic density (see section~\ref{sec:ROIC-radiation}).
Thus the matrix became a mix of analog columns designed with conventional analog methods and digital
columns synthesized with digital design tools. Digital processing and storage in the columns meant the periphery area could shrink
and the hit rate capability increased (section~\ref{sec:ROIC-architecture}).
The 3$^{\rm rd}$ generation chips are now expected to be essentially digital,
with embedded analog amplifiers kept to a minimum and logic complexity rivaling commercial microprocessors.
This opens the door to unprecedented functionality within the front end chip.

\begin{figure}
 \centering
 \includegraphics[width=0.9\textwidth]{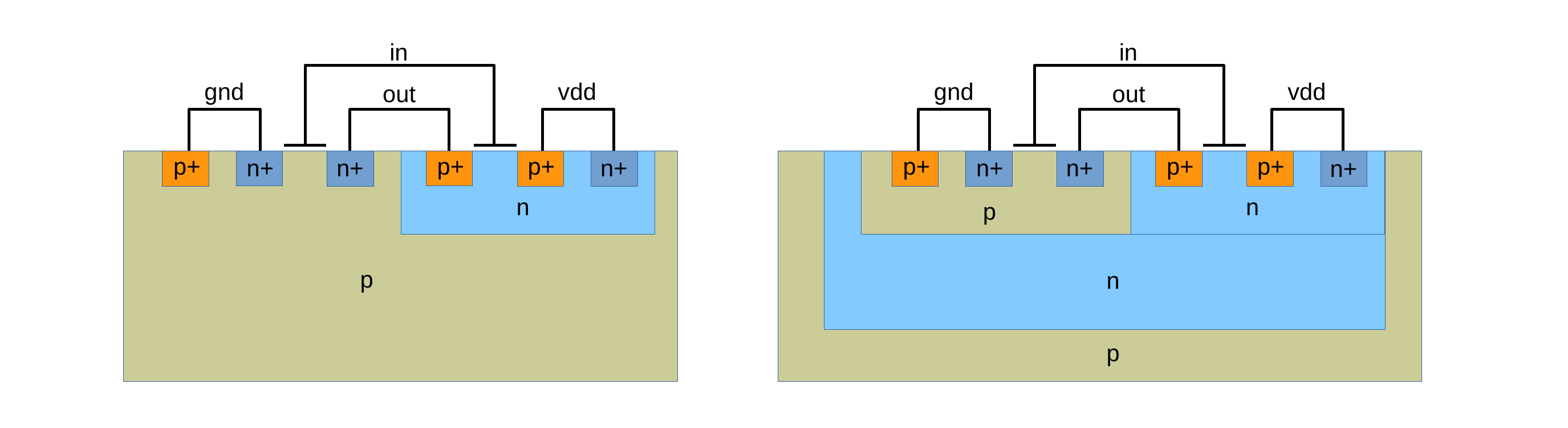}
 \caption{\label{fig:triplewell} Schematic representation of a conventional CMOS inverter (left) and the triple well process implementation (right).}
\end{figure}

A key element in the transition to essentially digital chips has been isolation of circuits within the same chip. Digital switching produces
current spikes. Even if analog and digital circuits use different power supplies and grounds, return currents induced in the silicon substrate can inject parasitic signals into sensitive analog circuits. 
This can be mitigated by isolating circuits from the bulk substrate, using options provided in many deep submicron processes. 
Figure~\ref{fig:triplewell} compares schematically a logic inverter in conventional CMOS with the triple well version. 
It can be seen that the NMOS device and the local ground are isolated from the substrate in the triple well process. Some processes 
offer more sophisticated implant structures that allow further isolation of the N wells, involving separately biased additional wells 
and/or insulating trenches. Silicon on Insulator (SOI) processes have not been popular for high radiation tolerance because they
contain a thick oxide layer under the circuitry that accumulates trapped charge with radiation dose, giving rise to
undesirable back-gating effects in fully depleted devices (FDSOI). 
The format of isolation structures exploited varies with vendor and technology.
Designs also vary between isolating only analog circuits, only digital circuits, or
both from the true bulk substrate.
While such isolation strongly attenuates noise coupling, it does not completely eliminate it.
Therefore, in addition to isolation, controlling power consumption transients or changes during chip operation is
increasingly important.
Isolation implants have also been exploited for monolithic active pixel designs, covered in chapter~\ref{sec:cmos}.

Logic density is most important in a triggered system (see section~\ref{sec:ROIC-architecture}). This is the case
for the highest rate detectors in a collider geometry, where the raw hit volume is too high to allow for read-all operation
with an acceptable mass of data cables. In a triggered system the ROIC function is to remember the position, arrival time,
and charge of all hits until a trigger arrives.
The rate of hits depends on the rate of particles impinging on the detector and, since each hit requires a
certain number of bits for digital storage, 
it can be characterized as a specific bandwidth in bits per second per unit area,
as was done in table~\ref{tab:generations}. Note that this does not depend on pixel size to first order, 
but on particle flux (one incident particle makes one pixel cluster regardless of pixel size, and
with the use of region architectures (section~\ref{sec:ROIC-architecture}) the effect of cluster size is small). 
Given a particle flux, the total memory needed depends on the trigger latency, which is typically a few microseconds.
The achievable memory per unit area is limited by the CMOS process logic density.
Thus, for a triggered system, higher hit rate (and/or longer latency) requires higher logic density.
This requirement, not pixel size or radiation, 
has driven pixel ROICs for high rate and radiation to ever deeper submicron CMOS processes 
(e.g. 65\,nm instead of 130\,nm).
This is not immediately evident, because pixel size also scales down with increasing hit rate.
However, simply making smaller pixels would not require a new process with smaller transistors (smaller feature size), 
since the analog functionality uses a small number of relatively large transistors. 
The requirement that does demand smaller transistors is increased logic density.
The reason pixel size must also be reduced with increasing hit rate is to avoid in-pixel pileup. After registering a hit,
a pixel needs a certain recovery time before being able to record another hit. 
To avoid losing efficiency to in-pixel pileup, 
the average time between hits in any given pixel (which scales inversely with pixel area) 
has to be much longer than the recovery time.
An alternative solution is synchronous
operation (where every pixel resets before each beam collision). This has been investigated,
but so far it has not been popular because, as such a front end must be faster, it consumes more power.

\subsection{Increasing circuit density through 3D integration}

The use of TSVs in hybrid assemblies has been discussed in section~\ref{sec:3D-integration} in the 
context of assembly integration. Very small TSVs can also provide
an alternate path to increase logic density (in a given footprint)
by stacking and vertically interconnecting multiple tiers of circuitry.  
Stacking of circuit layers this way results in a linear increase in logic density with number of layers, 
whereas feature size reduction results to first order in a
quadratic increase with feature reduction factor. 
In the commercial sector, industry has so far preferred Moore's Law scaling over 3D integration. 
Only for future system-on-chip architectures, below the 10\,nm feature size or equivalent,
do industry leaders favor 3D integration~\cite{eetimes-intel}.
One big advantage of 3D integration is reduction of power. This is because power scales with distance between switching elements,
and in 3 dimensions more switching elements can be placed in close proximity than in 2 dimensions. In fact, the main actual and growing use of
3D integration in commodity complex devices is to put memory as close as possible to processors, by integrating a 3D memory stack directly
on top of a CPU chip~\cite{eetimes-nvidia}. This increases transfer speed and reduces power.
But since high volume introduction of TSV in 2006, the fastest growth of 3D integration has been in
DRAM memory and image sensors, where either simple and regular structures (DRAM) or heterogeneous structures (image sensors) are being stacked.
The solid state RAM use case is obvious: it is not an alternative to Moore's Law scaling, but an augmentation of it. 
The DRAM levels being stacked
already have the highest available 2D logic density and 3D integration is the only way to fit even more into a given footprint.
In High Energy Physics, 
the use of 3D integration for higher logic density ROIC fabrication was explored (with 130\,nm feature size CMOS),
in parallel to the exploration of plain (not 3D) 65\,nm feature size.
Working devices were produced and good results were obtained~\cite{Deptuch:2010, Deptuch:2016},
but so far, as in industry, going to a smaller feature size process has been the more effective path to high logic density.

\subsection{Radiation tolerance of readout integrated circuits\label{sec:ROIC-radiation}}

\begin{figure}
\centering
\includegraphics[width=0.50\textwidth]{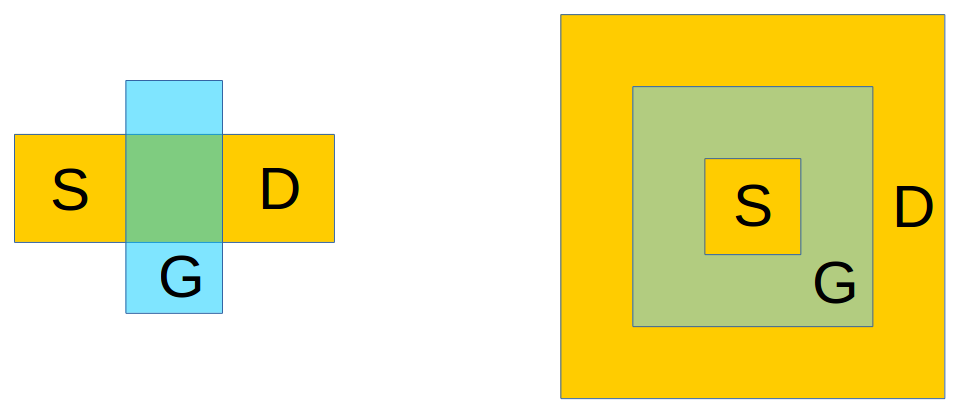}
\caption{\label{fig:ELT} Conceptual comparison of (left) typical linear transistor layout  and (right) an Enclosed Layout Transistor (ELT).}
\end{figure}

\begin{figure}
\centering
\includegraphics[width=0.60\textwidth]{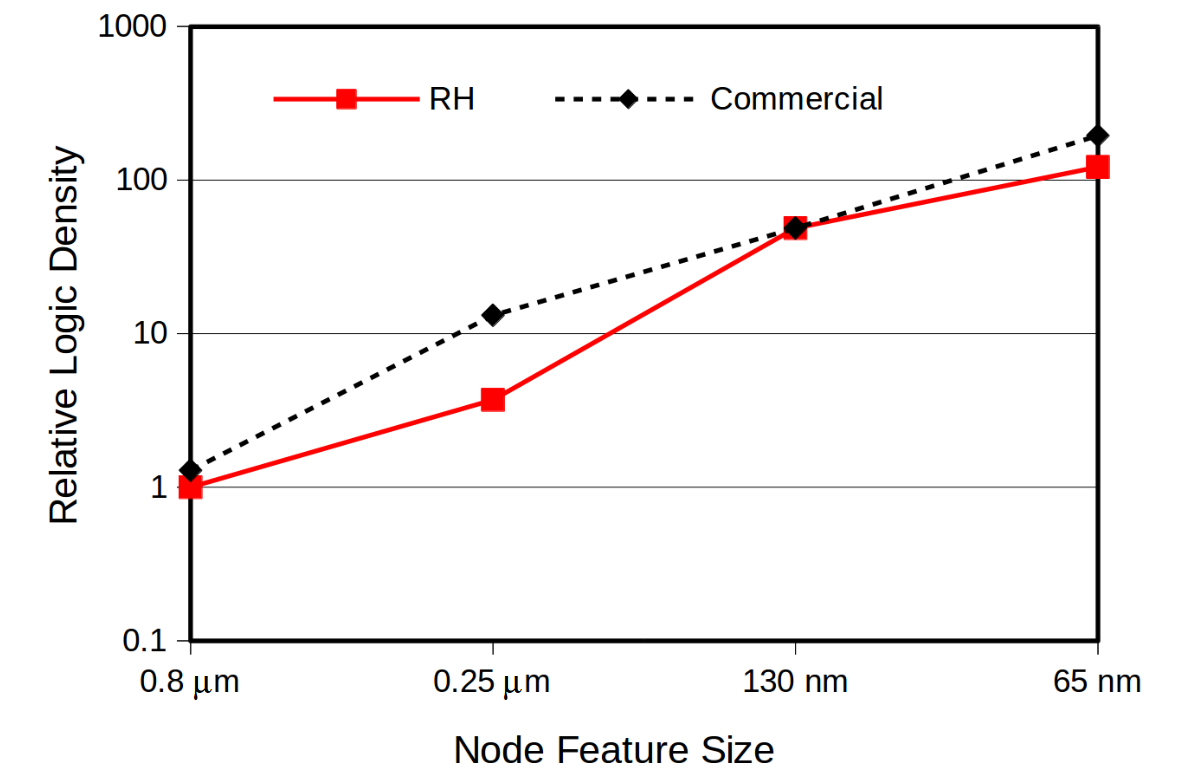}
\caption{\label{fig:logicdensity} Relative density of radiation hard (RH) and commercial logic in different CMOS technology nodes. In each node, we compared a memory latch from a commercial 7-track standard cell library to the smallest 
memory latch in the radiation hard library used for LHC pixel chips. 
The points have no error bars since each corresponds to a single latch.}
\end{figure}

First generation radiation hard ROICs were only possible in commercial 0.25\,$\upmu$m
CMOS processes with custom enclosed layout transistors (ELT). They achieved radiation tolerance above 50\,Mrad~\cite{Nowlin:2005}.
In an ELT the gate completely surrounds the source (drain) and in turn the drain (source) surrounds 
the gate (fig.~\ref{fig:ELT}). 
This avoids some channel edge effects (discussed later) since the channel is edgeless, but uses more real estate 
than a traditional linear transistor.  
ELT in 0.25\,$\upmu$m CMOS represented a 2.5-fold increase in logic density over 0.8\,$\upmu$m feature size military grade
radiation hard technology used in the 1990's, and this enabled sufficient logic density to produce the triggered
ROICs for ATLAS and CMS. 
Early development of pixel ROICs using 0.8\,$\upmu$m radiation hard processes 
did not succeed in incorporating all the needed functionality in a uniform pixel matrix;
the logic density was simply not high enough. 
Because of the ELT area penalty, 
radiation hard 0.25\,$\upmu$m lagged behind the Moore's Law scaling of their commercial counterparts (fig.~\ref{fig:logicdensity}).
Starting with the 130\,nm node, the minimum size linear transistor standard cell logic used commercially exhibited high total dose radiation tolerance. The ability to use linear transistors translated into a faster than Moore's Law increase in radiation hard logic density,
which went hand-in-hand with the transition to $2^{\rm nd}$ generation chips with synthesized logic in the pixel matrix.
In addition to high logic density, the good performance of linear transistors enabled the out-of-the-box 
use of commercial logic libraries. 
These have been perfected and extensively validated by large scale commercial manufacturers of consumer electronics
and come well integrated with powerful design and simulation tools.
However, the use of linear transistors comes with some risks and side effects, 
which, once again, make radiation hard circuits in 65\,nm lag their commercial counterparts in logic density.

The radiation hard design approach of $1^{\rm st}$ and 2$^{\rm nd}$ generation ROICs was 
to make or find devices (transistors) that tolerate the required dose without significantly changing their response. 
The design process typically included increased margin on the specifications, 
but otherwise relied on the device models available from the manufacturer. 
After fabrication, the ROIC was tested to measure the actual radiation tolerance. 
This approach is still followed for analog circuits. 
Optimization of analog performance would suffer if one had to consider a wide range of properties for each transistor in the circuit, so working with fixed properties is desirable. 
Fortunately, since for analog circuits the transistor dimensions and geometry are fully under the designer's control, 
it is possible to choose transistors that suffer little change with radiation, 
which means larger than minimum dimensions in the case of deep submicron CMOS (section~\ref{sec:ROIC-damage}).

On the other hand, for high density synthesized logic 
designers do not have the freedom to select devices that show little change with radiation. 
Instead, digital design must increasingly
rely on accurate modeling of radiation damaged transistors. 
Conventional digital design already offers several different models for the same transistor: 
a {\em typical} model plus so called {\em corner} models, which simulate operation with 
different temperatures, voltages, and/or variation of fabrication process parameters within an allowed range. 
The same transistor will be faster or slower depending on temperature, and so on. 
The synthesis tools produce circuits to function in all the selected corners. 
The new design approach is to use additional corner models that represent the radiation damaged transistors.
Such models can be custom made by parameterizing measurement data.
In this way, logic can be synthesized to work both before and after radiation damage, even though 
logic gate propagation delays, setup and hold times, etc. can change very significantly: a factor of 2 or more.
This is to be compared to changes of just tens of percent between the operating temperature extremes.  
Such large changes will limit the achievable clock speed, 
but most pixel ROIC applications require clocks of order 100\,MHz,
rather than the typical GHz speeds of microprocessors in the same technology. 
Radiation damaged transistor models are also used to simulate analog and mixed signal
circuits in order to confirm the design prior to fabrication and radiation testing, which still provides the ultimate validation.

\subsection{Total dose radiation damage in CMOS transistors \label{sec:ROIC-damage}}

Radiation damage in CMOS circuits is entirely due to charge carriers generated by ionization in the dielectric layers of the process, and not to bulk damage of the silicon lattice. 
Ionizing dose is delivered at hadron colliders by a combination of minimum ionizing particles 
(mainly pions) and background X-rays. 
The doping concentrations in CMOS transistors are high ($10^{15}$\,cm$^{-3}$ and higher), compared to which
the defect density introduced by bulk radiation damage is negligible~\cite{Radu:2015} 
(below $10^{14}$\,cm$^{-3}$ for HL-LHC inner layers after 3000\,fb$^{-1}$).  
However, there are many dielectric structures
in a modern CMOS process and each one leads to its own radiation effect due to ionizing dose. 
It is not by accident that radiation tolerance requirements
have kept pace with the logic density evolution in the ROIC generations. 
The reason is that both hit rate and radiation dose scale with particle flux. 
Required radiation tolerance went from 50\,Mrad for the $1^{\rm st}$ generation, to 250\,Mrad for the $2^{\rm nd}$,
to 1\,Grad in the $3^{\rm rd}$. 1\,Grad corresponds to about 50 minimum ionizing particles crossing every Si lattice cell.
Not all effects from charge generation in the dielectrics are equally important. As radiation dose increases,
understanding and managing previously negligible effects becomes necessary. 
The importance of each effect also depends on transistor geometry and size. 
In this respect, the 130\,nm CMOS technology node represented a \glq sweet spot\grq\ for which commercial logic
libraries could be used out-of-the-box up to doses well in excess of the $250$\,Mrad requirement (double or perhaps triple).
In contrast, in the 65\,nm node it is necessary to select or customize logic cell designs depending on the desired radiation tolerance, effectively trading off radiation tolerance for logic density. 
If the expected radiation dose is low, the out-of-the-box commercial logic can be used, 
while for higher expected doses, lower density logic cells must be substituted 
(The logic density point shown for 65\,nm in fig.~\ref{fig:logicdensity}
was appropriate for an expected dose of 500\,Mrad). 
These general statements should be seen as rough trends, and must be tempered by
process specificity and environmental effects discussed later. 
Beyond 65\,nm, which is at present well characterized, 
new research in high energy physics instrumentation is 
focused on understanding radiation damage in 40\,nm and 28\,nm feature processes~\cite{Mattiazzo:2017}, 
with detailed investigations expected over the coming years.

\subsubsection{Oxide charge and its effects}

\begin{figure}
\centering
  \includegraphics[width=0.7\textwidth]{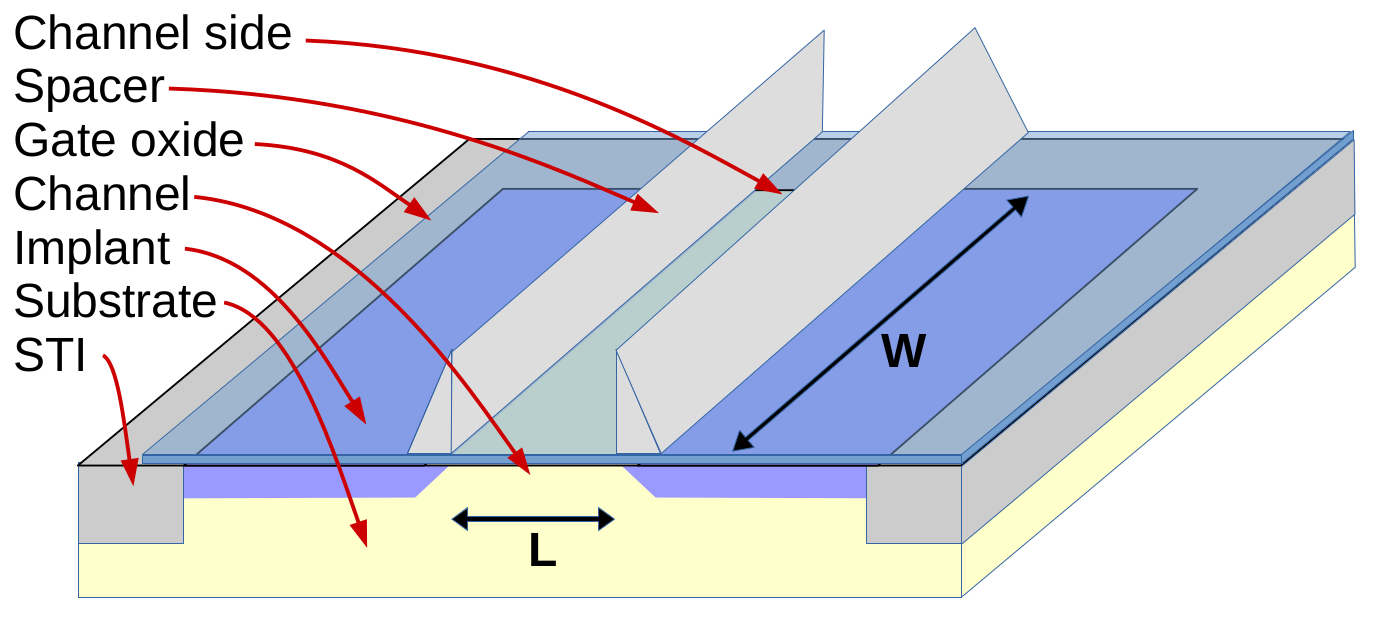}
\caption{\label{fig:transistor} Conceptual diagram of a MOS transistor of channel length L and width W, not to scale. 
A PMOS (NMOS) transistor would have N-type (P-type) substrate and P$^+$ (N$^+$) implants, where the + sign indicates
high doping. The gate conductor (not shown) would fill the region between the spacers. 
Spacers are used to control the doping at the ends of the channel. The spacers and the STI (Shallow Trench Isolation) oxides
are responsible for radiation damage effects. The STI closing off the front of the figure is not shown.
}
\end{figure}

Three oxide structures have been identified as producing radiation damage effects in the bulk CMOS processes used so far:
the gate oxide, the shallow trench isolation oxide (STI), and the gate spacers, all shown in fig~\ref{fig:transistor}.
Ionizing radiation generates charge carriers within these oxides, which can lead to an accumulation of positive static charge, because hole mobility in SiO$_2$ is 6 orders of magnitude lower than electron mobility. 
While charge in the gate dielectric would be most problematic, because it would directly cause a threshold shift,
gate oxide in 0.25\,$\upmu$m feature size and below cannot charge up. 
The gate oxides are thin enough (6\,nm and below) that quantum mechanical tunneling is an important effect, 
resulting in an effective electrical conductivity. 
Tunneling also means that there is a gate leakage DC current (unrelated to radiation), 
which is what eventually caused industry to move away from silicon dioxide gates in nodes below 45\,nm feature size.
The gate leakage in the nodes used so far for pixel chips is not yet large enough to result in significant 
standing power consumption, but it is visible in bias networks, which are high impedance nodes feeding large gate areas.

STI and spacers are both thick oxide structures and therefore will accumulate positive static charge.
However, not all the positive charge is truly static. 
Charge held in so-called deep traps is static at normal temperatures, 
but charge in shallow traps will gradually drift in an applied electric field (as present during chip operation)
and will thus reach the oxide-silicon interface. 
This is a complex region which undergoes manipulation during the fabrication process~\cite{Grove:1965}.
This manipulation involves the passivation of so-called dangling bonds,
which would otherwise manifest themselves as states within the energy band gap and spoil the semiconductor properties.
Positive charge reaching the interface reverses some of the passivation, bringing back some of the traps that were painstakingly neutralized during fabrication. 
However, these will then trap negative charge, and that will actually compensate some of the oxide
static charge. 
Remarkably, we would not be able to use commercial CMOS processes for ROICs, were it not for this 
fortunate cancellation of two detrimental phenomena: oxide charge and interface traps.  

At relatively low dose (up to a few Mrad) the net results can be complex depending on the time constants for positive
charge drift, the activation of interface traps, and their subsequent trapping of negative charge.
These time constants can be vendor-specific, depending on oxide structure formation chemistry and steps,
and are certainly affected by temperature. At high total dose the movement of charge and activation of traps have equilibrated, leading to an approximate, but not perfect cancellation of interface charge. 
A positive net charge always remains, and this degrades transistor performance. Since the sign of this net
charge at the oxide-silicon interface is always positive, it affects NMOS and PMOS transistors differently. 
The net positive interface charge results in parasitic lateral gating in the case of STI and in a modification of
source and drain in the case of spacers.
The former is known as Radiation Induced Narrow Channel Effect (RINCE)~\cite{Faccio-RINCE:2005}, while the latter
is known as Radiation Induced Short Channel Effect (RISCE)~\cite{Faccio-RISCE:2015}.

\subsubsection{RINCE}
RINCE or lateral gating affects the sides of the channel (see fig.~\ref{fig:transistor}),
causing them to be conductive in the case of an N-channel and non-conductive
in the case of a P-channel. 
This does not have a significant impact on a wide transistor, because a small fraction of the total
current flows near the sides. But in a very narrow transistor all the current is near the sides and the effect can be large, hence it is called Narrow Channel Effect. 
ELT's were used in 1$^{\rm st}$ generation ROICs to avoid this effect, as the channel in an ELT 
has no sides (fig.~\ref{fig:ELT}). This was necessary in 0.25\,$\upmu$m technology, 
because the radiation-induced parasitic standing current in linear NMOS transistors 
was measured to be too high for a useful design. 
But note that, while an ELT has no sides, it also cannot have a very narrow channel. The width of the channel must
be at least the perimeter of the source or drain, and so has to be many times the minimum feature size. 
Therefore, as long as
standing parasitic current is tolerable, making wide conventional transistors is just as good as making ELT's,
with the advantage that conventional transistors are standard and extremely well modeled. 
Thus, in 130\,nm and below, ELT's have not been used. 
This led to some surprises at low total dose in the case of the ATLAS IBL detector~\cite{Dette:2016}.
The parasitic standing current in NMOS transistors, while completely tolerable at high radiation dose,
can go through a transient period of being higher than tolerable depending on vendor and on temperature.
Accurate modeling of this varying leakage current behavior has been now developed~\cite{Backhaus:2017}.

RINCE affects PMOS and NMOS transistors in opposite ways. NMOS transistors develop parasitic standing current,
but this does not interfere with the transistor action, it simply adds to it. PMOS transistors are parasitically
turned off near the sides, which hinders the transistor action.
A critical question for PMOS is how far away from the sides is the channel affected by interface charge.
The effect can be visualized as a radiation dependent width change, such that the effective width is
$W_{\rm eff} = W_{\rm layout} - 2\Delta W_{\rm RINCE}{\rm (TID)}$, where $W_{\rm layout}$ is the width as drawn,
$\Delta W_{\rm RINCE}{\rm (TID)}$ is how far away from the STI-channel interface is the channel affected, and TID is
Total Ionizing Dose.
From the observation that minimum width PMOS devices in 65\,nm feature size are highly degraded at 500\,Mrad,
while those in 130\,nm feature size are mildly degraded, 
we can say that $\Delta W_{\rm RINCE}$(500\,Mrad)\,$\approx$\,30\,nm.

\subsubsection{RISCE}
RISCE affects PMOS and NMOS in a similar way, by impeding charge flow between source or drain and channel.
It can be roughly modeled as \glq adding\grq\ a certain length to the channel (a longer transistor conducts less current
than a shorter one). One may therefore think that making very short transistors
would be a good strategy against RISCE, because this would be a way to compensate. But the opposite is true:
longer transistors are less affected by RISCE, because the relative change in effective length is small if the
original device is long to begin with. Once again we can write
$L_{\rm eff} = L_{\rm layout} + 2\Delta L_{\rm RISCE}{\rm (TID)}$, where we are now adding effective length rather
than subtracting effective width. The magnitude of the effect is about twice as large in PMOS than in NMOS.
A 60\,nm long PMOS (NMOS) will experience a
70\% (30\%) reduction in full-on current after 500\,Mrad. Since transconductance scales as $1/L$, a 70\% (30\%)
reduction in current is equivalent to a factor of 2.5 (1.4) increase in length. 
\begin{minipage}{\linewidth}
For the original channel length of 60\,nm
this implies  $\Delta L_{\rm RISCE}$(500\,Mrad)$\approx$45\,nm ($\approx$12\,nm) for PMOS (NMOS).
\end{minipage}

\begin{figure}
\centering
\subfigure[Irradiation at $-15^{\circ}$C]{\includegraphics[width=0.49\textwidth]{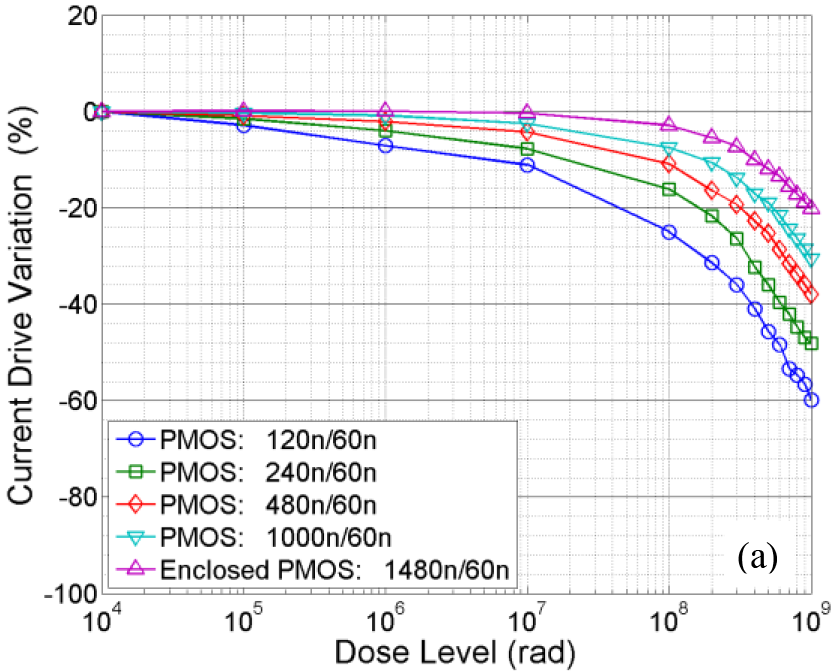}}
\subfigure[Annealing after Irradiation]{\includegraphics[width=0.49\textwidth]{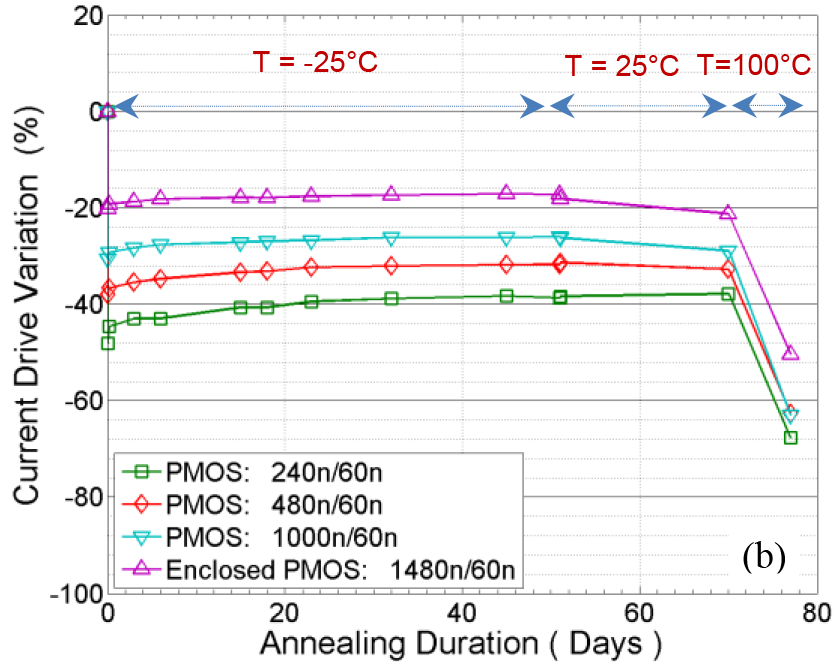}}
\caption{\label{fig:pmosrad} Current drive variation in 65\,nm technology PMOS devices of minimum length but varying width, 
as function of (a) total ionizing dose at $-15^{\circ}$C and (b) annealing duration and temperature under worst case bias \cite{Menouni:2015j}.}
\end{figure}

\subsubsection{Dependence on voltage and temperature}
The above $\Delta W_{\rm RINCE}$(500\,Mrad) and $\Delta L_{\rm RISCE}$(500\,Mrad) should be seen as
rough sketches of a more complex underlying behavior, and were given to provide intuition about the
magnitude and sense in which the transistors are affected, and why the effects become more important with decreasing feature size: effective width and length should not be regarded as a real model.

The RINCE and RISCE effects are modulated by transistor bias and by temperature. 
In general, both effects occur only when transistors are powered, 
which means there are electric fields in the STI and spacer oxides.
The larger the field (which depends on transistor bias conditions) the greater the effect, 
though there are quantitative differences between NMOS and PMOS. 
The effective width and length given above are for worst case bias.
Unpowered devices suffer little or no damage regardless of temperature - 
an important point to keep in mind when estimating how long a detector will last. 
 
For powered devices, high temperature increases damage during irradiation, 
and can decrease (anneal) or increase damage after irradiation. 
Figure~\ref{fig:pmosrad} shows results of a study~\cite{Menouni:2015j} in which different width PMOS devices were first
irradiated at low temperature and then annealed at different temperatures, always under power. All transistors are 
minimum length. When powered, devices can suffer very large damage in a matter of hours at 100$^{\circ}$C, but the same damage at 0$^{\circ}$C
would take over 10 years. This suggests that activation of deep traps is responsible.
As an oxide can contain multiple traps with different activation energies, the effect can be complex and process-dependent, which is confirmed by observations. 
Since the combination of power and high temperature is explicitly excluded by interlocks in all pixel detectors,
the main impact of high temperature annealing damage is to complicate qualification. The high integrated dose on a detector
is delivered gradually over years. Past qualification protocols
emulated this with high dose rate irradiation followed by high temperature annealing. 
However, application of such protocols is no longer straightforward in 65\,nm. Activation
energies must be precisely determined before thermal acceleration tests can be confidently applied.

\subsection{Single event upsets and mitigation \label{sec:ROIC-SEU}}  

In addition to long term degradation due to accumulated dose, 
energy loss by ionizing particles leads to instantaneous soft errors 
called Single Event Upsets (SEU). The most common SEU is the flipping of a stored bit in a memory. SEU also can produce
voltage transients on signal or control lines that can result in accidental operations - 
for example a single level asynchronous line to reset logic or memory would be vulnerable to SEU. 
Protection against SEU involves hardening of memory cells, avoiding designs with vulnerable control signals or 
hardening control signals where their use cannot be avoided, and circuit triplication. These techniques have been
in use since 1$^{\rm st}$ generation ROICs and have not seen significant changes in the 2$^{\rm nd}$ generation. 
However, as collider rate continues to increase and higher logic density translates into lower deposited charge 
needed for upset, 
these techniques will no longer be sufficient. An approach being introduced in 3$^{\rm rd}$ generation ROICs is to 
design for reliable operation while a significant level of upsets is taking place. Fundamentally this is 
abandoning the idea of circuit hardening as a solution to the SEU problem, and instead designing all functions such that SEU is not a problem to begin with. In practice, a combination of hardening and SEU-friendly functionality will be used. 

\begin{figure}
\centering
\includegraphics[width=0.6\textwidth]{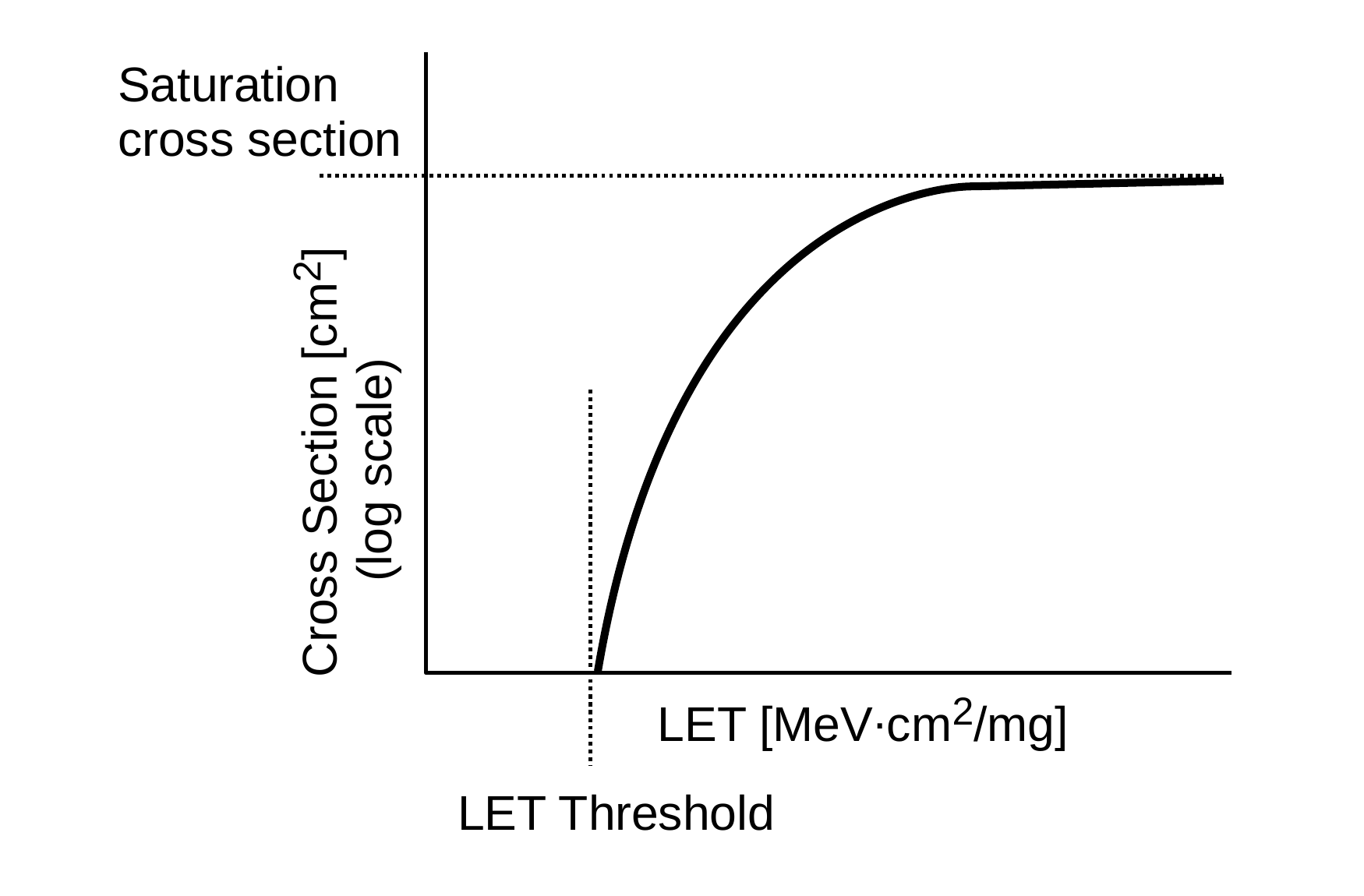}
\caption{\label{fig:SEULET} Conceptual plot of Single Event Upset (SEU) cross section vs. Linear Energy Transfer (LET) for a typical memory cell.}
\end{figure}

Extensive literature and experience exist on SEU of memory cells in the context of electronics used in space. 
This is not directly applicable to particle physics pixel detectors, but nevertheless a good starting point. 
SEU of a given circuit, like an SRAM cell, latch, or flip-flop, depends on the amount of energy 
deposited by an impinging ion, which is characterized by a Linear Energy Transfer (LET). It is important that 
this is meant to describe non-relativistic ions, which lose energy approximately uniformly 
along their path through electromagnetic interactions. 
The upset rate is characterized by the cross section for causing a bit flip 
(SEU cross section). Cross section vs. LET is typically fit with a Weibull function, 
resulting in a threshold and saturation cross section as shown in fig.~\ref{fig:SEULET}. 
In submicron technologies, 
the typical LET threshold pretty much regardless of memory cell type is of order 1\,MeV$\cdot$cm$^2$/mg. 
Saturation cross section varies with cell design, but is of order $10^{-7}$ to $10^{-8}$\,cm$^2$ for common SRAM,
latches and flip-flops. However, an energetic proton (or pion) has a LET of order 0.01\,MeV$\cdot$cm$^2$/mg, which
immediately signals that it cannot upset memory cells by the same energy loss mechanism as ions 
(it is far below the LET threshold). 
Upsets in this case are due to nuclear interactions. This can be seen from the fact that SEU 
cross sections are about the same for energetic neutrons and protons~\cite{Roche:2010}. 
There is thus a kinematic threshold depending on the nuclei in the material, rather than 
a LET threshold. Typical SEU cross sections for protons are of order $10^{-13}$\,cm$^2$~\cite{Roche:2010}. 
At relatively low energies, an adequate 
model has the proton imparting momentum to a nucleus which then becomes a traditional high LET heavy ion.
But at the GeV energies of the LHC, inelastic collisions can produce showers of high LET particles 
affecting a large area of silicon. This is important for hardening techniques. 

SEU hardening of memory cells exploits redundant connections or storage separated by some distance. 
This is more effective for heavy ions, which deposit energy very locally, 
but still useful for relativistic protons (or pions) which can produce extended deposits~\cite{Menouni:2008zz}.
A type of memory called Dual Interlocked CEll (DICE)~\cite{Calin:1996} is extensively used as 
its state can only be changed by switching two
physically separate voltages in coincidence, which gives it a lower SEU cross section than common latches. 
Special layout techniques, such as {\em interleaving} can be used to carefully separate sensitive 
elements and prevent charge sharing~\cite{Amusan:2007}.
Table~\ref{tab:SEUtable} compares the SEU cross sections of common structures~\cite{Menouni:2017u} 
in a high energy proton beam, for 65\,nm technology.

\begin{table}
\small
\renewcommand{\arraystretch}{1.2}
\centering
\begin{tabular}{ | l | c | }
\hline 
\textbf{Cell type} & \textbf{Cross section in cm$^2$} \\
\hline
Standard latch &  $2.8 \times 10^{-14}$ \\ \hline
DICE latch with interleaved layout &  $3.1 \times 10^{-15}$ \\ \hline
TR standard Latch &  $1.2 \times 10^{-16}$ \\ \hline 
TRL with error correction and control triplication &  $2.3 \times 10^{-17}$ \\ \hline 
TRL with error correction, control triplication & \\
and separation of sensitive nodes &  $6.8 \times 10^{-18}$ \\ \hline 
\end{tabular}
\caption{SEU cross sections of different memory structures in 65\,nm technology with 24\,GeV protons~\cite{Menouni:2017u}.
TR stands for triple redundant and L for latch.}
\label{tab:SEUtable}
\end{table}

One is not always free to choose the memory type with the lowest SEU cross section. The almost 4 orders of magnitude gain 
from a standard latch to maximal use of triple redundancy in table~\ref{tab:SEUtable} comes with an associated 
footprint increase of a factor of 10. Such cells can therefore not be used in the high density logic of the pixel matrix. 
Suppose every pixel has 8 bits of configuration and the acceptable fraction 
of corrupted pixels during an 8\,h data run is 1\%,
with 1\,GHz/cm$^2$ particle flux. The run-integrated fluence is $2.5 \times 10^{13}$\,cm$^{-2}$ and therefore the pixel SEU 
cross section must be 1\%/($2.5 \times 10^{13}$\,cm$^{-2}$) or less. As there are 8 bits per pixel, the single bit 
cross section must be $5 \times 10^{-17}$\,cm$^2$ or less. 
Even without considering added design margin, tab.~\ref{tab:SEUtable} shows that this requires triple redundancy, 
which may not fit due to its large footprint. A new solution to this problem in 3$^{\rm rd}$ generation ROICs 
is to implement a control protocol that allows continuous writing of configuration data during data taking, 
instead of configuring in a distinct operation mode. If configuration values are constantly 
being refreshed from outside, one does not need to rely on \glq long term memory\grq\ within the chip. 
For example if the time between consecutive writes is 1\,s instead of the duration of the data run. 
this relaxes the SEU cross section requirement by 4 orders of magnitude. 
It should be noted that 
hit data have never required SEU protection, because the time that data bits spend in chip memory 
before readout is very short, given by the trigger latency. 
Thus the relevant time scale for hit data is of order 10\,$\upmu$s for triggered, 3$^{\rm rd}$ generation ROICs. 
Even with a significant error amplification from data encoding\footnote{For example, 
a single bit flip corrupts and entire 64 bit packet},
the fraction of data lost due to SEU will be negligible with storage 
in standard latches\footnote{Since no detector is 100\% efficient, 
data losses well below the 1\% level are insignificant.}.

\subsection{Analog front end and ADC \label{sec:ROIC-fend}}

\begin{figure}
\centering
\includegraphics[width=0.9\textwidth]{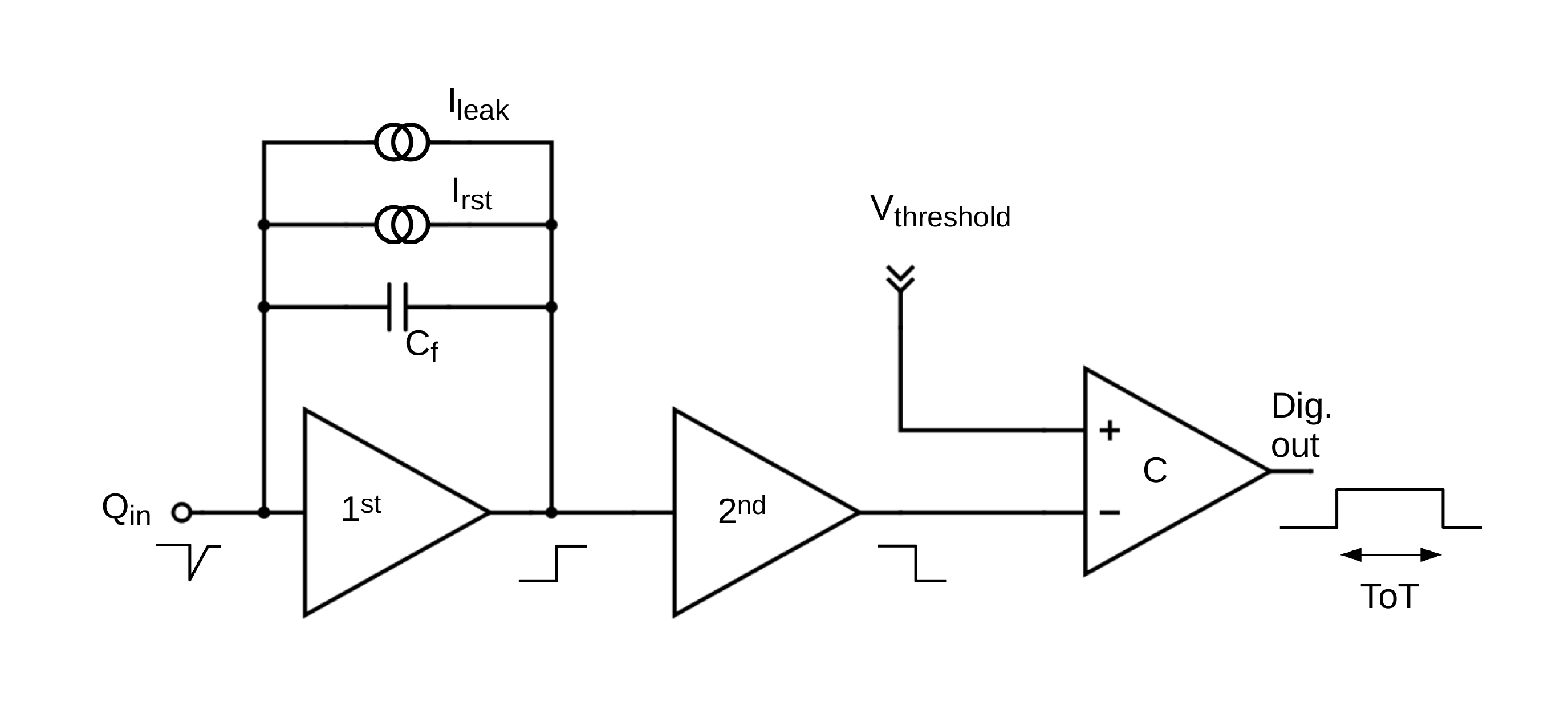}
\caption{\label{fig:frontend} Schematic diagram of a typical pixel analog front end.
Signal polarities and Time over Threshold (ToT) are indicated along the bottom.}
\end{figure}

The analog front end elements are shown schematically in fig.~\ref{fig:frontend}.
The front end design has universally consisted of a Charge Sensitive Amplifier (CSA),
followed by a 2$^{\rm nd}$ stage to provide additional voltage gain,
and a comparator (C) to carry out the pulse height discrimination.
The CSA has a 1$^{\rm st}$ stage amplifier with capacitive feedback (C$_{\rm f}$), a continuous reset (I$_{\rm rst}$)
and a low bandwidth feedback to compensate for sensor DC leakage current (I$_{\rm leak}$).
This is needed because hybrid pixels have so far been DC coupled to the readout and
sensor leakage current can be significant after irradiation 
(for development of AC-coupled pixels see section~\ref{sec:sensors_planar}). 

\begin{figure}
\centering
\subfigure[Straight cascode]{\includegraphics[width=0.45\textwidth]{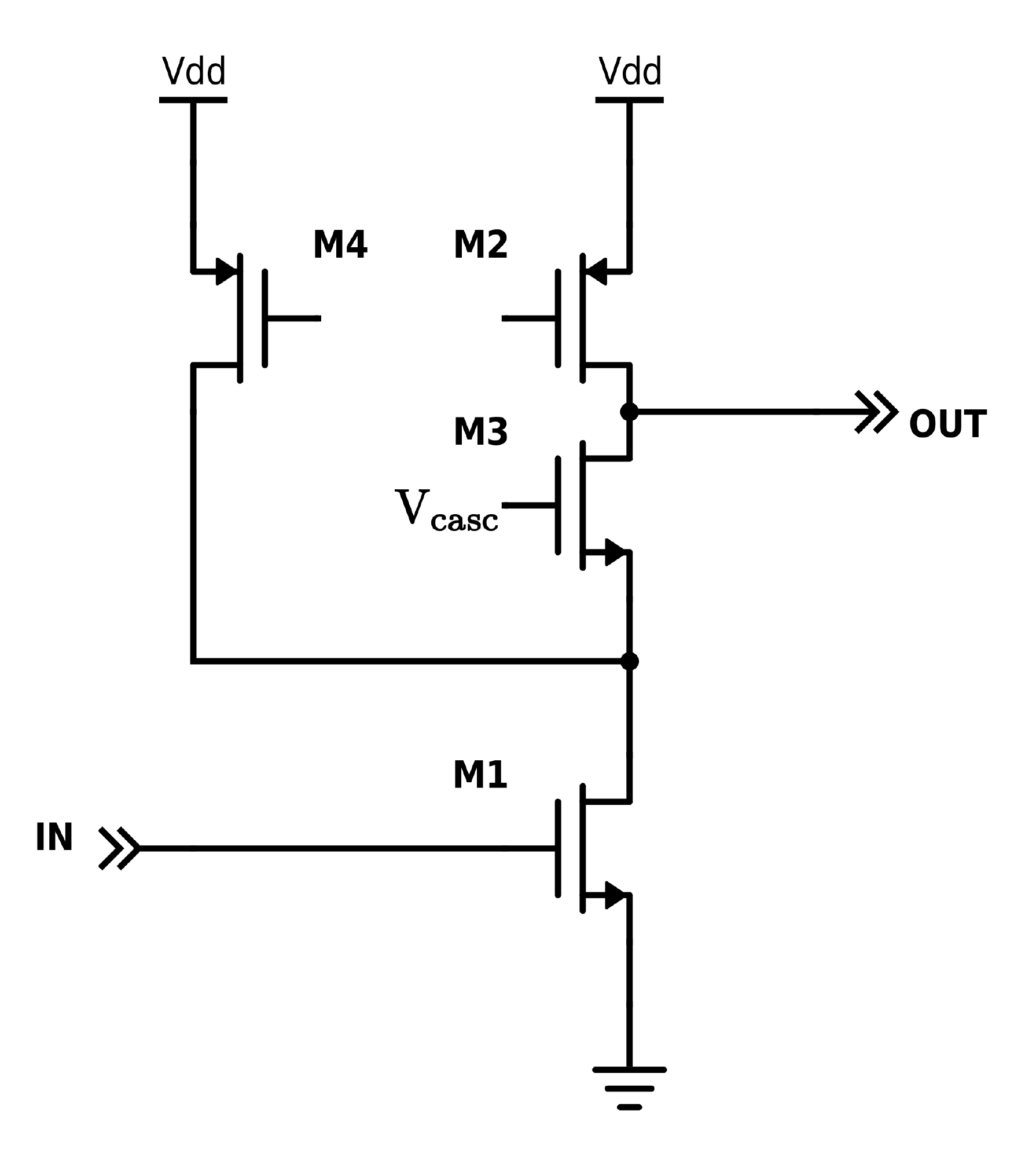}}
\subfigure[Regulated cascode]{\includegraphics[width=0.45\textwidth]{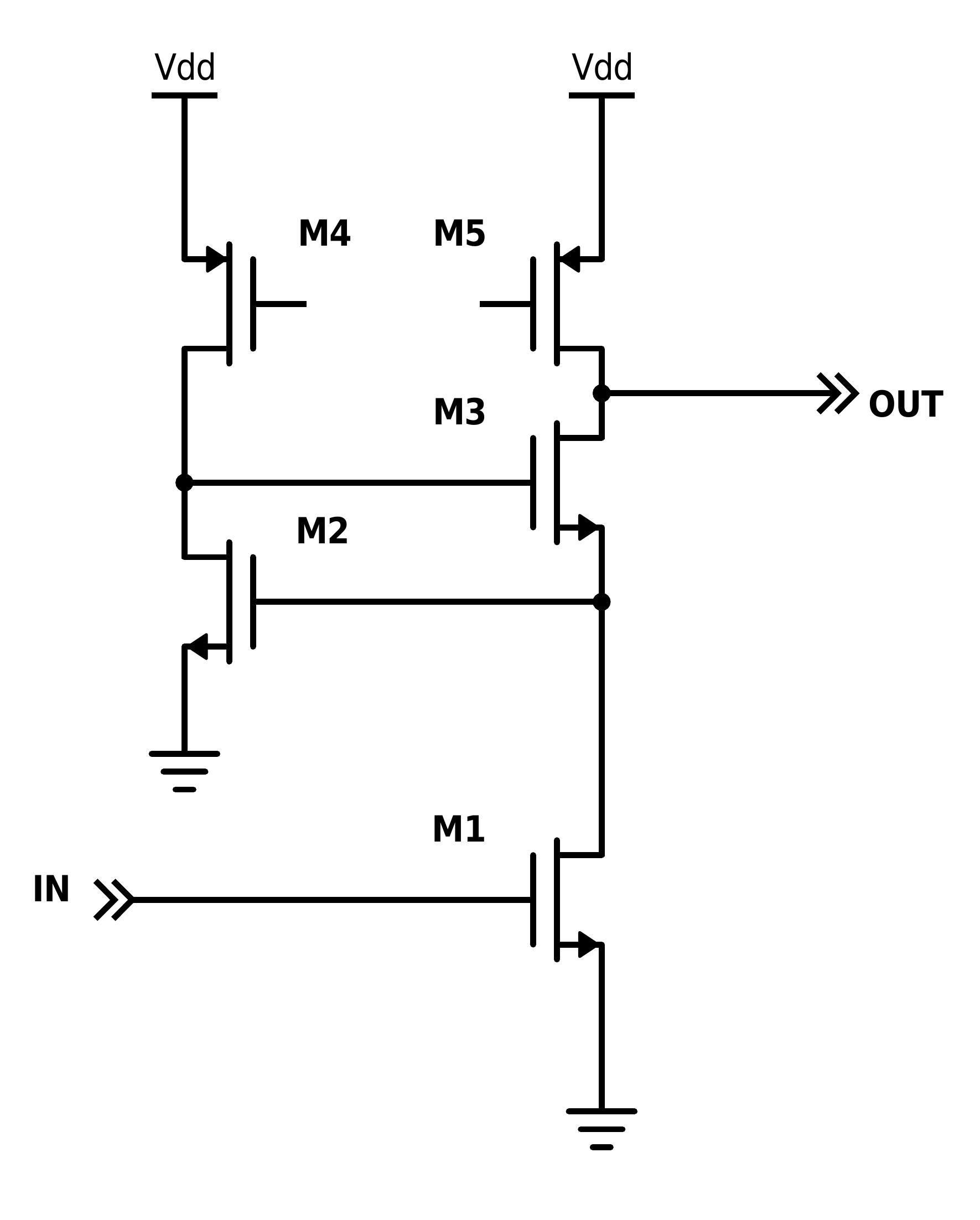}}
\caption{\label{fig:cascodes} Cascode circuit variants with NMOS input transistor, as typically used for 1$^{\rm st}$ stage pixel amplifiers.
The bias V$_{\rm casc}$ and the gates of M2 and M4 in (a) and of M4 and M5 in (b) are held at DC voltages, 
typically generated by bias current mirrors (not shown).}
\end{figure}

On-chip signal amplitude digitization has used ToT of the comparator output.
ToT is a simple digitization method that counts clock cycles while the comparator is high (meaning above threshold). 
The most important function of the pixel front end has been to
discriminate true hits from noise with the correct timing to within one bunch crossing of the accelerator, which
for the LHC means 25\,ns. This requires a fast leading edge response (high slew rate), 
which can be achieved with high input gain. 
The input charge to output voltage gain is
given by the inverse of the feedback capacitor C$_{\rm f}$ in fig.~\ref{fig:frontend} 
and the characteristics of the first stage amplifier.
On the other hand, 
the charge transfer efficiency from the sensor to the integrator is given by 
(C$_{\rm det}$\,+\,$G \times$C$_{\rm f}$)/C$_{\rm det}$,
where C$_{\rm det}$ is the detector load capacitance and $G$ is the 1$^{\rm st}$ stage open loop gain.
Clearly, a high open loop gain is needed in order to have both high gain (small C$_{\rm f}$) 
and good charge transfer efficiency.  
However, 
the specific transconductance 
(g$_{\rm m}/I_{\rm D}$, were $I_{\rm D}$ is the drain current) 
of CMOS transistors is only of order
10\,S/A, which means that a simple (i.e. single transistor) 
inverting amplifier can only achieve $G\sim10$ for 1\,$\upmu$A current 
and 1\,V supply, since $G \approx V_{\rm D}$g$_{\rm m}/I_{\rm D}$, where $V_{\rm D}$ 
is the supply voltage. 
Two simple amplifiers in series would achieve an open loop gain proportional to g$_{\rm m}^2$ 
(so $G\sim100$), but with low bandwidth. 
The solution in pixel ROICs has always been to use a cascode topology for the
1$^{\rm st}$ stage, which achieves an open loop gain proportional to g$_{\rm m}$(M1) $\times$ g$_{\rm m}$(M3) without sacrificing bandwidth. The labels M1 and M3 refer to fig.~\ref{fig:cascodes}, where two cascode variants are shown.
In the straight cascode configuration (a), 
the current source M4 is larger than the current source M2, and their combined current
passes through the input transistor M1. A small change in the input will cause a change in the M1 drain voltage
according to the transistor's gain. This in turn shifts the source of M3, leading to a significant V$_{\rm gs}$ change
for M3, which amplifies the change in the output voltage. The M3 gate voltage is held at a fixed bias, 
V$_{\rm casc}$. In contrast,
in the regulated cascode (b), the current source M4 is smaller than the current source M5, and the additional transistor
M2 shifts the M3 gate voltage in response to a change of M1 drain voltage, leading to additional output gain (bigger V$_{\rm gs}$ change for M3).

While the front end configuration and 1$^{\rm st}$ stage cascode architectures have remained, there have been important
quantitative changes between 1$^{\rm st}$ and 3$^{\rm rd}$ generation ROICs. 
The specific transconductance of CMOS transistors has nearly doubled going from 
16\,S/A in the 0.25\,$\upmu$m process to 22\,S/A in 130\,nm to 28\,S/A in 65\,nm.
The higher hit rate requirements have led to smaller pixels (with lower capacitance per pixel) and faster shaping
to avoid in-pixel pileup. While some 1$^{\rm st}$ generation ROICs used PMOS input transistors for better substrate isolation and better
1/f noise performance, with the faster pulse shaping and common substrate isolation techniques, NMOS
has become the input device of choice for amplifiers in 130\,nm CMOS and below~\cite{Manghisoni:2007}. The decrease in
pixel capacitance (by about a factor of 4) and increase in transconductance have translated into higher gain and speed.
The effect of time-walk is therefore significantly reduced in 3$^{\rm rd}$ generation ROICs, even before making use of
digital processing to correct the time of hits near threshold\footnote{Time-walk is the variation in relative 
delay between front input pulse and comparator firing, as a function of pulse amplitude above threshold.}. 
While so far all detectors have used ROICs with
continuous reset as shown in fig.~\ref{fig:frontend}, 
there has also been R\&D on front end designs that reset before every bunch crossing
(these are referred to as synchronous front ends)~\cite{Fahim:2014nux}. 
In this case there is by construction no time-walk, but the power consumption is higher 
due to the fast shaping and reset settling needed.
So far this approach has not been adopted for pixel detectors. 

The need for very high speed front ends with sub-ns time stamp resolution capability has been spearheaded by
NA62 and its Gigatracker detector. The Gigatracker ROIC~\cite{Martin:2010}, implemented in 130\,nm CMOS, 
uses a straight cascode front end with RC feedback. The RC time constant is 5\,ns, 
which gives the device its 5\,ns peaking time.
Two differential gain stages are then used in front of a constant fraction discriminator to achieve a 200\,ps time 
stamp resolution.
Future fast timing detectors discussed in section~\ref{sec:time-point} will continue to push this type of 
design to finer time resolution.

A pixel ROIC with free-running front ends is an inherently metastable circuit. Every pixel can fire at any time, and the act of firing
switches logic that would not otherwise switch. Any coupling from digital switching to analog front end can be a positive feedback
mechanism. For low enough threshold and/or enough pixels firing at the same time, this positive feedback can set off a chain reaction causing
all pixels to fire. For a given instantaneous hit occupancy (fraction of pixels firing at the same time due to an external stimulus),
there will be a minimum stable threshold. 
Each ROIC is different in this respect, but in general the minimum stable threshold was around 2500 electrons (e$^-$)
in 1$^{\rm st}$ generation ROICs, whereas it will be around 500\,e$^-$ for the 3$^{\rm rd}$ generation.
This reduction has been deliberate: required by decreasing input signal values. Large pixels ($2 \times 10^4$\,$\upmu$m$^2$),
thick sensors ($>200$\,$\upmu$m), and moderate sensor radiation damage for 1$^{\rm st}$ generation detectors translated into
expected signals of order 10\,ke$^-$, while small pixels ($0.25 \times 10^4$\,$\upmu$m$^2$), thinner sensors (100\,$\upmu$m),
and heavier sensor radiation damage will lead to signals as low as 2\,ke$^-$ at the HL-LHC. 
The minimum stable threshold is reduced by exactly the
same factor, since the important figure of merit for pixel detectors is not signal to noise ratio, 
but rather signal to threshold ratio.
The front end noise does set a lower bound to the threshold, but does not determine how far above this lower bound is the
minimum stable threshold. Assuming a noise occupancy of 10$^{-4}$ is acceptable, this corresponds to the Gaussian 1-sided tail fraction
for $3.7\sigma$. Inverting the question, for a 500\,e$^-$ minimum threshold, the noise must be less than 500/3.7=135\,e$^-$ equivalent
input charge. But this is just an upper bound. How far below this bound the noise needs to 
be depends on how the threshold varies with time and how from pixel to pixel. 

One can express
the front end noise requirement as
\begin{equation}\label{eqn:fenoise}
{\rm ENC} < \sqrt{ (T/3.7)^2 - T_{\rm RMS}^2(\vec{x}) - T_{\rm RMS}^2(t) }
\end{equation}
where ENC is the equivalent input charge noise, $T$ is the threshold, and $T_{\rm RMS}$ is the threshold variation as a function of
position ($\vec{x}$) or time ($t$).
$T(\vec{x})$, or the pixel to pixel threshold variation, is known as dispersion. The main cause of dispersion is 
fabrication process mismatch between
transistors and passives of identical design. While the mismatch amount is process dependent, and the
translation from mismatch to dispersion is design dependent, 
a typical dispersion value is in the range of 300-500\,e$^-$. As this is
clearly unacceptable, all ROICs so far have compensated for mismatch by programming a different threshold voltage
in each pixel, such that in units of input charge all pixels have an equalized (or {\em tuned}) threshold value.
With such tuning, a value of 40\,e$^-$ dispersion is typically achieved. The cost of this technique is circuit area
in the pixel in order to implement the needed DAC and memory. However, tuning only controls $T_{\rm RMS}(\vec{x})$,
and not $T_{\rm RMS}(t)$. 
The main source of $T_{\rm RMS}(t)$ has been coupling of transients from other circuits to the front end, which
has in turn been the main limitation on minimum stable threshold. This has been mitigated with circuit isolation, and
control of power transients, as already explained. Still other sources of $T_{\rm RMS}(t)$ may gain importance
in the future. In particular, short term radiation damage can spoil tuning and cause threshold shifts~\cite{Carney:2016}.
If the radiation dose in between tuning cycles is high enough, $T_{\rm RMS}(t)$ could become dominant for reasons unrelated to circuit
isolation. This has not been a problem until now, but will need attention as accelerator intensity increases and windows of
opportunity for tuning decrease. Techniques for near real-time equalization,
which would compensate both $T_{\rm RMS}(\vec{x})$ and $T_{\rm RMS}(t)$ simultaneously,
include periodic capture of the baseline level in each pixel, or auto-zeroing~\cite{Monteil:2016}, and self-adjustment in the
pixel to equalize to the same noise occupancy~\cite{Heim:2017}.

\subsection{Readout architecture \label{sec:ROIC-architecture}}

Section \ref{sec:datatransmission} gives an overview of the hit rate impinging onto the ROIC as well as the bit rate
coming out of the data links.
Readout architecture refers to the transfer of data from the pixels, where it is generated, to the ROIC output(s), including
storage and buffering along the way. There has been much evolution in this area from 1$^{\rm st}$ to 3$^{\rm rd}$ generation. The most
significant requirement for the readout architecture is whether the system is full readout or triggered readout. 
In a full readout ROIC
every hit from every pixel must eventually make its way to the data output.
In the language of queuing theory this is a single server queue with random customer arrival time
(each pixel is a {\em customer}, while the ROIC data output stage is the single {\em server}),
which is characterized by eq.~\eqref{eq:mm1}, discussed below.
A full readout system may
not require high logic density, because, provided the data output stage has high enough bandwidth, there is no need
to store hit data on chip for any length of time. The problem with a full readout system is that it may not be possible to implement sufficiently
high output bandwidth to handle the pixel hit rate. In a triggered system, on the other hand, all pixel hits must wait a
pre-defined {\em latency} until a trigger decision determines which ones to read out. In this case the data output bandwidth only has to
match the trigger rate, and is decoupled from the primary pixel hit rate. But the full incoming hit rate must be stored during the 
trigger latency, and storage is limited by the process logic density.

\begin{figure}
\centerline{\includegraphics[width=\textwidth]{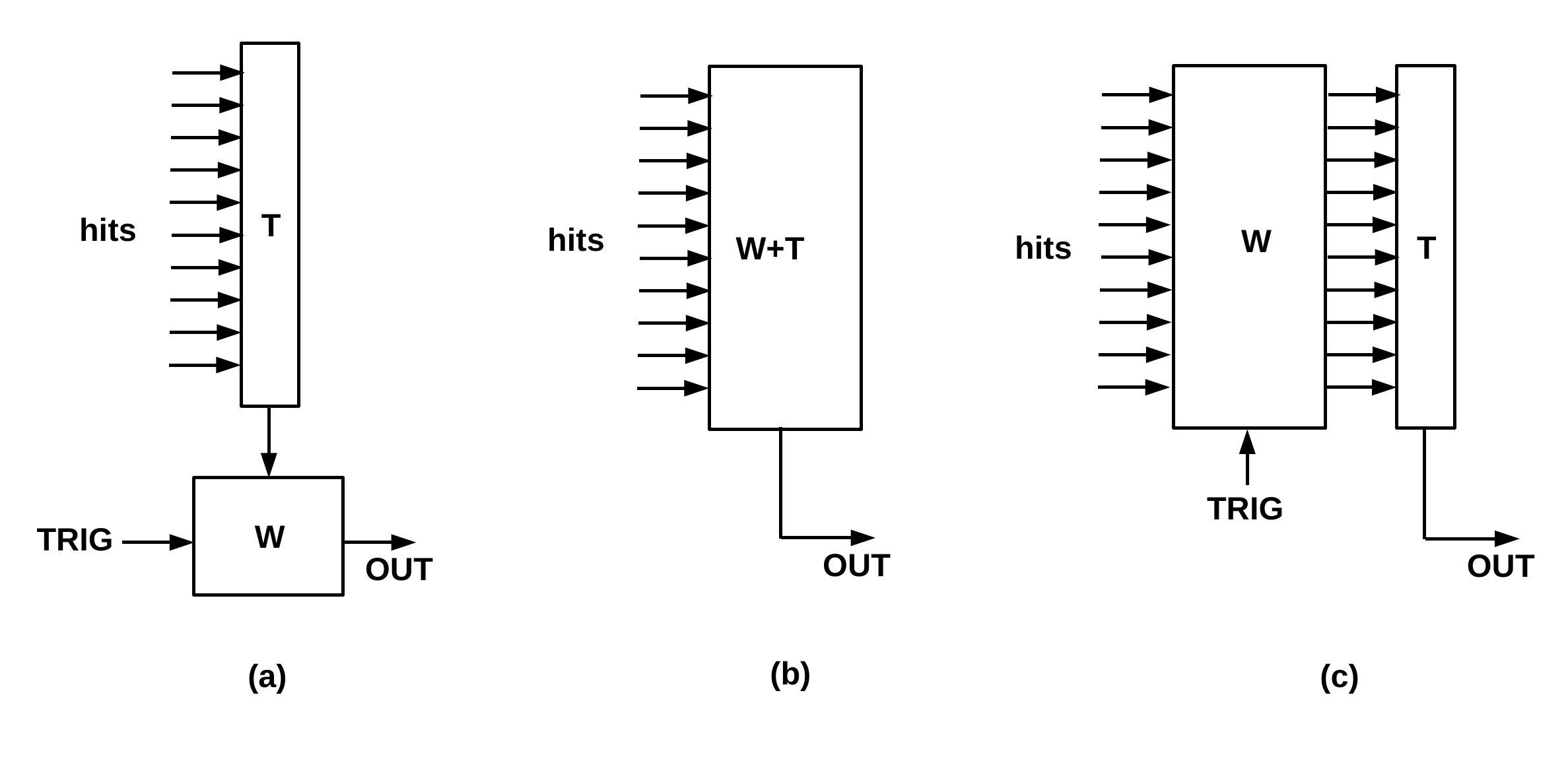}}
\caption{\label{fig:readarch1} Diagram of (a) classic Column Drain readout with triggered readout, 
(b) shift register implementation of column drain with full readout, and (c) Local Buffer triggered readout.
W stands for wait and T strands for transfer.}
\end{figure}

Figure~\ref{fig:readarch1} compares schematically three readout architectures, where the buffered storage of hit data
is represented by W for wait, and the motion of hit data by T for transfer.
In the classic {\em column drain} architecture of 1$^{\rm st}$  generation chips (fig.~\ref{fig:readarch1}\,(a)\,),
all pixels in a column share a data bus that only one pixel at a time can use.
As soon as a pixel is hit, it tries to grab the bus and transmit its data.
The bus is arbitrated so each hit pixel must wait its turn. This column bus is itself a single server queue,
just like the case of full chip readout. The bandwidth of the bus must exceed the incoming hit rate for this
to be viable. Equation~\eqref{eq:mm1} shows the probability for waiting a time greater than $t$ in such a simple 
single server queue~\footnote{In Kendall notation~\cite{Kendall:1953} this is an M/M/1 queue},
 \begin{equation}\label{eq:mm1}
P(W>t) = {\frac{\lambda}{\mu}}\,{\displaystyle e}^{\displaystyle -(\mu-\lambda)t}
\end{equation}
where $\lambda$ is the input hit rate and $\mu$ is the output bandwidth. For a pixel hit rate
$r$, $\lambda=Nr$, where $N$ is the number of pixels served by the bus. Hits
must be transferred within a short time, $t_s$, as there is no place to store them within the pixel array. Therefore,
to avoid hit loss, $P(W>t_s)\approx 0.001$, which means that $(\mu-\lambda)t_s\approx 6$. Since $t_s$ is small,
one must have $\mu \gg \lambda$: the well known condition that column drain needs a high bandwidth.
Implementation of a trigger buffer in 1$^{\rm st}$  generation chips was at the chip bottom, requiring
all hits to be transferred out of the pixel matrix. Only after storage in the periphery was the trigger selection applied.

The LHC-b Velopix readout chip implements a read-all architecture using a shift register (SR) for transferring hit data instead of a bus.
This is represented by fig.~\ref{fig:readarch1}\,(b).
The SR not only transfers data, but is also temporary storage. The output bandwidth $\mu$ of the SR is
simply the clocking speed (assuming one pixel hit is read out per clock cycle). But unlike an arbitrated bus,
each pixel sees a different output bandwidth.
The very first pixel feeding the SR sees the full bandwidth, but the next pixel sees less bandwidth because occasionally
it will be blocked by data from the previous pixel already in the SR.
Thus, pixel $i$ sees an output bandwidth $\mu_i = \mu-\sum_j{r_j}$, where $r_j$ is the hit rate of the $j^{\rm th}$ pixel.
Because the SR stores hits, the short transfer time requirement of column drain is removed, and one just
needs $\mu_i > r_i$ (as opposed to $\gg r_i$). If all pixels have the same hit rate $r$, one needs $\mu > \lambda$,
where $\lambda=Nr$, as before. Since the last pixel sees the smallest output bandwidth, its waiting time is the longest,
so one can choose $\mu$ such that the last pixel waiting time is less than the time between hits.
This architecture is particularly well suited to the highest rate chips in the Velopix detector, because the hit rate is {\em not}
the same in all pixels. The pixel nearest the beam line has the highest rate, 
and also happens to be the pixel farthest
from the readout-- that is, the first pixel in the SR. The last pixel in the SR has the lowest hit rate. Thus, the pixel dependent hit rate $r_i$ nicely follows the pixel dependent output bandwidth $\mu_i$.

Starting with the 2$^{\rm nd}$  generation triggered ROICs, high logic density permitted the implementation of local hit storage within the pixel matrix (fig.~\ref{fig:readarch1}\,(c)\,).
The input rate to the column bus is reduced as $\lambda' = \lambda \times h \times t_{BX}$,
where $h$ is the trigger rate and $t_{BX}$ is the bunch crossing period. If the readout time can be long,
then one can have $\mu \gtrsim \lambda'$, which makes for a very relaxed column bus transfer bandwidth requirement.
In addition to higher hit rates, this relaxed requirement has been exploited to make larger ROICs (more pixels per column bus).
Instead of being limited by the transfer rate, the hit rate that can be handled is limited by the amount of memory
that can fit in the pixel matrix. All hits must be stored for the entire trigger latency, so the memory needed
scales like $\lambda \times t_L$. The longer the trigger latency the smaller the hit rate that can be handled.

\begin{figure}
\centerline{\includegraphics[width=0.75\textwidth]{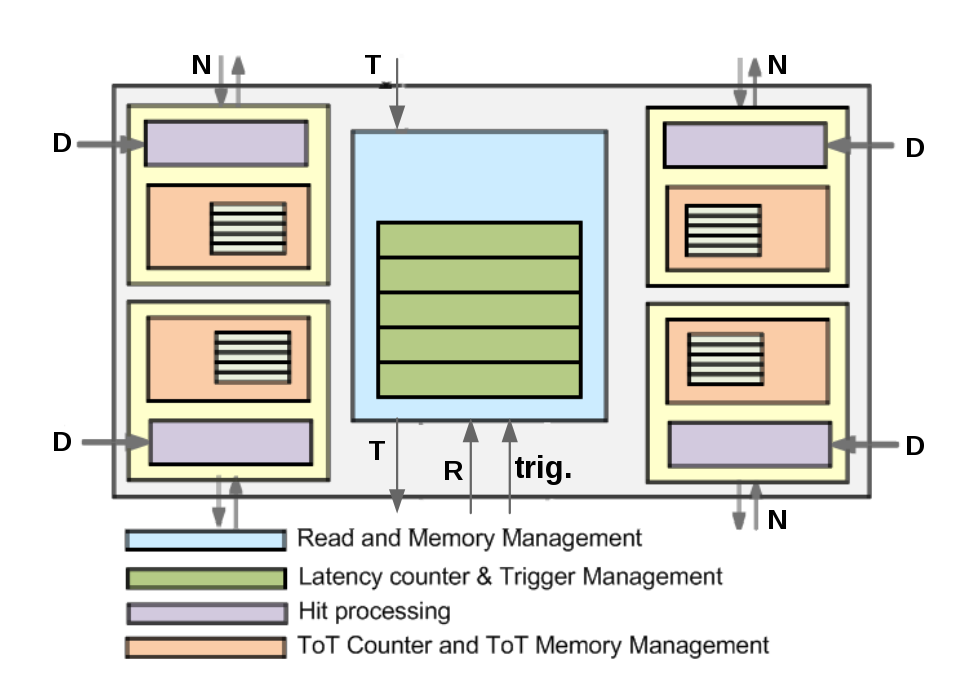}}
\caption{\label{fig:region} Block diagram of 4-pixel region logic on FE-I4 ROIC. N stands for neighbor, 
D for front end discriminator output, T for readout token (Adapted from \cite{Hemperek:2009xra}).}
\end{figure}

The local storage of hit data does not make efficient use of memory when done at the individual pixel level. 
This is because the mean time between single pixel hits
is longer (but not much longer than the trigger latency). 
For example for 50\,kHz single pixel hit rate and 5\,$\upmu$s trigger latency, the
average number of hits per latency period is 0.25. Using Poisson statistics, 
one would need to store up to 3 hits per pixel in order to
keep 99.9\% of hits. We refer to this number (3 in this example) as the {\em buffer depth}.
Instead, grouping pixels into regions with shared storage significantly reduces memory needs.
Consider grouping 4 pixels together. If the 4-pixel region average hit rate were 1 hit per latency period, then, using Poisson statistics,
one would need a buffer depth of 5 region-hits per latency period in order to keep 99.9\% of hits.
That is only 5/4 = 1.25 buffer locations per pixel instead of 3 as before. But the advantage
is even greater, because in a pixel detector the hits among neighbor pixels are correlated
(charged particles produce clusters).
The correlation between pixels means that the region hit rate is less than the single pixel hit rate times the number of
pixels per region. The amount of correlation depends on sensor and detector geometry. 
Charge information, on the other hand, must be stored for every hit pixel regardless of how the pixels are grouped,
and this reduces the advantage of having many pixels per region.  
The simplest approach is to store a fixed number of charge bits per pixel times the buffer depth.
Suppose in the above example one stores 10 bits of region hit arrival time and some flags, plus
4 bits per pixel charge information. 
If storage is independent per pixel, each pixel must store 14 bits per hit times buffer depth of 3,
or 42 bits per pixel.
Conversely, in a 4 pixel region, one must store $10 + 4 \times 4 = 26$ bits per region hit times buffer depth of 5, 
or 32.5 bits per pixel.
The FE-I4 ROIC architecture is based on a 4-pixel region along these lines~\cite{Hemperek:2009xra},
shown schematically in fig.~\ref{fig:region}.

To take full advantage of larger regions (or to efficiently store higher precision charge information) 
one needs a more sophisticated storage mechanism that allocates memory only to hit pixels, and 
does not store zeros for pixels below threshold in the same region.  
Effectively, hit information must be transferred as soon as it is produced from the
pixels to a central storage. This should evoke the column drain architecture of fig.~\ref{fig:readarch1}\,(a).
A large region with arbitrated storage can be labeled as \glq region drain\grq\ and can be analyzed the
same way as column drain. 
The  main problem of column drain was the high bandwidth requirement for the column bus, $\mu \gg \lambda$.
But given a bus bandwidth $\mu$, one can always make $\lambda$ small enough by reducing the number of pixels on the bus, 
$N$, since $\lambda=Nr$ ($r$ being the single pixel hit rate).
In other words, column drain can accommodate high hit rate for a small enough ROIC.
Therefore, a large, high hit rate ROIC could be built as a matrix of \glq little chips\grq\, each
with column drain architecture. It is interesting that the architecture evolution moved away from 
column drain because it could not scale to larger size (or to higher hit rate for fixed size), 
in favor of region architecture, for which size and hit rate are decoupled.  
But, evolving to large regions eventually leads to the same point as replicating a column 
drain architecture many times, as a way to achieve both large total size and high rate.

\begin{figure}
\centerline{\includegraphics[width=0.9\textwidth]{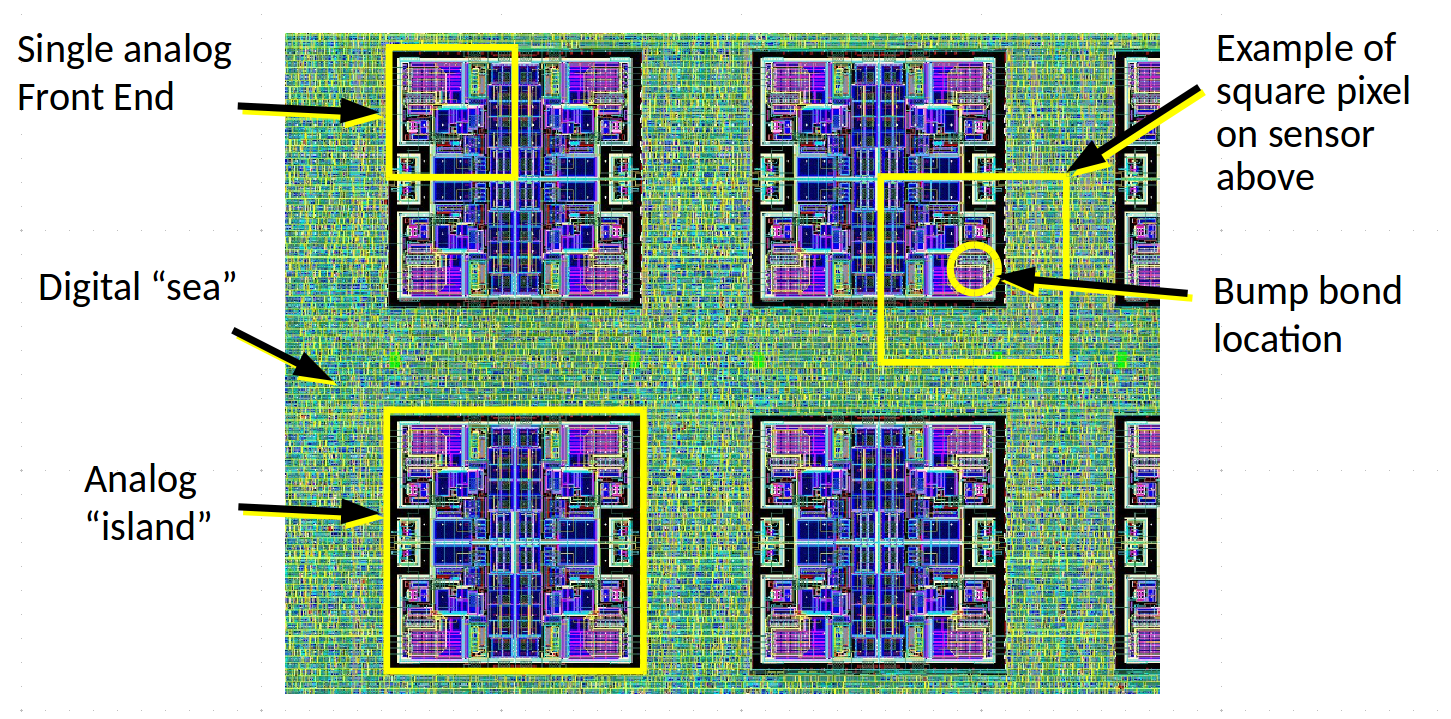}}
\caption{\label{fig:digitalsea} Layout detail from RD53 Collaboration illustrating the concept of islands of analog circuitry (blue) embedded in 
a \glq digital sea\grq\ of synthesized logic (green).}
\end{figure}

Third generation ROICs will continue to expand region architectures. 
While the basic 4 pixel region described above
probably represents a sweet spot for efficient memory usage, 
the use of larger regions is being explored~\cite{Gaioni:2016ihj}. 
Already in the 2$^{\rm nd}$ generation an important change took place,
in which a ROIC core went from being a collection of pixel circuits, stepped and repeated, to a collection of regions,
stepped and repeated. A 3$^{\rm rd}$ generation ROIC will be a collection of digital cores, stepped and repeated.
A core is not the same as a region and can in fact contain many regions. A core is simply the smallest stepped and repeated instance
of digital circuitry. A relatively large core allows one to take full advantage of digital synthesis tools to implement complex
functionality in the pixel matrix, sharing resources among many pixels as needed. Large cores can have 2 dimensional digital
connectivity, removing the constraint on all previous ROICs that communication could only take place up and down pixel columns,
but not along rows. Figure~\ref{fig:digitalsea} shows a layout detail from the RD53 Collaboration in which identical 4-pixel
analog front end islands can be seen completely surrounded by synthesized logic. The core logic has been dubbed
\glq digital sea\grq\ to stress that it results in a different and variable environment surrounding each analog island
(depending on where synthesis tools place gates and connections). This is a radical departure from the single pixel step and repeat,
perfectly symmetric environment of the 1$^{\rm st}$ generation, raising potential concerns about systematic variations within the
pixel matrix. The FE65-P2 prototype~\cite{Carney:2016},
which implemented 4 by 64 pixel cores, has shown that with modern isolation techniques
(see section~\ref{sec:ROIC-fend}) excellent uniformity can be achieved within a large synthesized core.

\subsection{Input hit rates and output data transmission (electrical) \label{sec:datatransmission}}

Hit rates and output bandwidth increase by an order of magnitude or more between between 
2$^{\rm nd}$ and 3$^{\rm rd}$ generation ROICs. For both past and planned detectors, 
output data are transmitted electrically for at least the first meter away from the ROIC
(optical links are covered in section~\ref{sec:ROIC-optical}).
The LHC-b experiment is implementing a full triggerless readout system in the upgraded VELO detector 
using the Velopix IC~\cite{Velopix-ASIC-2015}.
They are able to do this because the hit rate, while very high, is not extreme, and the experiment has a fixed target geometry,
allowing data cables to be placed outside of the physics acceptance. The ATLAS and CMS experiments, on the other hand,
must contend with an order of magnitude higher hit rate per ROIC and have nearly 4$\pi$ physics acceptance, so most data cables must pass through
the active volume. They will therefore use triggered readout. However, the trigger rates will be an order of magnitude higher than
today, which combined with busier events will make for a 20- to 40-fold readout bandwidth increase for 3$^{\rm rd}$ generation ROICs.
The LHC-b hit rate is projected to be 1.6\,GHz/cm$^2$ for the most occupied pixel, but as the rate decreases almost quadratically with
radius and the modules are perpendicular to the beam, the average hit rate in the most occupied ROIC is 0.3\,GHz/cm$^2$.
In contrast, for the ATLAS and CMS upgrades all pixels (and therefore all ROICs) in the innermost barrel layer have similar hit rates,
projected to be near 3\,GHz/cm$^2$ for a 3.2\,cm layer radius.

\begin{figure}
\centering
 \includegraphics[width=0.70\textwidth]{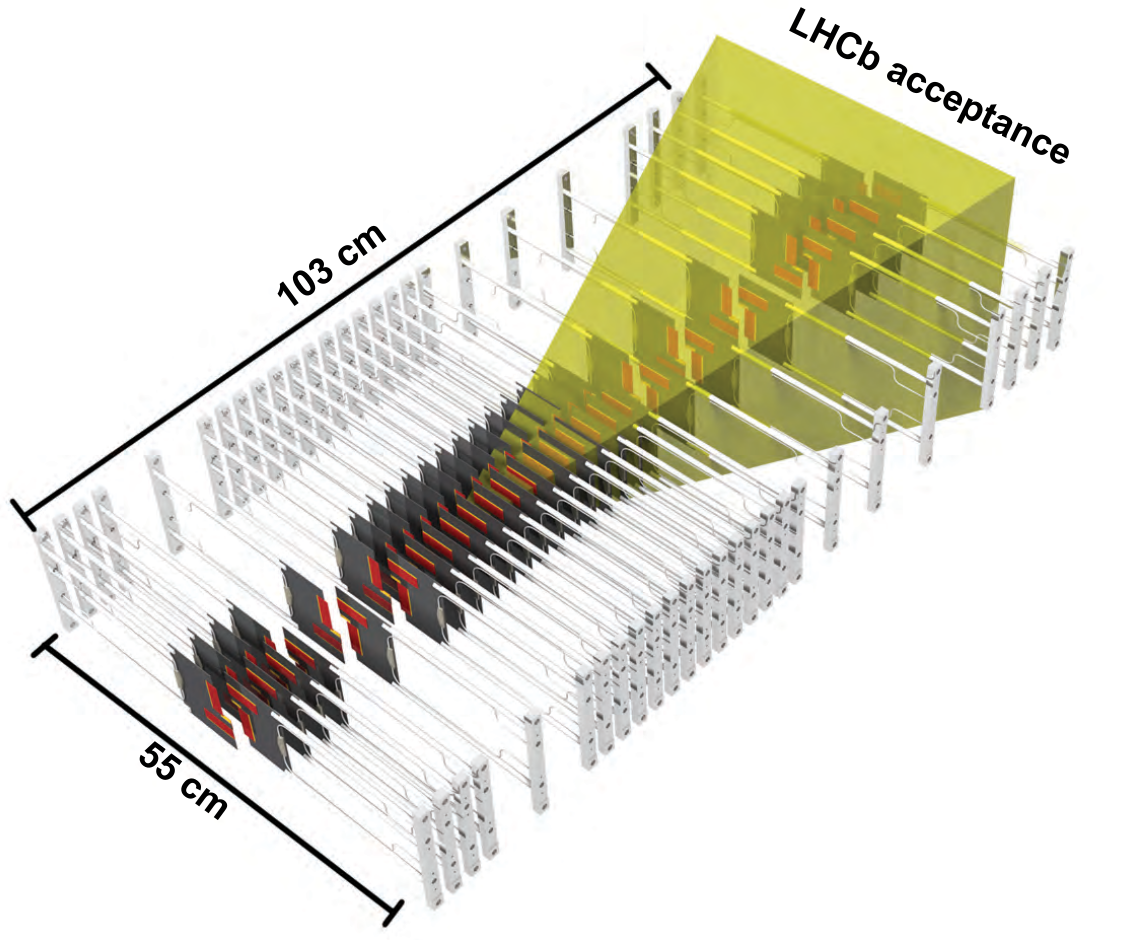}
\caption{\label{fig:LHCb-velo-upgrade} Geometry of LHC-b VELO detector. Beams circulate bottom left to top right and vice-versa, 
colliding near the center. The detector planes (dark grey and red) are perpendicular to the beam, and shown in 
the open configuration: during collisions they
will close onto the beam leaving a small square opening for the beam to pass through.
Only interactions boosted forward (towards top right) will be in the detector acceptance.
Taken from \cite{LHCb-velo-upgrade}.}
\end{figure}

Figure~\ref{fig:LHCb-velo-upgrade} shows the geometry of the LHC-b VELO upgrade.  
The fixed target geometry allows for high bandwidth data transmission with almost no cable mass in the active volume.
The Velopix IC supports 4 parallel differential outputs at 5.12\,Gbps each, fed by the shift readout architecture
described in section~\ref{sec:ROIC-architecture}. The data use a fixed frame custom format,
where each frame contains a time stamp and a bit map for a group of 8 pixels (no
charge information is recorded, just hit/no hit). Reading individual pixels instead would use 30\% more bandwidth, because 
it would not take advantage of the clustered nature of pixel detector hits. 

A theoretical discussion of lossless data compression in pixel detector readout can be found in~\cite{GarciaSciveres:2016nim},
which explains that the information content (entropy) is roughly proportional to the number of clusters and the achievable
compression depends on cluster size. The 30\% compression afforded by multi-pixel frames in Velopix is not the maximum possible
compression, but is adequate for LHC-b needs. The projected data packet rate for the most occupied chip is
520\,MHz, while the Velopix output consisting of 4 links at 5.12\,Gbps each will saturate at a 640\,MHz packet rate.

\begin{figure}
\centering
\includegraphics[width=0.60\textwidth]{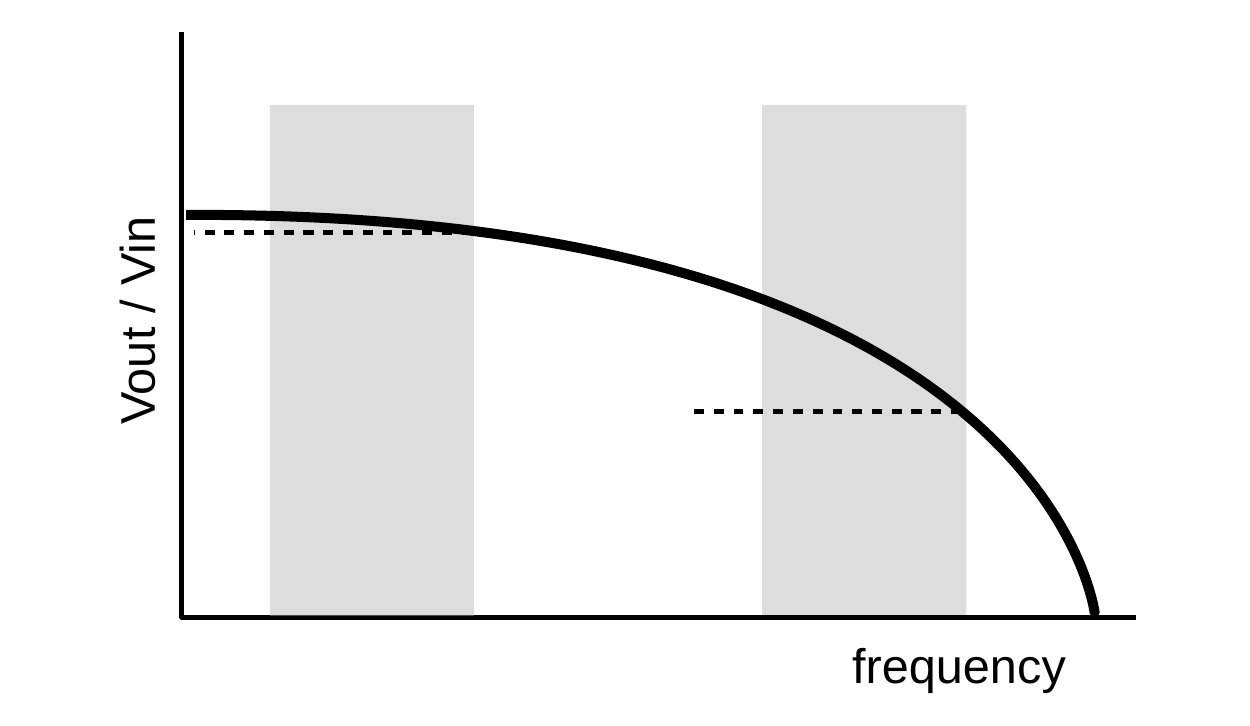}
\caption{\label{fig:cableloss} Conceptual sketch of attenuation vs.\, frequency in an electrical transmission line 
(solid line). The attenuation is given by the signal amplitude at the output divided by the input signal
amplitude (solid line, no specific functional form assumed). 
The the horizontal axis represents the frequency of a pure sine wave signal. 
The shaded areas represent the frequency content of a DC-balanced transmission if done at low
(left) or high (right) bit rate. The dashed lines represent the total attenuation of cable plus equalization,
which must be flat across the frequency band of interest to avoid signal distortion.}
\end{figure}

The ATLAS and CMS upgrades plan an output bandwidth of 5.12\,Gbps per ROIC, which is actually less than 25\% of the LHC-b maximum
because the planned ROIC size is slightly larger. A critical difference is that ATLAS and CMS must have longer cable runs and
at the same time minimize data cable mass, which means tolerating a high transmission line loss 
(meaning signal attenuation, not data loss).
Figure~\ref{fig:cableloss} shows conceptually the attenuation of a signal passing through a lossy transmission line.
Two vertical gray bands represent low and high bit rate transmission. These bands have a lower cutoff,
rather than extending to zero frequency, because DC-balance has been assumed.
DC balance avoids low frequencies by guaranteeing at least one transition every $n$ bits (e.g. for 
the 8b/10b encoding $n=5$). 
If the attenuation curve within a gray band is flat, that
means the signal shape will not be distorted: a good condition for error-free transmission.
This makes it obvious why DC balance improves transmission: the narrower the band, the smaller the variation of the attenuation within the band. However, for the band at high bit rate, 
even with DC balance there is a large variation of the attenuation within the band. 
This situation can be improved with equalization, indicated by the horizontal dashed lines.
Equalization attenuates lower frequencies more than higher frequencies 
(the opposite of what cables do) to achieve a flat response within the band. 
All this is well known textbook material~\cite{Proakis:2000},
but is only now playing a critical role in pixel detector design. 
The telecommunications industry and consumer electronics
have long ago optimized performance in lossy transmission situations. As a consequence, particle physics designs
are moving towards more sophisticated commercial protocols and solutions. State of the art equalization included in
commercial FPGA's can achieve reliable transmission with line losses as high as 28\,dB.
While prior ROICs (for example FE-I4) have used
8b/10b encoding, the RD53A ROIC will use an open source commercial protocol implementation of 64b/66b encoding, 
including a multilane version for balancing data over four 1.28\,Gbps outputs. 
Equalization will be an integral part of
future systems: RD53A drivers will have pre-emphasis capabilities, 
which boost high frequencies at the transmitter 
(sending a purposely distorted signal to counteract the cable distortion) 
and receivers with different types of equalization are
also being developed~\cite{Wallangen:2016, Flores:2016}. 
Receiver equalization has the advantage that it can be more easily made
adaptive, so that it does not have to be manually tuned for every link or 
every time environmental conditions change.
Even with these techniques, because of various constraints including radiation damage, 
the ATLAS experiment aims to keep line losses at 20\,dB or less.

It is clear from fig.~\ref{fig:cableloss} that there is a limit beyond which the signal is completely lost and no amount of equalization
can recover it (recall the 28\,dB mentioned above).
The only remaining option to further increase the data rate on such a cable is to use multiple logic levels
at a given frequency. For example with 4 logic levels instead of two, one can send two bits per clock cycle instead of one.
The problem is that the available output voltage must then be divided among the logic levels. The voltage difference between two levels
is $V_{\rm out}/(n-1)$, where $n$ is the number of levels. Noise can cause confusion between levels,
so one can use multiple levels as long as $V_{\rm noise} \ll V_{\rm out}/(n-1)$. This also applies to standard equalized signals
with two logic levels ($n$=2), which means that one cannot go too far to the right in fig.~\ref{fig:cableloss}. But one can
go further to the right for $n=2$ than for $n=4$, so whether multi-level transmission is advantageous or not depends on how fast
the cable transmission cuts off. What should be clear from this discussion is that using higher voltage at the driver
(for a given noise level) will increase the achievable data bandwidth. 
Commercial IC interface logic operates at a higher voltage than the core transistors, thanks to thicker gate oxide. But
such thick oxide devices are much less radiation tolerant and therefore not adequate for pixel detector applications.
For 65\,nm the maximum operating voltage of core transistors is 1.2\,V. Future development could include multi-level
transmission using stacked core transistors, where the voltage difference between all levels remains around 1\,V instead of
1.2\,V/$(n-1)$. A final problem facing high speed drivers in pixel ROICs is that high speed logic is more vulnerable to radiation
damage than low speed logic, because switching speed depends on transistor gain. This could be a pixel-specific reason
favoring the use of multiple logic levels at lower clock speed over the standard two levels at higher clock speed.

While not strictly an ROIC issue, the quality of available cables is a critical point for data transmission.
Commercial data cables are not developed with low mass requirements and contain dielectric materials that are
not radiation hard. Therefore, development of custom cables for low mass, radiation hard, high speed transmission
is an ongoing activity. This includes flexible printed circuits~\cite{McFadden:2016qdz},
short run unshielded twisted pairs, long run shielded twisted pairs, and twin-axial~\cite{Shahinian:2016C}.

\subsection{Optical and wireless transmission \label{sec:ROIC-optical}}

The technical design report for the original ATLAS pixel detector~\cite{ATLAS-TDR-pixel} specified
optical links located on the module, less than 1\,cm away from the closest ROIC, with optical fibers
forming part of the module interconnect. Such an arrangement was never realized and designs have instead
moved further away from this model. There are three reasons why optical conversion is located
some distance away from pixel modules: radiation, geometry, and reliability. Fibers and active optical
elements both suffer radiation damage~\cite{Hall:2012C, Mazza:2013C} and are not currently suitable
for environments exceeding 100\,Mrad. For the laser drivers, but not the fibers, there have been
promising developments in silicon photonic systems using Mach-Zehnder interferometers rather than switching
lasers, following the trend in commercial high speed optical links~\cite{Seif:2015}.
But customization of such devices for particle physics has so far not been widely successful.
Geometrically, pixel detectors are compact, with highly constrained cable routing to achieve
hermetic coverage. Routing of fibers to all modules in a pixel detector would require complex manipulation
of unjacketed fibers, which are fragile and have a minimum bend radius that must not be exceeded.
Electrical cables are much more forgiving and can tolerate any bend radius. 

\subsection{Power distribution \label{sec:ROIC-power}}

The 1$^{\rm st}$ generation ATLAS and CMS pixel detectors achieved a power distribution efficiency of approximately 25\%. 
This means that 75\% of the power delivered by rack power supplies was dissipated in cable IR loss and voltage regulators.
Since the only current return path is thorough the detector, the power delivery efficiency is given by 
$V_{\rm L}/V_{\rm sup}$, where $V_{\rm L}$ (load voltage) is the voltage across the detector 
elements connected to each single power supply channel, and 
$V_{\rm sup}$ is output voltage at the power supply. 
Inefficient power delivery was recognized as a problem, 
but nevertheless the 1$^{\rm st}$ generation detectors could be built using simple, direct power 
delivery to each module within a material budget considered acceptable. The planned high luminosity upgrades, 
on the other hand, face a much greater power delivery challenge. Not only will the detectors be larger, but the 
deep submicron electronics used will operate at half the voltage yet with slightly more power per unit area.
This leads to a total detector current requirement an order of magnitude higher than present detectors. 

However, the current
that can be delivered from rack power supplies to the ATLAS or CMS inner detectors is limited by
external factors.
The resistance of the cable plant supplying the inner detector has a lower practical limit, 
$R_{\rm min}$, imposed by mass and space considerations. 
At the same time, the IR-loss heat load on the subdetectors traversed by these cables is limited
to $P_{\rm lim}$ so as not to degrade their performance. Therefore, the rack power supplies are only allowed 
to deliver at most $I_{\rm sup} = \sqrt{P_{\rm lim}/R_{\rm min}}$. 
Approximate values for ATLAS are $P_{\rm lim} \approx 10$\,kW and 
$R_{\rm min} \approx 0.5$\,m$\Omega$ (round trip), 
yielding $I_{\rm sup} \approx 5$\,kA. The only free parameter left to control the power
supplied to the detector is the load voltage $V_{\rm L}$. 
For an expected 3$^{\rm rd}$ generation ROIC power of order 0.6\,W/cm$^2$, 
a 10\,m$^2$ pixel detector would need a load voltage of 12\,V to deliver the 60\,kW total power using 5\,kA 
of rack supply current. Since the ROIC operating voltage $V_{\rm ROIC}$=1.2\,V, directly powering 
the ROICs is out of the question: power conversion at the load is required. 

From the above discussion, the voltage conversion ratio required for HL-LHC 
pixel detectors is of order 10.
Two conversion methods have been extensively investigated: DC to DC converters and serial power. 
The DC-DC conversion technologies investigated include magnetic switching 
using air core inductors~\cite{Michelis:2007, Faccio:2015dc} 
\footnote{Ferromagnetic cores cannot be used in the strong tracker solenoid fields.}, 
piezo-electric converters~\cite{Lui:2006}, 
and capacitor charge pumps~\cite{Denes:2009}. While discrete DC-DC converters have now been developed 
to be compatible with LHC 
experiments~\cite{Faccio:2015dc}, their use on pixel detector modules 
would result in too much added mass due to the high current per unit area required. 
Fully on-chip DC-DC conversion has been investigated~\cite{krieger:2014} as this would not add any mass, 
but has not yet matured enough for consideration in the construction of HL-LHC detectors. 

\begin{figure}
\centering
 \includegraphics[width=\textwidth]{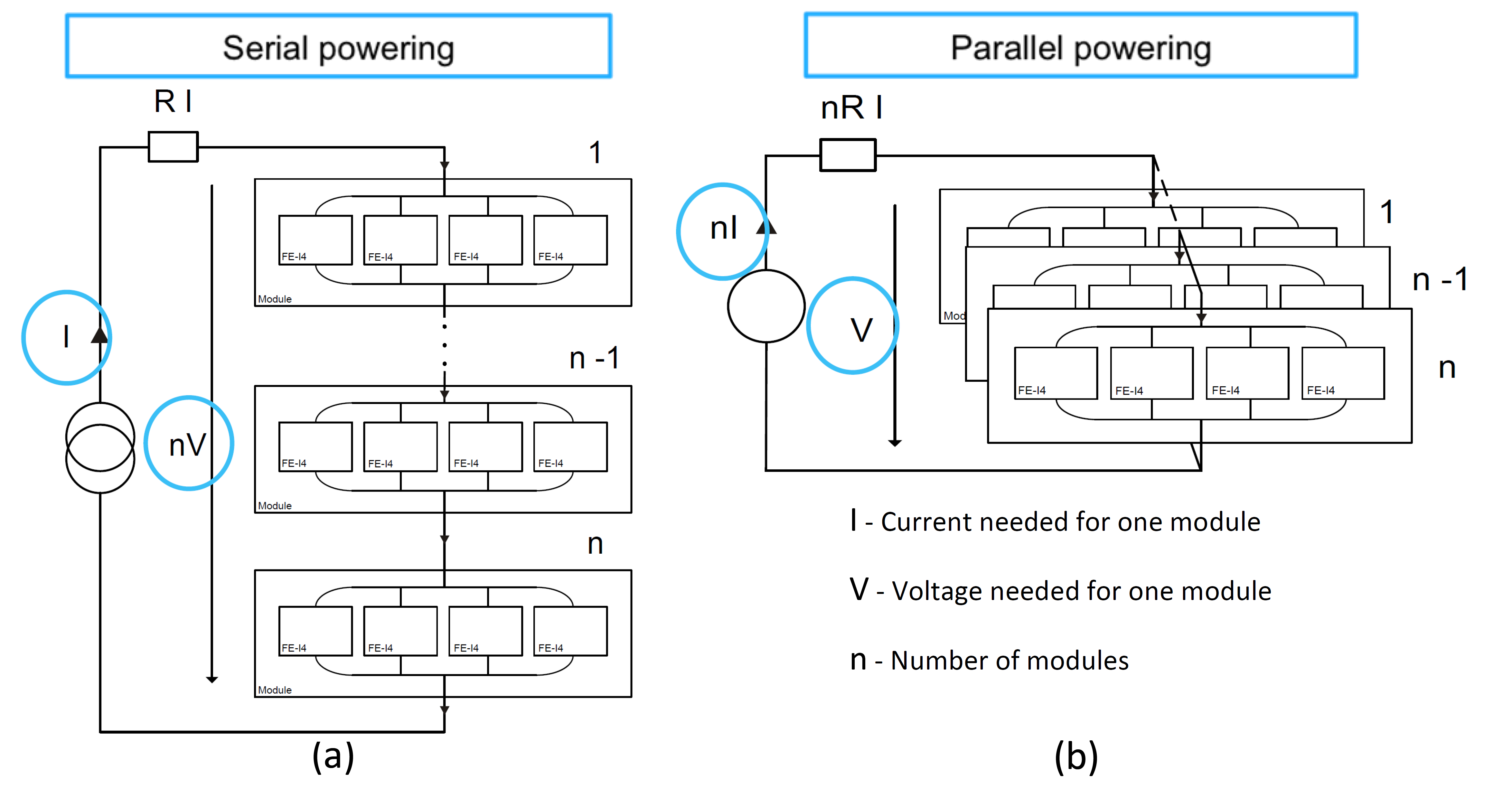}
\caption{\label{fig:serialpower} Schematic comparison of serial (a) vs. parallel (b) powering
of a set of 4-chip modules. Each module has 4 ROICs that are always
in parallel within the module, while the modules themselves can be placed in series (a) or in parallel (b). 
The load voltage for the parallel configuration is equal to the ROIC operating voltage V, 
whereas in the configuration with $n$ modules in series the load voltage seen by the remote power supplies 
is $n \cdot$V. The voltage drop in the cables is shown above discrete resistors of value R (same 
R in both cases)}
\end{figure}

Serial power achieves power conversion without added mass by connecting devices in series and 
operating with constant current instead of constant voltage power supplies.  
Figure~\ref{fig:serialpower} compares 4-chip modules in serial vs. parallel power configurations.
The serial configuration only needs two conductors and one power supply channel to serve $n$ modules. 
Because $V_{\rm L} = n \cdot V_{\rm ROIC}$, the power delivery efficiency ($V_{\rm L}/V_{\rm sup}$) 
can be increased as needed by increasing $n$. Furthermore, constant current supplies are 
better suited to delivering power over resistive cables than the more common constant voltage 
supplies. Constant current powering 
is used commercially with very long, resistive cables, for example in undersea 
communications lines~\cite{Chesnoy:2015}. 
However, the requirements of pixel detector operation are very different and a custom serial
power technology solution had to be developed.   
Serial power has been found to be reliable to implement and compatible with 
pixel~\cite{Stockmanns:2005, Ta:2006zf, Filimonov:2017vdi}
(and strip~\cite{Matheson:2011}) detector operation, and is the baseline for both ATLAS and CMS 
HL-LHC pixel detector designs. Implementation requires specialized constant current regulators, which have been 
developed~\cite{Karagounis:2009} and included in the FE-I4 ROIC and the RD53A ROIC design. 
Prior efforts used simpler shunt voltage regulators~\cite{Peric:2006km}.  

\subsection{Future development \label{sec:ROIC-future}}

Even as monolithic solutions mature, hybrid technology with special purpose ROICs will continue to be necessary for the highest
rate capability and the implementation of new functionality, such as fast timing. Looking to industry we find that the
monolithic CMOS sensors in modern smartphones are actually 3D integrated devices. To the extent that such technology becomes open to
low volume third party customization, monolithic devices for particle physics could also consist of multiple layers with
processing and readout in small feature size CMOS exactly as conventional hybrid ROICs. In the coming years ROIC R\&D will
explore smaller feature sizes, with 28\,nm being the likely candidate to follow 65\,nm for 4$^{\rm th}$ generation ROICs
within a ten year horizon~\cite{Zhang:2016epfl}.

The 4$^{\rm th}$ generation ROICs will add even more sophisticated digital functionality with minimal,
but high speed analog front ends to enable fast timing.
The operation of the individual pixels will likely be continually monitored and adjusted in real time with no need for storage of
configuration data or lengthy calibrations. Improvements in communication and data compression will continue.
Single lane output switching frequency will likely not increase beyond 5\,Gbps due to issues with radiation tolerance of high speed logic,
but data rate could be increased by making use of multi-level logic,
essentially following industrial trends to maximize data rates in existing infrastructure
originally designed for lower bandwidth. The point in particle physics is not to fit into existing infrastructure but to use the lowest
possible mass cables. It is also possible that improvements in data cables themselves will be realized (higher bandwidth for lower mass),
for example by replacing copper conductors and/or aluminum shields with carbon nanotube \glq rope\grq\ or graphene foil, respectively.
Some use of wireless technology is possible, if nothing else to facilitate testing by being able to communicate during detector construction/integration without physical connections. 
Finally, on-chip power management will surely increase, dynamically managing core voltage levels and
possibly adding on-chip DC-DC conversion with high ratio (4 or more) providing an alternative to serial power.

%% file: supports.tex
\section{\label{sec:supports} Advanced materials: new possibilities for supports and cooling}

The original ATLAS and CMS pixel detectors achieved an amount of material around 3.5\% of a radiation length per layer 
at normal incidence~\cite{ATL-PHYS-PUB-2015-050, ATLAS-TDR-IBL}.  
Of this, 0.5\% was due to silicon (sum of ROIC and sensor). 
Since then, significant advances in mechanical support (section~\ref{sec:SupSupport}) 
and cooling (section~\ref{sec:SupCooling}) have taken place, 
and future ATLAS and CMS detectors are projected to be more than 50\% lighter, between 1\%  and 1.5\% per layer, 
out of which 0.3\% will be silicon. 
Still lower mass is achieved with monolithic instead of hybrid technology, of order 0.3\% per 
layer~\cite{Abelevetal:2014dna,Schambach:2014uaa}. The reason 
is not only less silicon, which can at most save 0.3\%, but less stringent cooling requirements for 
monolithic sensors with small pixels and small depleted volume (and therefore small leakage current and no thermal runaway issues). 
However, monolithic active sensors compatible with high rate and radiation are still under development 
(see chapter~\ref{sec:cmos}). 

Some critical developments have been CO$_2$ evaporative cooling, 
new composite materials, new methods for structure design and fabrication, and serial power distribution. 
Serial power distribution, which is an electronic system development, 
has been covered in section~\ref{sec:ROIC-power}. 

\subsection{\label{sec:SupCooling}Cooling}

The high power density and large extent of pixel detectors at the LHC demands a coolant that can transport a large 
amount of heat with low mass flow. Since evaporation can remove more heat for the same mass flow as single phase 
liquid cooling, it is preferred. Gas phase (air) cooling has been used for the STAR Heavy Flavor 
Tracker~\cite{Wieman:2009},
but that detector is compact (20\,cm active length and 0.2\,m$^2$ active area) 
and based on CMOS sensors that do not need to be kept cold. Nevertheless,
0.2\,W/cm$^2$ was removed with 10\,m/s simple air flow (no fins or heat transfer enhancement features). 
An important advantage of evaporation is the absence of a temperature gradient along cooled structures. 
With an evaporative cooling system, the coolant temperature is fixed, regardless of heat load (for load less than maximum capacity). 
The heat is absorbed by the liquid to gas phase transition, so the difference between inlet and outlet 
of a cooled structure is the liquid fraction (high at the inlet and lower at the outlet). The 
maximum load capacity is reached when all the liquid has evaporated - a condition called dry-out, which should never 
be reached in a properly working system. 

\begin{figure}
\centering
\subfigure[]{\includegraphics[width=0.35\textwidth]{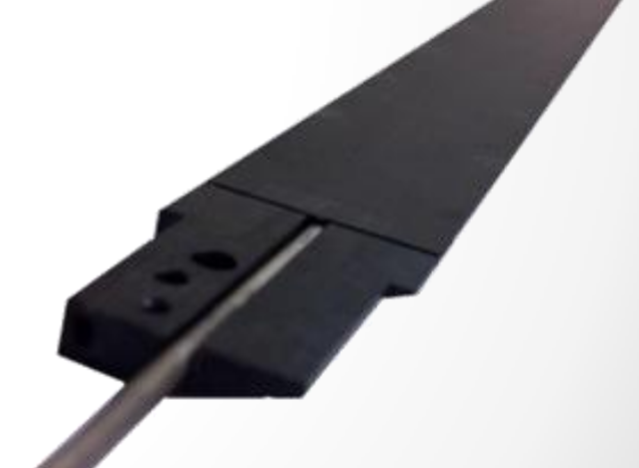}}
\hspace{0.3in}
\subfigure[]{\includegraphics[width=0.55\textwidth]{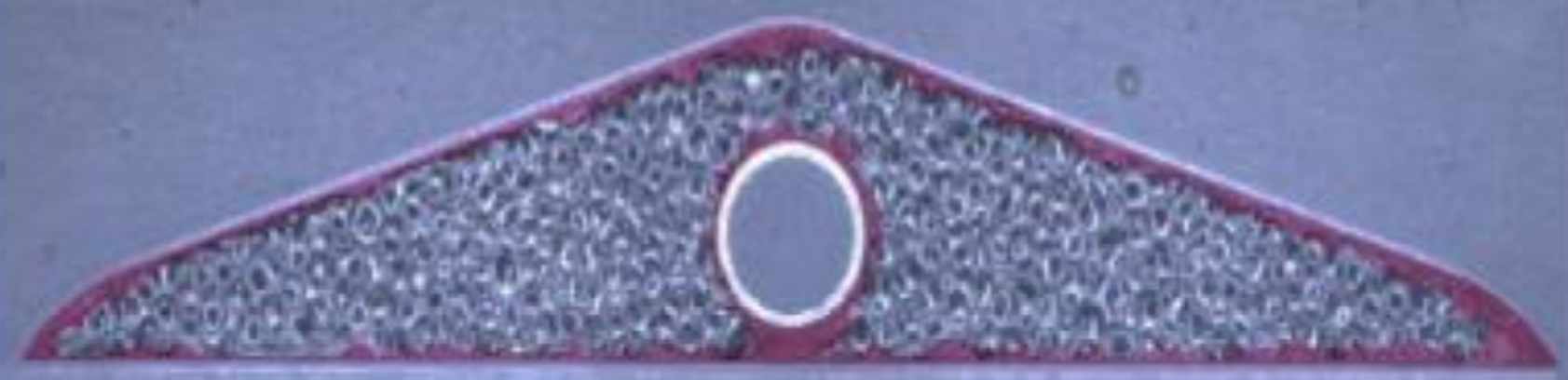}}
\caption{\label{fig:IBLstaves} ATLAS IBL stave support with 1.7\,mm diameter Ti pipe for CO$_2$ evaporative cooling.
Full stave view (a) and cross section (b), showing thermally conductive foam interior 
and 100\,$\upmu$m wall thickness pipe.}
\end{figure}

CO$_2$ is an ideal evaporative coolant for tracking detectors because it offers heat transfer 
coefficients an order of magnitude higher than traditional refrigerants, and 
its high evaporation pressure (around 50\,bar) means small produced vapor volume, resulting in small diameter tubing.
However, pressure safety is one of the challenges of using CO$_2$, requiring a rating of 200\,bar 
throughout. 
The use of evaporative CO$_2$ for tracker cooling was pioneered by the AMS experiment~\cite{Delil:2002}, 
and in colliders by the LHC-b experiment. 
It was later adopted for the ATLAS IBL upgrade (fig.~\ref{fig:IBLstaves}) and the CMS pixel upgrade. 
For a review see~\cite{Verlaat:2009zz}. The original ATLAS and CMS detectors used other coolants that were 
better established at the time and did not need very high pressure plumbing. 
ATLAS used evaporative C$_3$F$_8$~\cite{Attree:2008},
while the original CMS pixel detector used monophase cooling with C$_6$F$_{14}$.     

\begin{figure}
\centering
\includegraphics[width=0.4\textwidth]{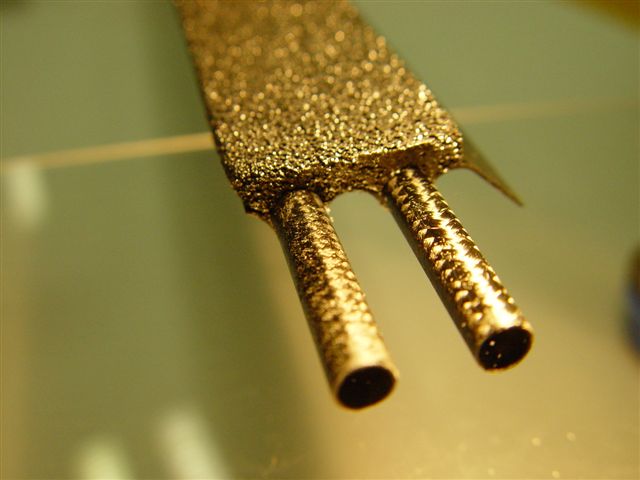}
\caption{\label{fig:cfpipe} Photograph of a prototype mechanical support built with braided carbon pipes, 
embedded in carbon foam.}
\end{figure}

While aluminum has been used for cooling pipes in more conventional (fluorocarbon) evaporative systems~\cite{Attree:2008}, 
it is very difficult to prevent corrosion in aluminum pipes during detector construction~\cite{Garcia-Sciveres:2009}, 
and weakening of pipe walls due to corrosion can be fatal for a high pressure system. 
Therefore, for many present CO$_2$ cooling systems the pipe material of choice is titanium, which is 
corrosion resistant and high strength (so can withstand high pressure with thin walls), yet relatively low mass.
The main reduction in mass from CO$_2$ cooling comes not from the pipe material, but from the very small diameter
pipes that can be used. Other materials have also been explored,
most notably carbon fiber, for which braided fiber tubes are a common industrial product and so it 
is possible to produce braided pipes, but so far this has not been adopted. Figure~\ref{fig:cfpipe}
shows a prototype support with 2\,mm diameter braided pipes embedded in carbon foam. For a review 
of pixel detector cooling and thermal management materials see \cite{Viehhauser:2015}. 

\begin{figure}
\centering
\includegraphics[width=0.8\textwidth]{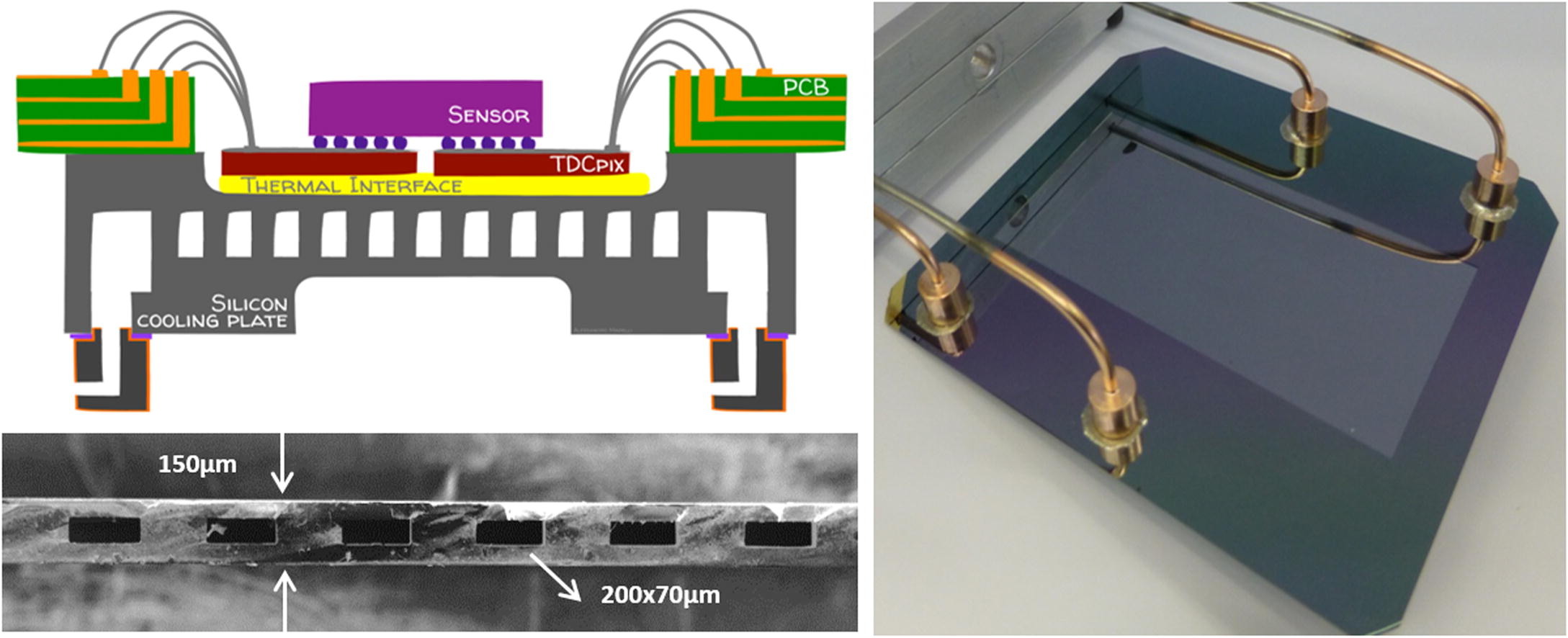}
\caption{\label{fig:GTKcooling} Photograph of a micro channel cooled substrate (right), 
with diagram of cross section (top left), 
and photo of sectioned substrate showing micro channels~\cite{Romagnoli:2015} (bottom left).}
\end{figure}

A more recent development has been the use of silicon micro-channels with CO$_2$ cooling, in order to achieve very
high cooling capacity with even lower mass than possible with separate mechanical supports cooled with metal pipes. 
This is suitable for fixed target geometries where mass can be placed immediately outside active elements, and has 
been pursued in the NA62 Gigatracker~\cite{Romagnoli:2015} (fig.~\ref{fig:GTKcooling}) 
and LHC-b~\cite{Aguilar:2015} upgrades. 

\begin{figure}
\centering
\includegraphics[width=0.8\textwidth]{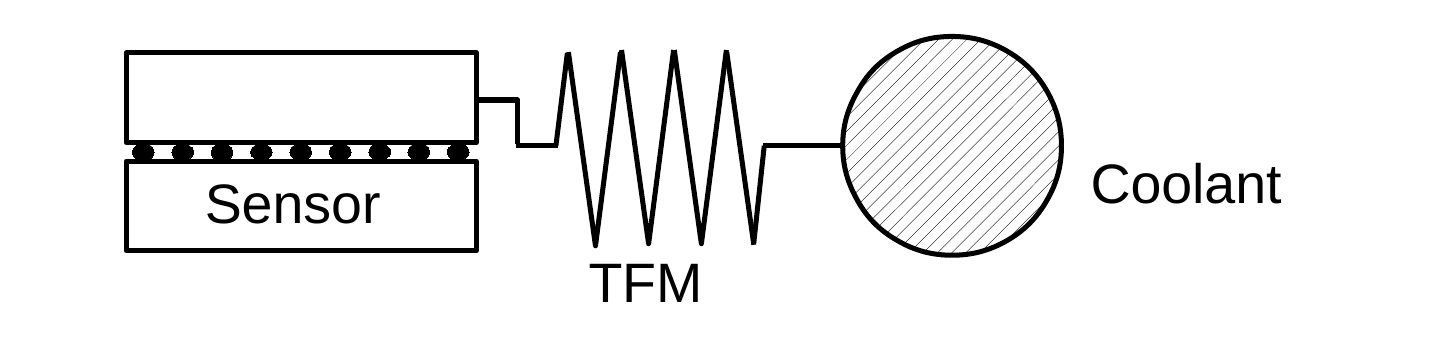}
\caption{\label{fig:TFM} Simple model of thermal behavior of hybrid pixel devices. The specific thermal resistance is known as Thermal Figure of Merit (TFM).}
\end{figure}

For hybrid pixels, detector lifetime and cooling performance are closely coupled. The irradiated sensor leakage current scales with 
temperature (rule of thumb is that current doubles every 7$^{\circ}$C), while the temperature depends on the power dissipated in the sensor,
which in turn depends on leakage current. 
This leads to the well known condition of thermal runaway above a certain temperature~\cite{Kohriki:1996}.
The behavior can be captured by a simple model with the cooling performance of mechanical supports represented 
by a specific thermal resistance (cm$^2 \times ^{\circ}$C/W), 
also referred to as Thermal Figure of Merit (TFM)~\cite{Sharma:2017a}, as shown in fig.~\ref{fig:TFM}. 
With an evaporative cooling system, the coolant temperature is fixed, so the sensor temperature rises linearly with 
power dissipation, and the proportionality constant is the TFM. Thus the TFM defines positive feedback that leads to thermal 
runaway. High TFM means large positive feedback and therefore early thermal runaway (short lifetime), while zero TFM would be ideal. 
The mechanical supports of 1$^{\rm st}$ generation ATLAS and CMS pixel detectors as built achieved a TFM 
of 30$^{\circ}$C\,cm$^2$/W,
while support structures prototyped for the HL-LHC upgrade detectors have achieved 
a TFM as low as 10$^{\circ}$C\,cm$^2$/W
(can expect 15$^{\circ}$C\,cm$^2$/W for as built detectors). 
This impressive advance has been obtained by utilization of new, more performant carbon composite 
materials, such as thermally conductive carbon foam~\cite{allcomp}. High thermal conductivity, low mass foam
allows one to take advantage of the CO$_2$ cooling by providing a way to efficiently collect heat from a large area module 
and couple it to a small diameter tube. Commonly used values of foam density and thermal conductivity are 0.2\,g/cm$^2$
and 40\,W/m/K, 
respectively\footnote{Foam thermal conductivity scales approximately with density, which can be adjusted in production.}. 

\subsection{\label{sec:SupSupport}Supports}

\begin{figure}
\centering
\subfigure[I-beam]{\includegraphics[width=0.36\textwidth]{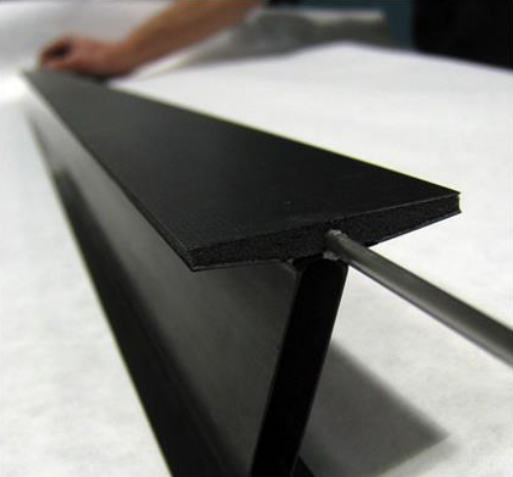}}
\hspace{0.3in}
\subfigure[Alpine]{\includegraphics[width=0.45\textwidth]{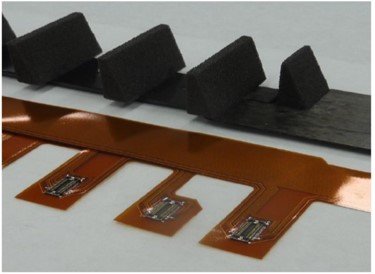}}
\caption{\label{fig:staves} Photographs of carbon composite pixel support prototypes: 
(a) \glq I-beam\grq\ geometry coupled layer and 
(b) \glq Alpine\grq\ design with thermally conductive foam pedestals to hold modules out of plane.
(From \cite{CERN-LHCC-2015-020}.)}
\end{figure}

Support structures must combine the excellent cooling capacity discussed in section~\ref{sec:SupCooling} with 
low mass and mechanical stability high enough to maintain module positions to within 10\,$\upmu$m. 
Carbon composite designs and methods have enabled substantial mass reduction. A successful design approach for inner pixel layers 
has been the use of coupled layer structures. Single layer supports spanning long distances (tens of cm) 
suffer gravitational and vibrational 
deformations that elements are too large without auxiliary stiffening, such as shells or frames, 
and the mass of these elements also contributes to the detector radiation length. By coupling layers 
in pairs, structures can be made stiff enough over long spans without auxiliary 
support structures, thanks to the large moment of inertia or the coupled layers.
The ALICE and STAR experiments used box beam structures for this purpose~\cite{Pepato:2006jr,Schambach:2014uaa}. 
In the case of STAR the approximately 30\,cm long structures were held cantilevered from one end only. A further 
development in this direction, providing similar mass and stiffness with additional usable space, 
higher nesting freedom, and 
monolithic construction uses an I-beam rather than a box beam shape~\cite{Hartman:2013ib} (fig.~\ref{fig:staves}~(a) ).
Other large moment of inertia structures use truss-like assemblies~\cite{Abelevetal:2014dna}.

New materials such as foam have also enabled development of supports with more complex geometry that position modules with some 
tilt angle relative to the colliding beam direction~\cite{Smart:2016}. 
An example of a support called Alpine~\cite{Delebecque:1516529} using foam pedestals 
is shown in fig.~\ref{fig:staves}~(b). Endcap pixel detectors have seen similar evolution, 
with CO$_2$ cooling and carbon composite 
developments allowing for lower mass and greater design freedom~\cite{CMS-TDR-pixel-upgrade}. 
Planned high luminosity upgrades of ATLAS and CMS will 
have greater acceptance coverage and therefore larger endcap systems~\cite{Smart:2016, Migliore:2016}.

%% file: cmos.tex
\section{CMOS active sensors: towards monolithic pixels}\label{sec:cmos}
The many advantages of hybrid pixel detector technology have been highlighted throughout this report. 
However, the hybrid choice also has some notable disadvantages. Hybrid pixels constitute a relatively large material budget,
typically more than 1.5\% X$_0$ per detector layer (ATLAS IBL), distributed among the module components 
and the cooling and support structures.
The module production including bump-bonding and flip-chipping is complex and laborious, leading to a large number of production steps.
Consequently, hybrid pixel detectors are comparatively expensive.

Modern CMOS imaging sensors instead make use of 3D integration to combine high resistivity and fully depleted charge collection layers with high
density CMOS circuitry, in order to achieve high speed and high collection efficiency (for low light operation). Such a combination
of fully depleted high resistivity silicon with CMOS readout sounds like a requirement from particle physics, not from consumer electronics,
but smartphone image sensors have independently evolved in this direction for different reasons and with different optimization.
For example, the pixels are very small ($<$\,20\,$\upmu$m$^2$) and the depleted layer is very thin (few $\upmu$m). Nevertheless, their
out-of-the-box use for radiation and particle detection is being explored~\cite{Whiteson:2015nj_arxiv}. Such high end
fabrication processes are currently not available to be customized for particle physics.

The following sections address current developments towards fully depletable, radiation tolerant, and high rate capable monolithic pixel devices, inheriting
in part from successful adaptation of CMOS camera type sensors with thick epitaxial layers. The development of fully depletable devices becomes feasible by exploiting
advances in the CMOS industry, in particular the availability of multi-well technologies for 
both low and high resistivity substrate wafers.

\subsection{From MAPS to DMAPS}\label{sec:DMAPS}
\begin{figure}
\centering
   \subfigure[MAPS]{\raisebox{0.2cm}
            {\includegraphics[width=0.40\textwidth]{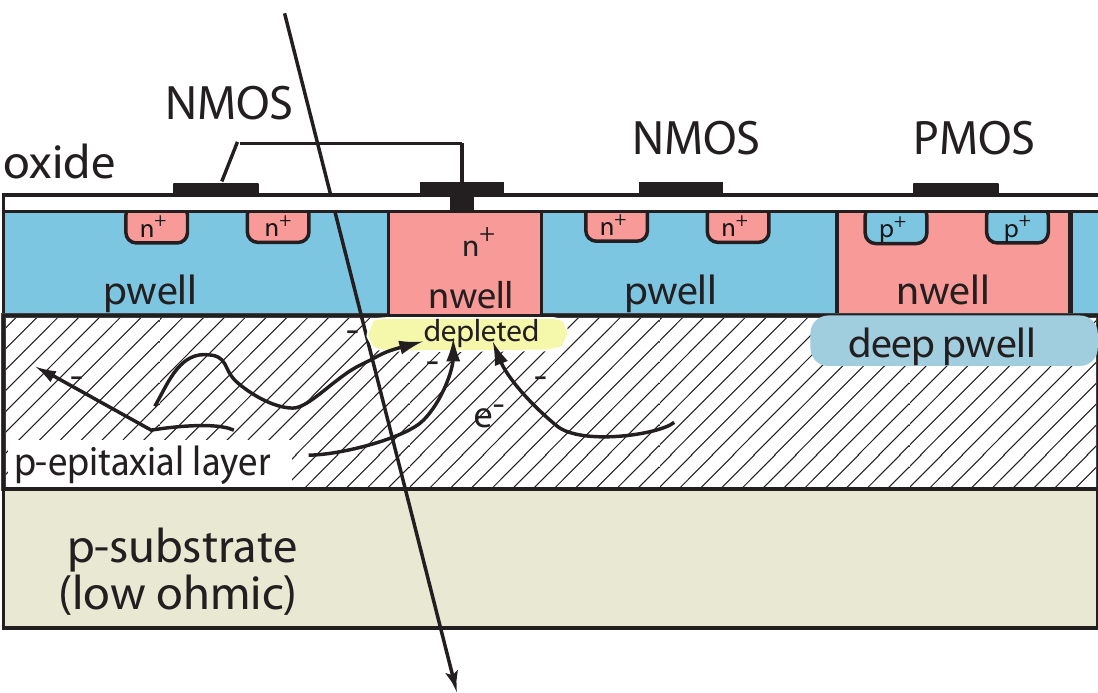}}\label{fig:CMOS_epi}}\hskip 0.5cm
   \subfigure[DMAPS]{
    \includegraphics[width=0.52\textwidth]{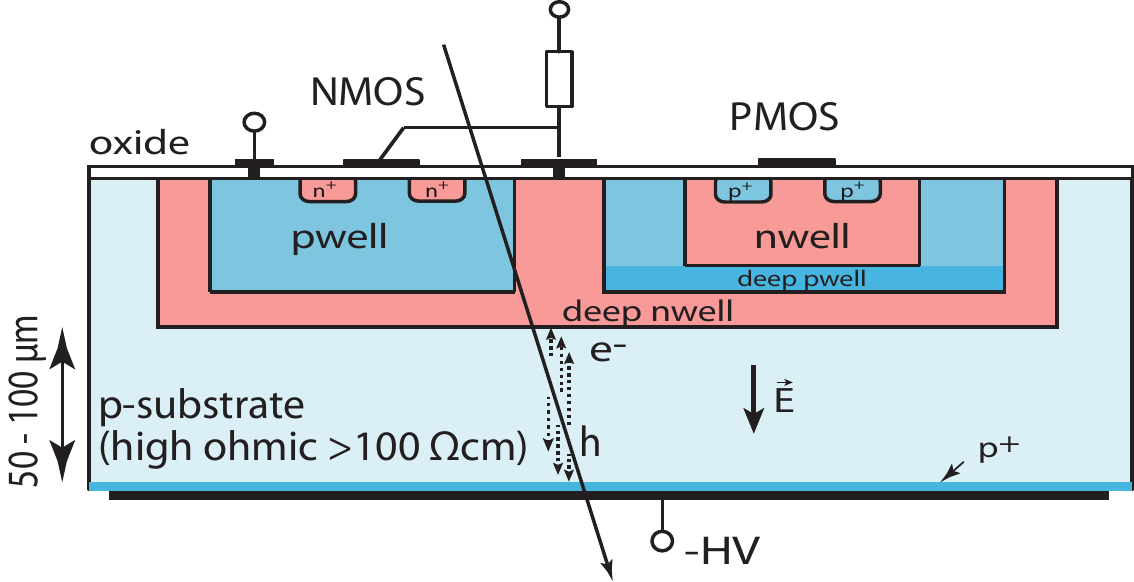}\label{fig:DMAPS_generic}}
\caption{(a) Conventional monolithic active pixel sensor (MAPS) with charge collection in an epi-layer mainly by diffusion. An \emph{n-well} acts as the charge collection node. Other n-wells in the circuitry
must be shielded (here by a p-well). (b) DMAPS (Depleted MAPS) structure. Multiple wells on high resistive substrate
allow complete embedding of the CMOS electronics layer in a charge collecting \emph{deep n-well}.}
\end{figure}

Employing commercial CMOS technologies to produce a monolithic (rather than hybrid) pixel detector
in which pixel sensor and electronics circuitry form one entity, has first been proposed \cite{parker1989} and realized \cite{KENNEY1994} in the early 1990s. Some years later, Monolithic Active Pixel Sensor (MAPS) detectors
were introduced~\cite{Dierickx1998,MAPS-epi_Turchetta:2001}, exploiting as the sensing volume
an epitaxial layer often grown on top of the lower quality substrate wafer and hosting the CMOS circuitry.
The thickness of this epi-layer typically is in the range of $1\mbox{--}20\,\upmu$m, where thicker layers are
often used in processes addressing CMOS camera applications (fig.~\ref{fig:CMOS_epi}).

For particle detection the charge deposited in the epi-layer can be as large as 4000\,e$^-$ for
a typical thickness of 15\,$\upmu$m. Since the epi-material usually has low resistivity and the allowed biasing voltages are low in CMOS technologies, the epi-layer usually is depleted only very locally around the charge collection node. The deposited charge of a traversing particle therefore is mostly collected by diffusion rather than by drift. This renders the signal generation slow and incomplete (not all charges arrive at the collection node) and is, besides the lower radiation hardness, the main reason why this original MAPS technology is not suited for high rate applications as needed in LHC pp-experiments. Furthermore, other n-wells, e.g. those hosting PMOS transistors, act as competing nodes for charge collection. The latter can and must be cured by additional deep p-well protections as is also shown in fig.~\ref{fig:CMOS_epi}.

MAPS pixel detectors have been successfully used in lower than LHC rate and low radiation experiments, such as the STAR experiment at RHIC \cite{RHIC_HSTD_2015,RHIC_Schambach_2015}.
Also the ALICE ITS Upgrade \cite{Abelevetal:2014dna} has chosen MAPS pixels based on the 180\,nm CMOS node offered by TowerJazz with 6 metal layers \cite{Mager:2016yvj}. Note, however, that the
expected radiation level for HL-LHC ion collisions is 700\,krad and $10^{13}$\,n$_{eq}$/cm$^2$, respectively, i.e.\,1500 times lower than for LHC pp-experiments.
%

\subsubsection{Depleted (D)MAPS}
In order to further develop CMOS pixel detectors for LHC type rate and radiation applications, improved development lines have been pursued leading to \emph{depleted monolithic active pixel sensors} (DMAPS) \cite{Peric:2014faa}\cite{Wermes:2016dav}.
The goal of this new development is to employ commercial CMOS technology with some modifications to obtain sufficient signal and fast timing in monolithic CMOS designs, while maintaining charge collection via charges drifting in an electric field inside the chip's substrate. The technology and the
sensing properties must survive the radiation environment at the HL-LHC, at least in the outer
layers, far enough from the interaction point, such that the radiation levels are similar to those presently encountered at the inner layers, i.e.~100\,Mrad ionisation dose and $2\times 10^{15} n_{eq}/$cm$^2$ particle fluence, respectively.

The development of such detectors much relies on recent advancements and freedom in CMOS technologies, offered in particular by vendors interested in market corners away from mass IC production and offering
process add-ons or modifications. The goal is to achieve some (50--100$\,\upmu$m) depletion depth
\begin{equation}\label{eq:depletion}
  d \sim \sqrt{\rho V}
\end{equation}
where $\rho$ is the substrate resistivity and V is the bias voltage,
yielding a reasonably large signal ($\sim$\,4000\,e$^-$) with fast and in-time efficient charge collection,
while avoiding long collection paths on which charges can be trapped after irradiation.
At the same time full CMOS functionality shall be maintained, i.e. equal and unconstrained usage of 
PMOS and NMOS transistors and no or little interference of electronics signals and detector signal pulses.

DMAPS detectors in particular exploit the following CMOS technology features:
\begin{itemize}
  \item High voltage technology add-ons (from automotive and power management applications) that increase the voltage handling capability and create a depletion layer in a well's pn-junction of depth in the order of 10$\mbox{--}$15\,$\upmu$m.
  \item Medium to high ($>$100$\,\Omega$cm) resistivity 8$''$ silicon substrate wafers, accepted and qualified by the foundry. A depletion layer develops due to the high resistivity with only moderate bias voltages applied from the electronics side or a (specially processed) backside contact.
  \item Multiple nested wells (see also chapter~\ref{sec:fechip}) that can be used to isolate transistors and shield deep well potentials in order to optimize charge collection. The foundry must accept some process or design rule changes in order to optimise the design for HEP applications.
  \item Backside processing add-ons allowing for example a backside biasing contact acting as an additional 
field shaping potential of the device.
\end{itemize}
Interest in CMOS pixels for HL-LHC has been aroused by its potential for low cost and
the feasibility of large area monolithic devices in outer tracker layers when stitching is employed, but also out of intellectual curiosity about whether a one-piece pixel detector can be developed for HL-LHC environments demands.

\subsubsection{DMAPS capacitances}
\begin{figure}
\centering
   \subfigure[DMAPS capacitances]{\raisebox{0.0cm}
            {\includegraphics[width=0.53\linewidth]{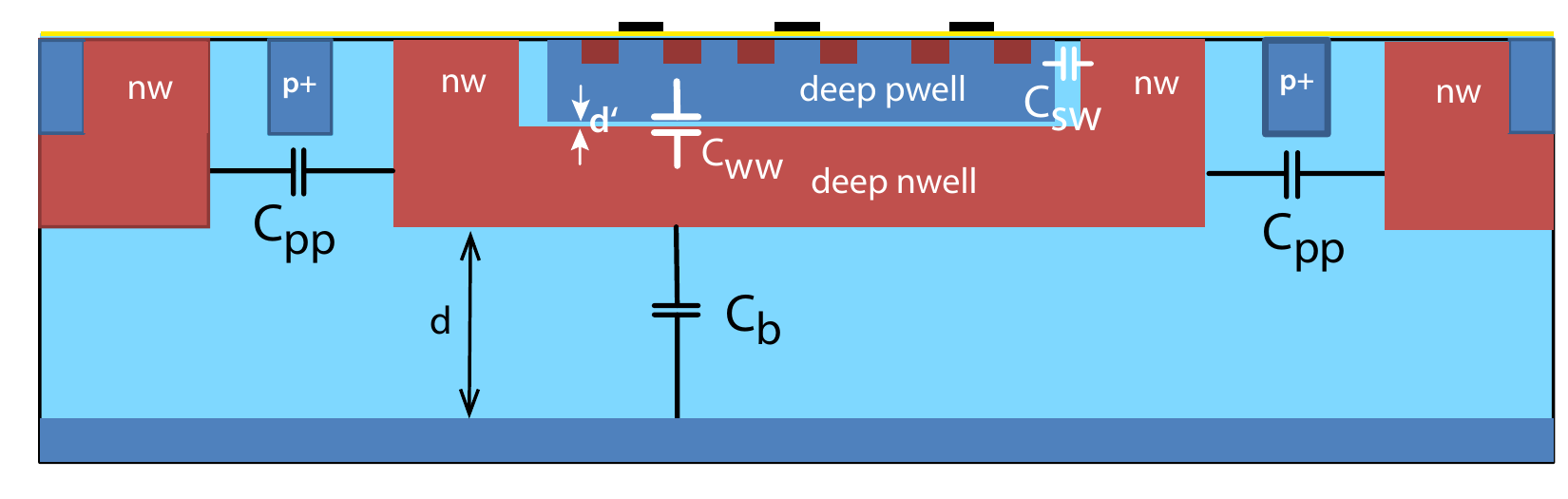}}\label{fig:DMAPS_caps_a}}\hskip 0.2cm
   \subfigure[Coupling]{\raisebox{0.3cm}
            {\includegraphics[width=0.43\textwidth]{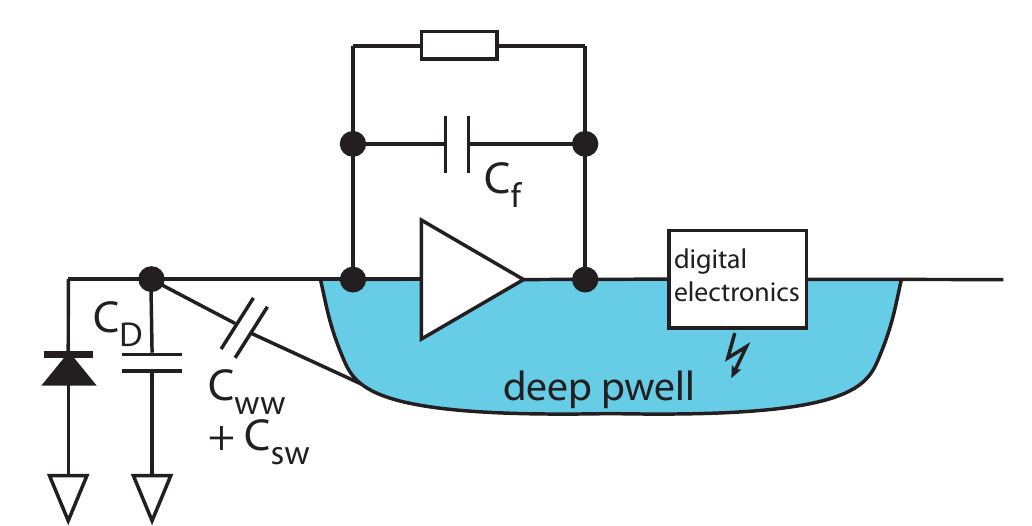}}\label{fig:DMAPS_caps_b}}
\caption{(a) Capacitance contributions of a typical CMOS pixel cell to the amplification node. Apart from capacitances to the backside and between pixels a non-negligible contribution comes from the close-by wells, deep n-well and p-well. (b) Through the capacitance between p- and n-well digital signals can couple into the sensor bulk if no particular care is taken to prevent this.}
\label{fig:CMOS_capacitances}
\end{figure}

Figure~\ref{fig:CMOS_capacitances} shows the main capacitances in a DMAPS design that contribute to the total
amplifier input capacitance of a pixel cell. In addition to the pixel-to-pixel (C$_{pp}$) and pixel-to-backside (C$_b$) capacitances, also present in any other sensor design, inter-well capacitances to the well sides (C$_{SW}$) and between
deep n-well and p-well (C$_{WW}$) play a significant role. In particular, if the deep n-well is large (large fill-factor, see below) C$_{WW}$ can achieve significant values (as large as 100\,fF for 10\,000\,$\upmu$m$^2$ pixel area) increasing the total capacitance.
This is to be compared to typical hybrid pixel capacitances of $C_D \approx$\,120\,fF (planar) and $C_D \approx$\,180\,fF (3D-Si) \cite{Havranek:2013xkn}.
A large amplifier input capacitance $C_D$ directly enters the thermal noise (and also 1/f noise) figures of a pixel detector with CSA readout and shaping as well as the detector response time $\tau_{CSA}$:
\begin{equation}\label{eq:pixel_noise}
  ENC^2_{thermal} \propto \frac{4}{3}\frac{kT}{g_m} \frac{C_D^2}{\tau_{sh}} \hskip 1.0cm {\rm and} \hskip 1.0cm  \tau_{CSA} \propto \frac{1}{g_m}\frac{C_D}{C_f}
\end{equation}
where $C_f$ is the feedback capacitance, $g_m$ the transconductance, $\tau_{sh}$ the shaping time, 
and $kT$ is the Boltzmann constant times temperature.

While it may be possible to cope with additional input capacitance as a noise factor, e.g.\,\,by increasing $g_m$ (which increases power), the inter-well capacitance
also couples the CMOS electronics with the sensor volume as illustrated in fig.~\ref{fig:DMAPS_caps_b}.
This requires careful circuit design and special measures to prevent digital activity from coupling to the sensor part, thus faking particle signals. An alternative is to place the digital logic away from the active sensor part (n-well), 
at the price of a smaller fill-factor (see below).

\subsubsection{Fill-factor choices}
\begin{figure}
\centering
   \subfigure[Large fill-factor]{\raisebox{0.0cm}
            {\includegraphics[width=0.45\linewidth]{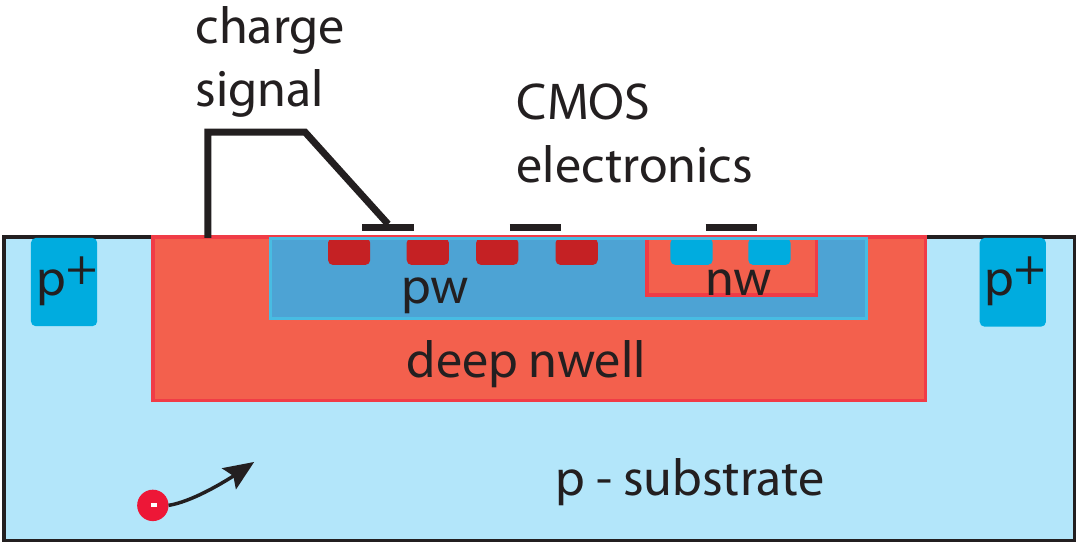}}\label{fig:large_FF}}\hskip 0.6cm
   \subfigure[Small fill-factor]{\raisebox{0.0cm}
            {\includegraphics[width=0.45\textwidth]{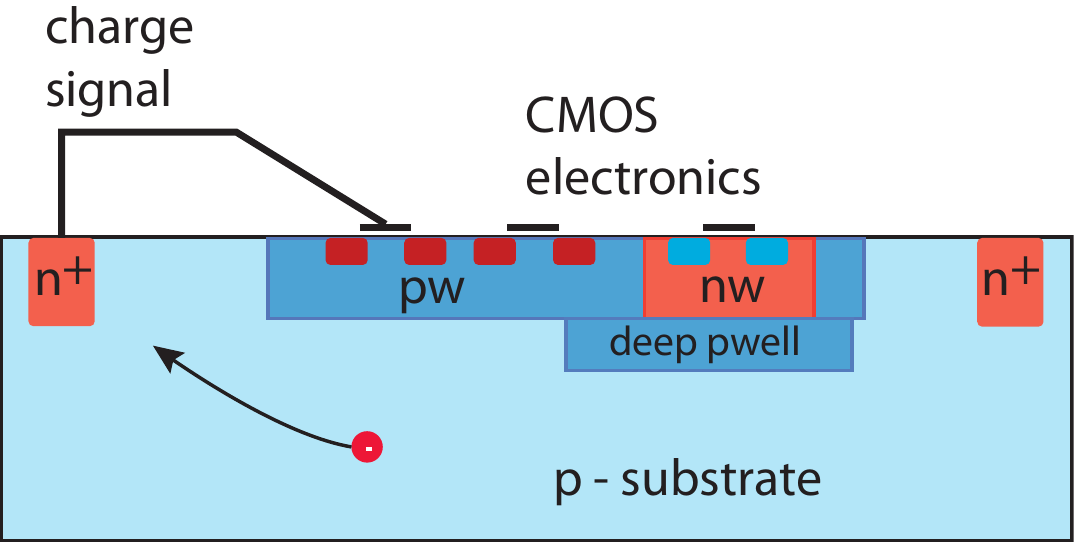}}\label{fig:small_FF}}
\caption{Two principal variants of CMOS cell geometries: (a) Large fill-factor: the charge collecting deep n-well encloses the complete CMOS electronics. (b) Small fill-factor: the charge collection node is placed outside the CMOS electronics area.}\label{fig:fill-factor}
\end{figure}
Figure~\ref{fig:fill-factor} shows two principal variants of a pixel cell arrangement. While in fig.~\ref{fig:large_FF} the entire CMOS electronics is enclosed in the deep, charge collecting n-well, in fig.~\ref{fig:small_FF} the deposited charge is collected at a small n-well located outside the electronics area.
A large fill-factor provides good charge collection properties over the entire pixel area with on average shorter travel distances and hence smaller trapping
probabilities after irradiation, but it suffers from a comparatively large inter-well capacitance contribution
($\sim$\,100\,fF) as discussed above leading to larger noise figures and slower timing. Increased power (compared to hybrid pixels) 
for the same pixel area is needed to cope with the larger capacitance.
Most developments so far have chosen this variant to minimize radiation hardness issues, the main challenge at the HL-LHC. 
Note that with the advancement of CO$_2$ cooling a change in the power bill does not linearly translate into material. 
Thermal conductivity of the components is the key parameter (see also section~\ref{sec:SupCooling}).
For example, for a cooling tube embedded in a carbon foam, increasing (decreasing) the power by a factor of 
2 leads to a material increase (decrease) of 10 or 20\%~\cite{EAndersen_oct2016}.

The small fill-factor variant on the other hand promises node capacitances of only 5--20\,fF and hence excellent
noise and timing performance. However, the radiation tolerance is an issue, given that on average the drift distances
of signal charges are longer for same cell size. Small pixel sizes are therefore beneficial for a small fill-factor design at the expense of power density.

\subsubsection{Substrate resistivity}
\begin{figure}
    \centering
            \includegraphics[width=0.7\textwidth]{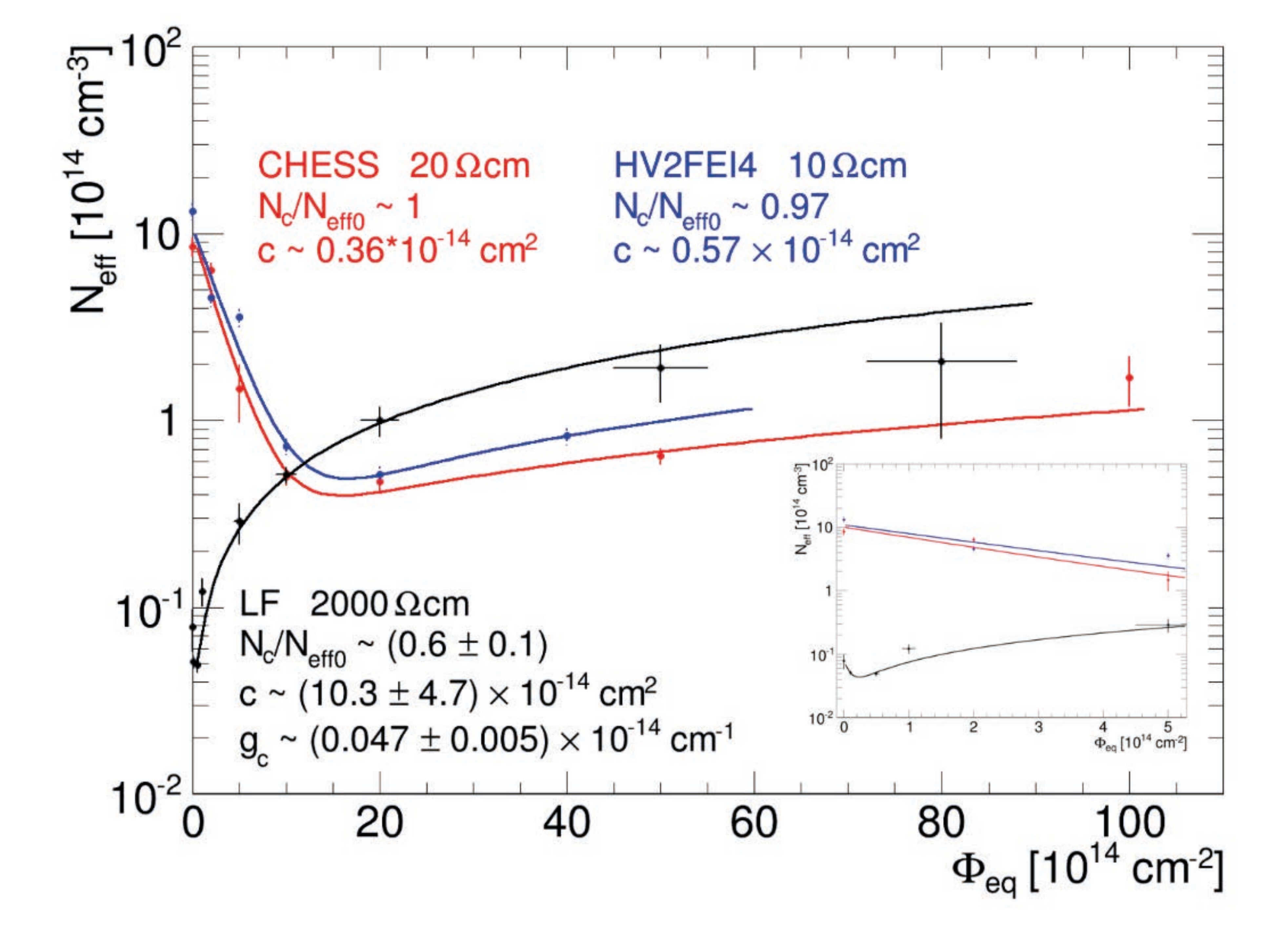}
    \caption{Effective space charge concentration measured after neutron irradiation for different starting materials
    varying from 10\,k$\Omega$cm to 2\,k$\Omega$cm as a function of the neutron fluence \cite{Mandic:2017tgh}. The insert shows the same curves at low fluences.} \label{fig:CMOS_resistivity}
\end{figure}
According to eq.~\eqref{eq:depletion} high resistivity is a means to obtain adequate signal sizes by sufficient depletion depth at moderate bias. Figure~\ref{fig:CMOS_resistivity} \cite{Mandic:2017tgh}, however, shows that
independent of the starting substrate resistivity, over a large range of resistivities
(10\,$\Omega$cm--2\,k$\Omega$cm) one observes that after a fluence of about 10$^{15}$\,n$_{eq}$cm$^{-2}$ the effective space charge concentrations approximate each other at a level of about $N_{\rm eff} \approx 10^{14}$\,cm$^{-3}$ corresponding to about 100\,$\Omega$cm.
This effect is commonly attributed to radiation induced acceptor removal, setting in earlier at higher
p-doping concentrations and eventually leading to similar $N_{\rm eff}$ concentrations.
Therefore, for high radiation applications, the starting resistivity is not very critical, and in fact an intermediate (100\,$\Omega$cm - 1\,k$\Omega$cm), not too high initial resistivity may be advantageous as this will lead to a fairly constant charge collection throughout the lifetime.

\subsection{Designs and technology variants}\label{sec:designs}
The high energy physics community has targeted different prototyping designs with various foundries providing
standard CMOS technology add-ons necessary to cope with the given demands. The emphasis at the start was either
on dedicated high voltage technologies \cite{Peric:2014faa} or on technologies accepting high resistivity substrate wafers for processing \cite{Havranek:2015}. It turned out that both, mid to high resistivity ($\lesssim 1$\,k$\Omega$cm) as well as sufficient bias ($\gtrsim$ 150\,V), are needed for good performance under irradiation.
The general approach of different groups has been in three prototyping steps.
\begin{enumerate}
  \item Simple prototypes to characterize technology features and charge collection performance.
  \item Large pixel arrays with stand-alone readout as well as readout via a dedicated pixel readout chip (usually FE-I4 \cite{FE-I4}) bonded to it.
  \item Large, fully monolithic CMOS pixel matrices including on-chip digital readout architecture for rates and occupancies expected at the HL-LHC.
\end{enumerate}
The development is currently progressing fast. One can say that steps 1 and 2 have been successfully carried out, and different designs for step 3 have been fabricated and are under characterization, but not yet published at the time of writing.  

\subsubsection{Hybrid active CMOS pixels}
A variant under step 2 exploits usage of a \glq smart CMOS sensor\grq\ coupled to a readout chip like FE-I4 \cite{FE-I4}. 
In this case the CMOS sensor provides a first stage of analog pulse processing (usually a CSA preamplifier plus discriminator) 
leading to an output voltage pulse which is DC or AC coupled to the FE-I4 analog input. 
This approach is commonly called \glq capacitively coupled pixel detector\grq , CCPD \cite{Peric:2010zz}).
While it allows for some extra freedom to explore new functionality such as the implementation of subpixel 
decoding described in chapter~\ref{sec:resolution}, it is basically an alternative hybridization approach,  
and so must be compared to traditional hybrid pixels in terms of material budget and power performance.

\subsubsection{DMAPS in HV/HR technologies with large fill-factor}
\begin{figure}
\centering
   \subfigure[CMOS w/ HV add-on]{\raisebox{0.5cm}
            {\includegraphics[width=0.46\linewidth]{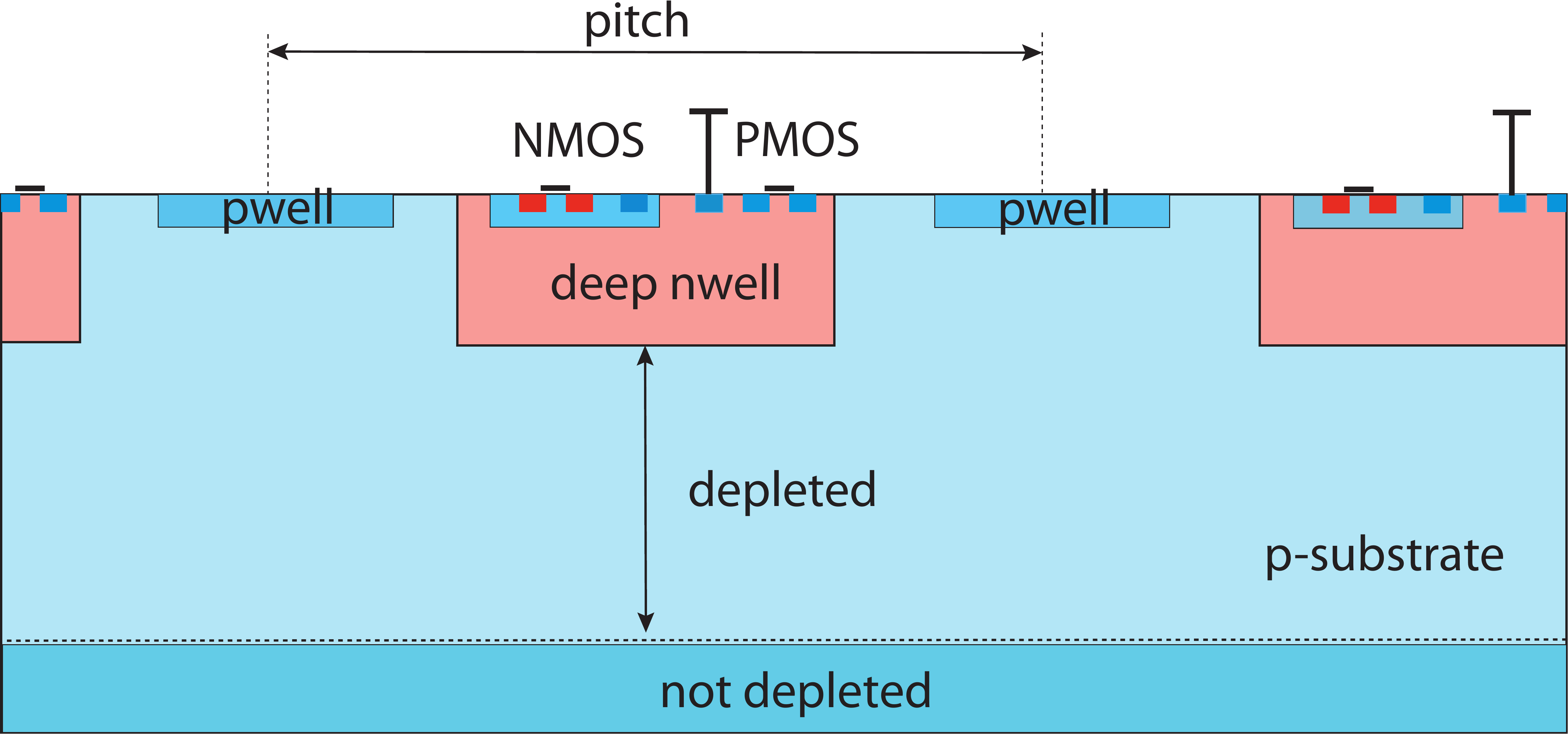}}\label{fig:AMS180}}\hskip 0.3cm
   \subfigure[CMOS on SOI]{\raisebox{0.0cm}
            {\includegraphics[width=0.51\textwidth]{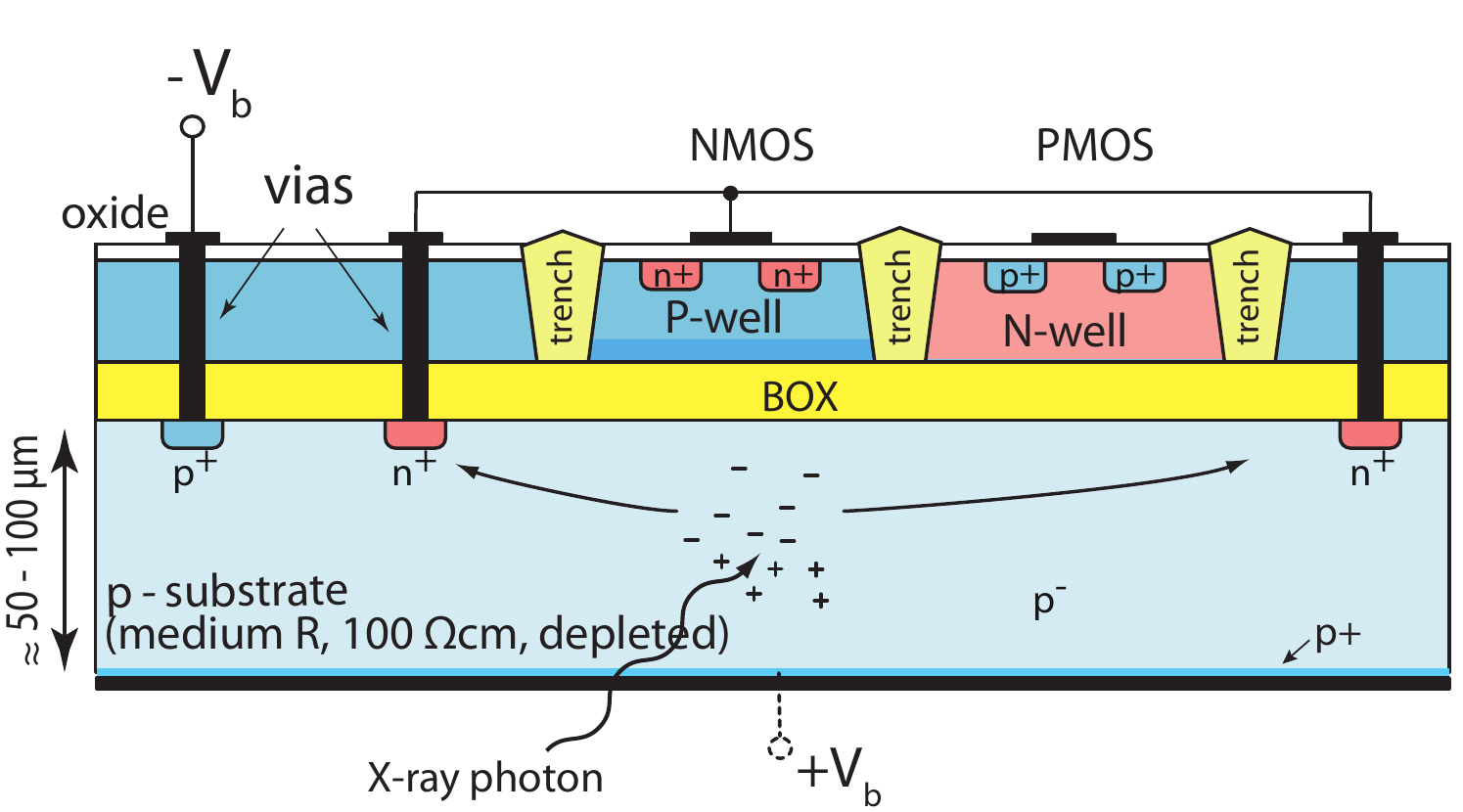}}\label{fig:CMOS_SOI}}
\caption{(a) Pixel layout in 180\,nm HV technology (AMS). The triple well technology contains both transistor flavours in a deep n-well acting as the charge collecting node at the same time. (b) DMAPS realisation employing SOI technology. A buried oxide layer (BOX) separates the depleted sensor substrate from the electronics layer. Vertical vias connect electronics to the bulk. Trench isolations shield individual transistors. Partially depleted (thick film) n-well and p-well regions prevent the \glq back-gate\grq\ effect.}\label{fig:designs}
\end{figure}

CMOS pixels with some depletion depth were first implemented in \cite{Peric:2007zz} using a dedicated HV technology (called HVCMOS) with 350\,nm and later with 180\,nm feature sizes (AMS H35 and H18) allowing for up to three nested wells. Both PMOS and NMOS transistors sit in a large deep n-well which at the same time acts as the charge collection node (large fill-factor, see fig.~\ref{fig:AMS180}). Because the PMOS transistors' n-well is identical to the collection node, care must be taken in its usage in the electronics design limiting somewhat the CMOS functionality.
The achievable depletion depth is around 15--30\,$\upmu$m for low resistivity substrates around 10\,$\Omega$cm. Higher
resistivities up to about 1\,k$\Omega$cm are under study.
HVCMOS prototypes have been tested to TIDs well above 100\,Mrad and fluences up to 5$\times$10$^{15}$ n$_{eq}$/cm$^2$ \cite{Benoit:2016vup}. The most critical performance metric
is the so-called in-time efficiency of hits after irradiation which includes both radiation damage characterizations as well as rate demands at the LHC. The term
\emph{in-time} means that a hit must be time-stamped within one LHC bunch-crossing window of 25\,ns.
Test beam characterization of prototype matrices coupled to the FE-I4 pixel chip via adhesive bonding yielded
efficiencies of above 97\% (99.7\% up to 1$\times$10$^{15}$n$_{eq}$\,cm$^{-2}$) and in-time efficiencies comparable
to those of FE-I4 hybrid pixel modules used in the same test beam 
($>$95\% within three beam bunch crossings of 25\,ns)~\cite{Benoit:2016vup}.

\begin{figure}
    \centering
            \includegraphics[width=0.7\textwidth]{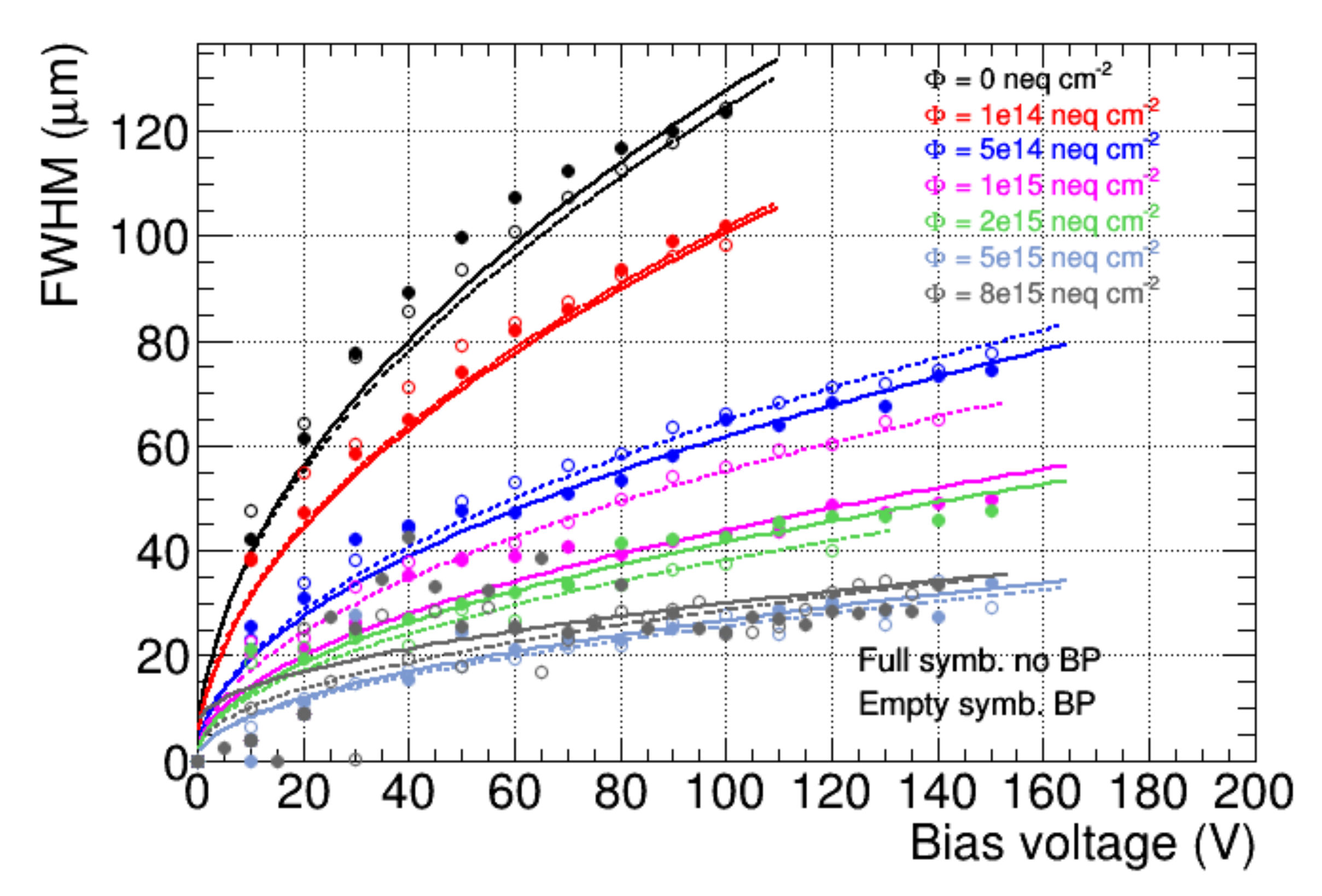}
    \caption{Depletion depth measured using edge TCT for high resistivity (2--3\,k$\Omega$cm) CMOS pixel devices \cite{Mandic:2017tgh}. Plotted is the extracted depletion depth as a function of bias voltage, in seven increasing irradiation steps (up to 8$\times$10$^{15}$n$_{eq}$\,cm$^{-2}$). Open symbols denote devices with and full symbols without backplane contact.} \label{fig:CMOS_depletion}
\end{figure}

Prototype matrices were tested up to fluences of 1$\times$10$^{15}$n$_{eq}$\,cm$^{-2}$) 
and TIDs of 50\,Mrad \cite{Hirono:2016zck} using an approach that differs from the above mainly in two aspects, 
(i) use of a quadruple well 150\,nm process (LFoundry LF15A) as in fig.~\ref{fig:DMAPS_generic}, 
and (ii) use of high resistivity substrate wafers (2--3\,k$\Omega$cm). 
This technology allows for bias voltages in the 150--200\,V range. 
The devices performed well in test beams, showing high time integrated efficiencies,
but not yet with fast enough timing (1 bunch crossing in-time efficiency $<$\,91\%) owing to the large capacitance
inherent in the large fill-factor approach. 
The depletion depth of these devices has been measured 
for increasing radiation fluence \cite{Mandic:2017tgh} (up to 8$\times 10^{15}$n$_{eq}$\,cm$^{-2}$, 
fig.~\ref{fig:CMOS_depletion}) using edge TCT (\emph{transient current technique}, 
see for example \cite{Kramberger:2009fua}), which measures the transient current generated by a laser pulse parallel to the 
CMOS sensor surface, entering the bulk through a diced edge. 
As a function of bias voltage, the square root shape of eq.\,\eqref{eq:depletion} is observed, 
clearly showing that depletion depth values of 30--50\,$\upmu$m are maintained even for fluences beyond 10$^{15}$n$_{eq}$\,cm$^{-2}$.
%

Fully monolithic versions in these technologies (AMS and LFoundry) have been fabricated and are currently being characterized \cite{Wang:2016dgi,Peric:2015ska}.

\subsubsection{CMOS on SOI}
The SOI technology provides a buried oxide layer (BOX) separating the
CMOS electronics from the substrate layer. Both parts are connected by vertical via structures reaching through the BOX and leading to an n-implant which acts as the charge collecting node. Monolithic SOI-based pixel structures have been developed for some time using the fully depleted (FD) SOI technology \cite{Arai:2011ara}, invented for high speed CMOS electronics with reduced (parasitic) capacitances.  The CMOS electronics layer is embedded
in depleted silicon. The developments in this FD-SOI technology, however suffer from effects inherent to the BOX oxide layer, most notably the so-called back-gate effect and from radiation effects \cite{Arai:2011ara} that can be compensated to some degree \cite{Miyoshi:2014qha} for moderate radiation doses, but not for the radiation conditions expected at the HL-LHC.

Thick film SOI, however, featuring trench isolation and doped, only partially depleted regions underneath CMOS transistors, thus shielding the transistors (cf.~fig.~\ref{fig:CMOS_SOI}), is free of these difficulties and can also sustain much larger radiation doses. SOI CMOS pixel detectors have been realised~\cite{Hemperek:2014yoa} and characterised~\cite{Fernandez-Perez:2015gua} in the lab and in test beams. The devices show impressive TID tolerances tested up to 1\,Grad. Also here the substrate resistivity initially increases with hadron fluence due to acceptor removal. The measured in-time efficiency also needs modest improvement in order to cope with LHC demands.

While the CMOS on SOI approach towards monolithic pixel detectors remains very interesting, not the least because of the attractive approach of separating sensing and electronics volume by a buried oxide layer, it is currently not
the main focus of the HL-LHC developments.

\subsubsection{DMAPS with small fill-factor}
%
\begin{figure}
    \centering
            \includegraphics[width=0.6\textwidth]{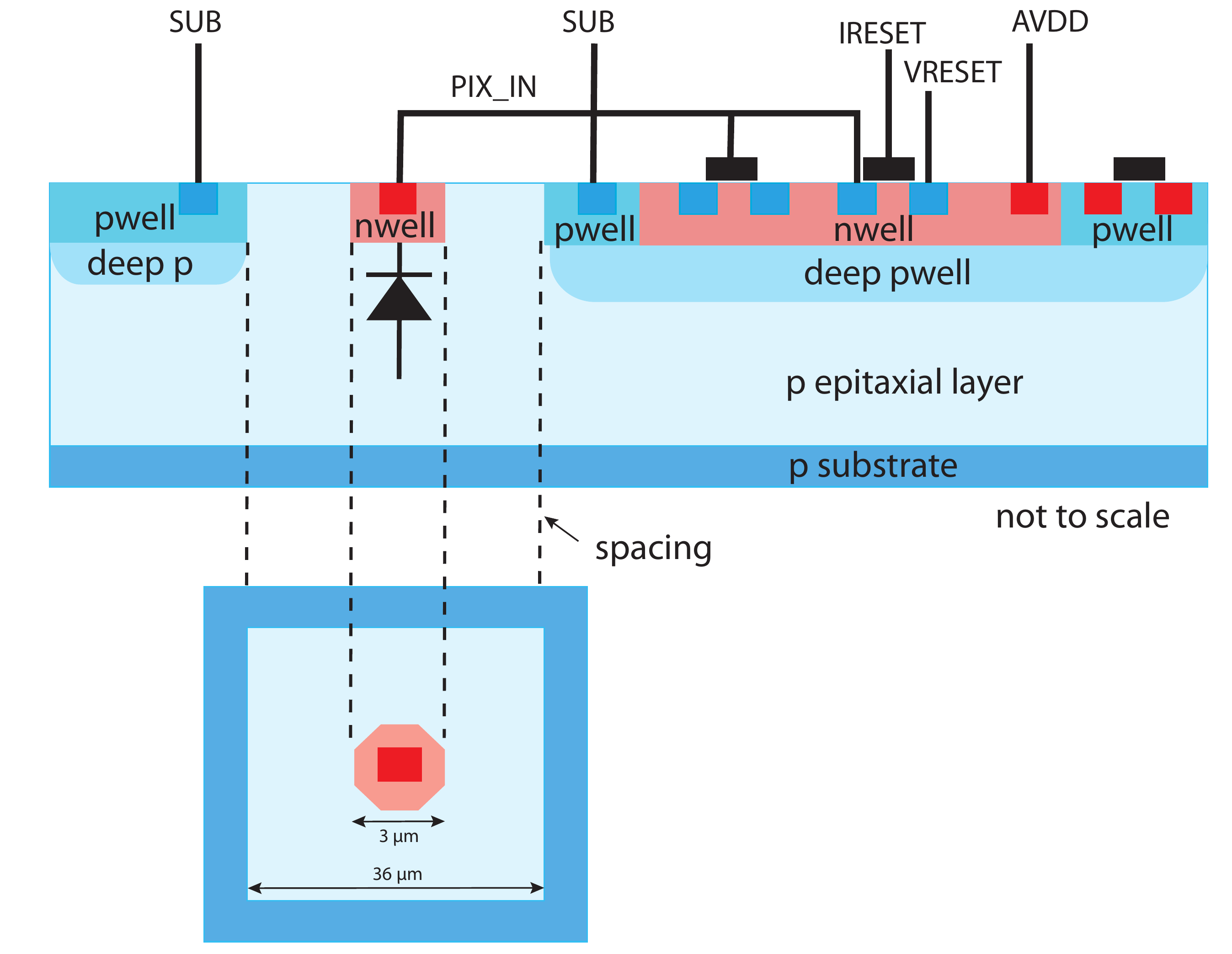}
    \caption{Small fill-factor design of a pixel cell in a 180\,nm quadruple well technology (TJ 180nm) \cite{Pernegger2017:yyy}. Note that the electronics circuitry is fully enclosed in a large deep p-well while charge is collected by a small n-well collection node.} \label{fig:TJ_cell}
\end{figure}
\sloppy
A recent development aims to achieve sufficient radiation hardness with a small fill-factor design,
promising low noise and fast timing due to the small resulting input capacitance. The development inherits from the
design of the ALICE ITS pixel chip (ALPIDE) development \cite{Aglieri:2013xma,Mager:2016yvj} introducing some technology improvements for better charge collection after irradiation \cite{Gao:2016zut}.
Figure~\ref{fig:TJ_cell} shows a principal cell layout.
The charge collection is obtained in a high ohmic ($>$\,1\,k$\Omega$cm) epitaxial layer. The
quadruple well features are very similar to those in fig.~\ref{fig:DMAPS_generic}. The main
difference is the layout of the charge collection node placed far outside the electronics area with minimal geometric
size. The capacitance is very low: 3--10\,fF \cite{Hoorne_PhD_2015}, depending on the geometry and pitch of the collection node. 
The pixel area must be small ($\lesssim 40\upmu$m) in order to limit the maximum path length for charge collection.

In a first prototype chip (TJ Investigator) the main features of such a design have been investigated \cite{Pernegger2017:yyy} confirming the low capacitance and the good timing precision ($\tau_{\rm rise}\sim 16$\,ns).
To obtain a fully depleted volume the TJ process was modified replacing the epitaxial p-layer by a
planar deep n$^+$-p junction \cite{Pernegger2017:yyy}. A fully monolithic chip including readout architecture has been recently submitted for fabrication with expected noise figures of 16--20\,e$^-$ and similar timing precision. 

\subsection{Outlook on DMAPS pixels}
The development of depleted CMOS active pixels currently has a large momentum in the high energy physics community.
Many groups are active in de\-velo\-ping DMAPS sensors and also DMAPS based modules.
The specific developments discussed above are not an exhaustive survey and we are aware of other manufacturers and competing efforts that have not yet published results. Significant new material is therefore expected within 1 year of this review. 
Whether the development can be brought to a level of maturity sufficient to be adopted for (parts of) an HL-LHC pixel detectors
remains to be seen. Regardless of the outcome for HL-LHC, researchers in this field expect that CMOS active pixels will be 
the preferred technology for future, large area, high rate and radiation trackers

%% file: conclusions.tex
\section{Summary and conclusions}\label{sec:conclusions}
With the planned high luminosity upgrades of the LHC machine and experiments,
pixel detectors operating in high rate and radiation environments are facing yet another
extreme challenge, surpassing that of the current LHC by roughly an order of magnitude.
While integrated (monolithic) pixel detectors have been developed already for some
time (and are in fact used in detectors under less hostile experimental conditions), 
as of today hybrid pixel detectors still
constitute the state of the art in tracking detector technology close to the interaction point of such
high rate and radiation experiments.

In this paper we have reviewed the fundamentals of track space-time point and direction measurements
and described development routes and choices to be made for the next generation HL-LHC experiments.
These address all components of the hybrid pixel technology, most prominently the sensor and the readout chip,
but also interconnection and 3D-stacking techniques, and lightweight support and cooling structures.
Beyond the traditional space point measurement, ways to extract and use additional information from pixel detectors are being explored,
such as space vector information or precision timing information with the promise of so-called 4D tracking.

Silicon sensors capable of good performance after fluences of $10^{16}$\,n$_{eq}$cm$^{-2}$ and beyond have been developed (sections~\ref{sec:sensors_planar} and \ref{sec:sensors_3D}). Both 3D-Si sensors and conventional planar sensors have been shown to be able to cope with the demands. 3-D sensors after further optimization following their successful performance in the IBL pixel upgrade of 
ATLAS~\cite{Darbo:2014kma}, and planar sensors after reducing thickness and optimizing guard ring structures. 

A major challenge is the development of a next generation of pixel readout chips which must cope with very harsh radiation conditions (total doses up to 1\,Grad), with particle rates $>3$\,GHz/cm$^2$, and with MHz trigger rates leading to data output bandwidths 
of ${\cal{O}}$(10\,Gb/s).
The development of a close to 10$^9$ transistor readout chip is jointly addressed by ATLAS and CMS 
through the RD53 collaboration \cite{RD53A-specs} using deep submicron 65\,nm technology. 
Narrow and short transistor channel effects lead to complex radiation damage behavior 
that demands sophisticated modelling to be properly addressed (section~\ref{sec:ROIC-damage}). 
An approach dubbed \glq analog islands in a digital sea\grq\ is employed to process the very 
large rates on chip. 
The RD53 chip will serve as the workhorse for the HL-LHC hybrid pixel detectors of the experiments.

Advances in evaporative cooling and in using lightweight yet stiff composite materials render possible a material budget reduction per pixel layer from previously above 3\% to below 1\%, even for power and cooling intensive high
rate and radiation applications (chapter~\ref{sec:supports}). Coolant choices have narrowed down to CO$_2$
as today's state of the art. Very small diameter pipes (micro channels) are an attractive development path
for high cooling capacity.

Finally, the monolithic pixel module ansatz is being revived by employing multiple well technologies on high resistivity substrates that have become more commercially accessible to the HEP community and now offer good 
performance even under HL-LHC requirements (chapter~\ref{sec:cmos}). 
This has also launched R{\&}D using CMOS technology lines for (passive) 
sensors in hybrid pixel detectors (section~\ref{sec:passive_CMOS}), offering advantages as low cost, 
capacitive coupling of electrodes as well as efficient routing between sensor and 
ROIC pixels by exploiting CMOS metal layers.

%% file: acknowledge.tex
\section*{Acknowledgements}
The authors are grateful to many colleagues for their excellent published work and ideas, and for many discussions and other contributions.
This work was supported by the 
Office of High Energy Physics of the U.S. Department of Energy under contract DE-AC02-05CH11231, 
by the Deutsche Forschungsgemeinschaft DFG, grant number WE 976/4-1, 
and by the German Ministry for Research BMBF under grant number 05H15PDCA9.